\documentclass[preprint, trackchanges]{aastex702} % linenumbers preprint
\usepackage{multirow, url, color}
\usepackage{bbding, amssymb} % symbols
\usepackage{ulem} % underline for emphasis
\usepackage{verbatim} % display LaTeX commands
\usepackage{enumerate}
\usepackage{amsmath} % \usepackage{txfonts}
\usepackage{booktabs} % Table Top/Mid/Bottom Rule
\usepackage{arydshln, bbding} % dashed line
\usepackage{graphicx, subfigure}
\usepackage[fleqn]{mathtools}
\usepackage{mathrsfs}
\usepackage[makeroom]{cancel}
\newcommand{\bs}[1]{\boldsymbol{#1}}

\shorttitle{Simulating the 2017 September 10 SEP Event}
\shortauthors{Liu et al.}
\received{September 30, 2026}
\revised{November 6, 2026}
\accepted{November 13, 2026}
\submitjournal{ApJ}
\begin{document}

\title{Comparative Analysis of Influences of Magnetic Field Inputs on Simulating\\ the 2017 September 10 Solar Energetic Particle Event}

\correspondingauthor{Weihao Liu}
\email[show]{whliu@umich.edu}

\author[orcid=0000-0002-2873-5688]{Weihao Liu}
\affiliation{Department of Climate and Space Sciences and Engineering, University of Michigan, Ann Arbor, MI 48109, USA}
\email{whliu@umich.edu}
\author[orcid=0000-0001-9114-6133]{Nishtha Sachdeva}
\affiliation{Department of Climate and Space Sciences and Engineering, University of Michigan, Ann Arbor, MI 48109, USA}
\email{nishthas@umich.edu}
\author[orcid=0000-0003-3936-5288]{Lulu Zhao}
\affiliation{Department of Climate and Space Sciences and Engineering, University of Michigan, Ann Arbor, MI 48109, USA}
\email{zhlulu@umich.edu}
\author[orcid=0000-0002-6118-0469]{Igor V. Sokolov}
\affiliation{Department of Climate and Space Sciences and Engineering, University of Michigan, Ann Arbor, MI 48109, USA}
\email{igorsok@umich.edu}
\author[orcid=0000-0001-7075-6530]{Hongfan Chen}
\affiliation{Department of Mechanical Engineering, University of Michigan, Ann Arbor, MI 48109, USA}
\email{chenhf@umich.edu}
\author[orcid=0000-0002-1604-3326]{Christina O. Lee}
\affiliation{Space Sciences Laboratory, University of California Berkeley, Berkeley, CA 94720, USA}
\email{clee@ssl.berkeley.edu}
\author[orcid=0000-0001-9016-8381]{Patrick A. Dunn}
\affiliation{Space Sciences Laboratory, University of California Berkeley, Berkeley, CA 94720, USA}
\email{patrickadunn@berkeley.edu}
\author[orcid=0000-0001-5191-1662]{Alessandro Bruno}
\affiliation{Heliophysics Science Division, NASA Goddard Space Flight Center, Greenbelt, MD 20771, USA}
\affiliation{Department of Physics, Catholic University, Washington, DC 20064, USA}
\email{alessandro.bruno-1@nasa.gov}
\author[orcid=0000-0002-3176-8704]{David Lario}
\affiliation{Heliophysics Science Division, NASA Goddard Space Flight Center, Greenbelt, MD 20771, USA}
\email{david.larioloyo@nasa.gov}
\author[orcid=0000-0002-3787-1622]{Kathryn Whitman}
\affiliation{Space Radiation Analysis Group, NASA Johnson Space Center, Houston, TX 77058, USA}
\email{kathryn.whitman@nasa.gov}
\author[orcid=0009-0009-2176-6017]{Xianyu Liu}
\affiliation{Department of Climate and Space Sciences and Engineering, University of Michigan, Ann Arbor, MI 48109, USA}
\email{xianyu@umich.edu}
\author[orcid=0000-0001-8459-2100]{G\'abor T\'oth}
\affiliation{Department of Climate and Space Sciences and Engineering, University of Michigan, Ann Arbor, MI 48109, USA}
\email{gtoth@umich.edu}
\author[orcid=0000-0001-9360-4951]{Tamas I. Gombosi}
\affiliation{Department of Climate and Space Sciences and Engineering, University of Michigan, Ann Arbor, MI 48109, USA}
\email{tamas@umich.edu}
\author[orcid=0000-0002-9672-3873]{Meng Jin}
\affiliation{Lockheed Martin Solar and Astrophysics Laboratory, Palo Alto, CA 94304, USA}
\email{jinmeng@lmsal.com}
\author[orcid=0000-0002-7388-173X]{Robert F. Wimmer-Schweingruber}
\affiliation{Institute of Experimental and Applied Physics, Christian-Albrechts-University Kiel, 24118 Kiel, Germany}
\email{wimmer@physik.uni-kiel.de}
\author[orcid=0000-0003-0458-4050]{Robin Ramstad}
\affiliation{Laboratory for Atmospheric and Space Physics, University of Colorado, Boulder, CO 80303, USA}
\email{robin.ramstad@lasp.colorado.edu}
\author[orcid=0000-0001-5258-6128]{Jasper S. Halekas}
\affiliation{Department of Physics and Astronomy, University of Iowa, Iowa City, IA 52242, USA}
\email{jasper-halekas@uiowa.edu}
\author[orcid=0000-0001-9127-7133]{Claudio Corti}
\affiliation{Heliophysics Science Division, NASA Goddard Space Flight Center, Greenbelt, MD 20771, USA}
\affiliation{Universities Space Research Association, Washington, DC 20024, USA}
\affiliation{Physics and Astronomy Department, University of Hawaii at Manoa, Honolulu, HI 96822, USA}
\email{corti@hawaii.edu}
\author[orcid=0000-0002-7056-3517]{Yoshifumi Futaana}
\affiliation{Swedish Institute of Space Physics, SE-98128 Kiruna, Sweden}
\email{futaana@irf.se}
\author[orcid=0000-0001-5030-6030]{Davin E. Larson}
\affiliation{Space Sciences Laboratory, University of California Berkeley, Berkeley, CA 94720, USA}
\email{davin@ssl.berkeley.edu}
\author[orcid=0000-0001-8830-1200]{~Donald M. Hassler}
\affiliation{Solar System Science and Exploration Division, Southwest Research Institute, Boulder, CO 80302, USA}
\email{hassler@boulder.swri.edu}
\author[orcid=0000-0002-5956-5722]{Bent Ehresmann}
\affiliation{Solar System Science and Exploration Division, Southwest Research Institute, Boulder, CO 80302, USA}
\email{ehresmann@boulder.swri.edu}
\author[orcid=0000-0002-7463-9419]{Shannon Curry}
\affiliation{Laboratory for Atmospheric and Space Physics, University of Colorado, Boulder, CO 80303, USA}
\email{Shannon.Curry@lasp.colorado.edu}

\begin{abstract}
Physics-based solar energetic particle (SEP) models simulate the ambient solar wind, coronal mass ejections (CMEs), and SEPs using photospheric magnetic field maps as the inner boundary condition. In this study, we select the 2017 September 10 SEP event, which produced the first ground level enhancement observed on the surfaces of both Earth and Mars, to examine how these inputs influence the modeled results. 
We use the SOlar wind with FIeld lines and Energetic particles (SOFIE) model suite, driven independently by the GONG and ADAPT-GONG magnetic field maps, to simulate the solar wind, CME, and SEPs from the low corona to beyond Mars' orbit. 
The two maps yield different open flux regions and hence different solar wind conditions, both globally comparable to \textit{in situ} observations at Earth, STA, and Mars, with the GONG-driven results performing slightly better. We find that in the ADAPT-GONG-driven run, a more energetic CME flux rope propagates into a slower, wider stream, producing broader CME expansion and a more extended shock, which reaches both planets within $\sim$4 hours of the observed arrival, while the GONG-driven run misses both arrivals. 
Both simulations reproduce the observed SEP fluxes at Earth within a factor of $\sim$3, while the ADAPT-GONG-driven results compare better with observations at Mars, which connects magnetically to the shock flank. 
Therefore, for this event, the relative performance of the two simulations depends on the observer and observable, with SEP fluxes shaped by CME propagation through the ambient solar wind and by magnetic connectivity to the evolving shock. 
\end{abstract}

% See \url{https://astrothesaurus.org/concept-select/} for \keywords{...} when needed.}

\section{Introduction} \label{sec1:intro}

    \subsection{Space Weather and Solar Energetic Particles in the Sun--Earth--Mars System} \label{sec1.1:sepsw}

        Solar eruptive activity drives space weather throughout the inner heliosphere \citep{schwenn2006space, temmer2021space}. Flares deliver intense X-ray and extreme-ultraviolet radiation to planetary atmospheres within minutes \citep[e.g.,][]{tsurutani2009brief}; coronal mass ejections \citep[CMEs;][]{chen2011coronal, manchester2017physical} drive interplanetary shocks and trigger intense geomagnetic storms and substorms; and solar energetic particles (SEPs) accelerated by CME-driven shocks and in association with solar flares spread penetrating radiation across the heliosphere \citep{reames1999particle, desai2016large}. The effects of a single eruption typically reach a planet on three timescales: flare photons within minutes, the first near-relativistic protons within tens of minutes at a well-connected observer and later elsewhere, and the CME-driven shock and ejecta one to a few days later. 
        
        Among these space weather phenomena, SEPs are the main radiation exposure concern, as energetic particles can penetrate spacecraft shielding and planetary atmospheres and deliver substantial doses over short timescales \citep{minow2020space, lowe2025nowcasting}. Protons $\gtrsim 100$ MeV can generally penetrate spacecraft shielding, while the highest-energy SEP protons, typically $> 500$ MeV \citep[e.g.,][]{maurin2015neutron}, can generate secondary particle cascades in Earth's magnetosphere and atmosphere that are detected at the ground as ground level enhancements \citep[GLEs\footnote{\url{https://gle.oulu.fi} \label{url:gle}}; e.g.,][]{poluianov2017gle, crosby2024very}. The radiation hazard caused by an SEP event depends strongly on its intensity, duration, and energy spectrum, all of which can vary greatly among events and locations \citep[e.g.,][]{cane1988role, lario2013longitudinal, richardson201425, papaioannou2016solar, liu2023AGU, liu2024AGU}. 
        Major, long-duration SEP events are often associated with fast and wide CMEs capable of driving shocks that are efficient accelerators of particles, resulting in enhanced particle fluxes that can last for several days \citep[e.g.,][]{reames2013two, klein2017acc}. 
        
        In the past, SEPs as space weather hazards have been studied extensively from the Sun to Earth, where strong relationships have been established between the observed particle profiles, flares, CMEs, and magnetic connectivity \citep[e.g.,][and references therein]{anastasiadis2019solar, papaioannou2025exploring}. Multi-spacecraft measurements further show that the same solar eruption can produce very different SEP intensities and onset times at different heliolongitudes and heliocentric distances (see statistical analysis by, e.g., \citealp{reames1996spatial, lario2006radial, lario2013longitudinal, dresing2014statistical, richardson201425, richardson2026solar, paassilta2018catalogue, xie2019statistical, posner2024sep}, and case studies by, e.g., \citealp{kallenrode1993multi, malandraki2009energetic, gomez2015circumsolar, kollhoff2021first, kouloumvakos2022first, lario2022influence, lario2026major, palmerio2024improved, trotta2025variability, liu2026counterintuitive}). 
        Extending these studies beyond the Sun--Earth system and to heliocentric distances larger than 1 astronomical unit (au) is thus important for characterizing SEP radiation environments throughout the inner heliosphere and for evaluating potential radiation risks encountered by future missions to the Moon, Mars, and interplanetary space. 
        
        Mars is the primary destination beyond 1 au for studying space weather in the inner heliosphere. Orbiting the Sun between 1.38 and 1.67 au, observations at Mars provide an opportunity to investigate space weather effects well beyond Earth while also being a major target for future crewed exploration. Unlike Earth, however, Mars has no global intrinsic magnetic field, and its thin atmosphere ($\sim$20 g$\;$cm$^{-2}$) provides only limited shielding against energetic particles \citep[e.g.,][and references therein]{acuna1998magnetic, hassler2014mars, guo2021radiation}. 
        Radiation exposure by energetic particles is therefore an important consideration for both the transit to Mars and operations on the Martian surface. Decades of orbital and surface measurements have provided direct observations of the Martian radiation environment \citep[e.g.,][]{zeitlin2004overview, zeitlin2013measurements, hassler2014mars, lee2017maven, semkova2023observation, zhang2026constraint}, while models have been developed to connect the interplanetary particle environment to that at Mars orbit and on its surface and subsurface \citep[e.g.,][]{saganti2004radiation, schwadron2010earth, matthia2017radiation, sanchez2019origin, rostel2020subsurface, liu2023modeling, zhang2023radiation}. Together with observations near Earth and elsewhere in the inner heliosphere, these measurements make it possible to examine the same solar eruption at widely separated longitudes and heliocentric distances \citep[see][and references therein]{hapgood2019mars, lee2023heliophysics}. 

        The 2017 September 10 event stands out as one of the most exceptional SEP events observed over the past few solar cycles. It produced GLE \#72 at Earth, with ground-based neutron monitor count rates increasing by up to $\sim$10\%--15\% above the pre-event levels \citep[e.g.,][and Figure~\ref{fig1:overview}(d) below]{mavromichalaki2018real, mishev2018first}. In geostationary orbit around Earth, the Geostationary Operational Environmental Satellites (GOES) also recorded peak proton fluxes in the 110--900~MeV channel about 2--3 orders of magnitude above pre-event levels \citep[Figure~2(a) of][]{guo2018modeling}. 
        On the Martian surface, a strong particle enhancement was measured by the Radiation Assessment Detector \citep[RAD;][]{hassler2012radiation} on board the Mars Science Laboratory \citep[MSL;][]{grotzinger2012mars} on the Martian surface, making it the first SEP event identified as a GLE at both Earth and Mars. 
        It has also been reported that energetic protons with energies of $> 150$ MeV are required to penetrate the Martian atmosphere and produce the observed radiation enhancement at the surface \citep{guo2018generalized, hassler2018space}. During this event, the penetrating proton flux to the Martian surface increased by a factor of $\sim$3--30 and the absorbed dose rate rose up to 2--3 times the corresponding pre-event levels \citep[][and Figure~\ref{fig1:overview}(h) below]{ehresmann2018energetic, zeitlin2018analysis}. 
        This event, as detailed in Section \ref{sec2:overview}, provides a valuable opportunity to examine the same extreme SEP event at two widely separated planetary locations.

    \subsection{Physics-based Modeling and Magnetogram Inputs} \label{sec1.2:mapinput}
    
        Current SEP forecasting methods include both empirical/probabilistic approaches \citep[e.g.,][]{laurenza2009technique, nunez2015real, richardson2018prediction, papaioannou2022prob, ali2024predicting} and physics-based approaches that explicitly describe particle acceleration and transport \citep[e.g.,][]{ng1994focused, sokolov2004new, luhmann2007heliospheric, zhang2009propagation, hu2017modeling, wijsen2019modelling, chen2025evidence, shane2025agu}; see, e.g., \cite{whitman2023review} and \cite{kasapis2026review} for a broader overview. 
        Physics-based models are based on the simulation of the physical processes occurring in the development of SEP events and can be applied to different locations and times throughout the heliosphere. They are therefore necessary for predicting CME-driven SEP conditions from the Sun to Earth and Mars, where the evolving particle environment needs to be described across different heliocentric locations. 
        
        A Sun-to-Earth-to-Mars simulation involves modeling of the solar wind, the CME and its driven shock, and energetic particles in the solar corona and the inner heliosphere coupled within one model chain. These physics-based models are driven by observations of the solar magnetic field, with photospheric magnetograms providing the inner boundary condition for modeling the corona and solar wind and, in many cases, the magnetic field from which the CME is initiated. 
        Different global magnetic field data products are available for this purpose, ranging from standard synoptic maps assembled from observations over a solar rotation \citep[e.g.,][]{harvey1996global, bertello2014uncertainties, hughes2016hmi, pevtsov2022NSOHMI} to surface flux-transport maps that evolve the photospheric field in time while assimilating new observations \citep[e.g.,][]{schrijver2003photospheric, arge2010air, hickmann2015data, caplan2025hipft}. More recent approaches also incorporate magnetic field observations from multiple viewing angles to improve the representation of the global photospheric field \citep{jeong2022improved, loeschl2024first, moreno2024comparison}. 
        
        Here, we focus on two magnetic field map products based on observations by the Global Oscillation Network Group (GONG): the National Solar Observatory's GONG (NSO/GONG\footnote{\url{https://gong.nso.edu/data/magmap/} \label{url:gong}}) synoptic map \citep{harvey1996global}, assembled as different longitudes rotate across the visible disk, and the Air Force Data Assimilative Photospheric Flux Transport map driven by GONG data \citep[ADAPT-GONG\footnote{\url{https://nso.edu/data/nisp-data/adapt-maps/} \label{url:adaptgong}};][]{arge2010air}, which evolves the photospheric field between observations and assimilates newly available measurements. 
        The NSO/GONG synoptic map is updated hourly, whereas ADAPT-GONG maps are produced every 2 hours as an ensemble of 12 realizations to represent uncertainties in the photospheric flux-transport model. Here, both products are based on zero-point corrected GONG observations. 
        For active regions near the solar limbs, both magnetic map products can be subject to large uncertainties. 
        For an east limb eruption, the active region has only recently rotated into view, so the latest direct observations are obtained at large viewing angles and are therefore less reliable. Although ADAPT-GONG can evolve previously observed magnetic flux across the far side and assimilate new measurements as they become available, substantial flux emergence or active region evolution occurring on the far side may still be poorly constrained. 
        Even for a west limb eruption, such as the 2017 September 10 event studied here (described later in Section~\ref{sec2:overview}), the magnetic field of the active region in a synoptic map is dominated by observations obtained several days earlier, near its central meridian passage, and thus may not capture its subsequent evolution, whereas ADAPT-GONG can assimilate more recent measurements and evolve the field toward the eruption time. 
        % This distinction becomes critical for eruptions from active regions near the solar limbs. 
        % For an east limb eruption, the active region has only recently rotated into view, so its magnetic field is still poorly constrained by the latest observations, which are obtained at large viewing angles and are subject to greater uncertainties. 
        % Even for a west limb eruption, such as the 2017 September 10 event studied here (described later in Section~\ref{sec2:overview}), the magnetic field of the active region in a synoptic map is dominated by observations obtained several days earlier, near its central meridian passage, and thus may not capture its subsequent evolution prior to the eruption onset. 
        
        In this study, the ambient solar wind, CME, shock, and SEP calculations are linked through the SOlar wind with FIeld lines and Energetic particles (SOFIE) model suite \citep{zhao2024solar}, built within the Space Weather Modeling Framework \citep[SWMF\footnote{\url{https://github.com/SWMFsoftware} \label{url:SWMF}};][]{toth2005space}. SOFIE has been validated against SEP observations at 1 au \citep{zhao2024solar, liu2025physics} and tested in a simulated operational setting \citep{liu2026testbed}, and thus provides a model chain in which the consequences of a single input choice of the photospheric magnetic map can be traced from the low corona to interplanetary space. 
        % \modify{The photospheric magnetic field map sets the inner boundary for the corona and solar wind, determines the open magnetic flux and hence an observer's field-line connection, and provides the active-region field used to specify the CME flux rope. Here we consider two GONG-based products. The standard Global Oscillation Network Group synoptic map \citep[GONG;][]{harvey1996global, hill2018global} is assembled from observations taken as each longitude crosses the central meridian. For a source region near the west limb, the map represents the active region as it appeared nearly one week earlier, while longitudes east of the central meridian are inherited from the previous rotation. The Air Force Data Assimilative Photospheric Flux Transport model applied to GONG data \citep[ADAPT-GONG;][]{arge2010air, arge2011improving, hickmann2015data} instead evolves the field through differential rotation, meridional circulation, and supergranular diffusion \citep{worden2000evolution} while assimilating new observations. Previous work has compared map-driven coronal solutions for a solar eclipse \citep{liu2026simulating} and examined the sensitivity of CME evolution to the background solar wind \citep{sachdeva2026evolution}. For a west-limb event, the source region field can differ substantially between a synoptic map and an assimilative map.}
    
    \subsection{Motivation and Paper Outline} \label{sec1.3:outline}

        This work extends previous studies using the SOFIE model suite in two directions. First, the influence of the magnetogram choice has so far been evaluated mainly on the steady-state corona and solar wind \citep[e.g.,][]{riley2013application, hayashi2016comparison, sachdeva2023solar, liu2026simulating}, magnetic connectivity \citep[e.g.,][]{lario2017solar, badman2020magnetic, li2021comparison}, as well as CME eruptions within the corona \citep[e.g.,][]{jin2022assessing, ledvina2023modeling, sachdeva2026evolution}. Here, we follow the influence of the input magnetogram through the ambient solar wind, CME and its driven shock, and eventually SEP distributions over one single eruptive event. Second, we extend model--observation comparisons from 1 au to Mars within the SOFIE model suite, providing an additional test at a different heliocentric distance and magnetic connectivity to the evolving shock. 
        Using GONG and ADAPT-GONG maps as separate inputs for the 2017 September 10 event simulations, we provide, to our knowledge, the first end-to-end comparison of how the input magnetogram affects CME--SEP modeling in the Sun--Earth--Mars system, with profound implications for space weather modeling and predictions. 
        
        This paper is organized as follows: Section~\ref{sec2:overview} provides an observational overview of the 2017 September 10 SEP event; Section~\ref{sec3:model} describes the SOFIE model suite and the computational setup in this work; Section~\ref{sec4:map} compares the input magnetograms; Section~\ref{sec5:result} presents the simulation results and their comparisons with the observations; and Section~\ref{sec6:summary} summarizes our findings and implications.

\section{Overview: The 2017 September 10 SEP Event} \label{sec2:overview}
    
\begin{figure*}[tp!]
    \centering {\includegraphics[width=1.0\hsize]{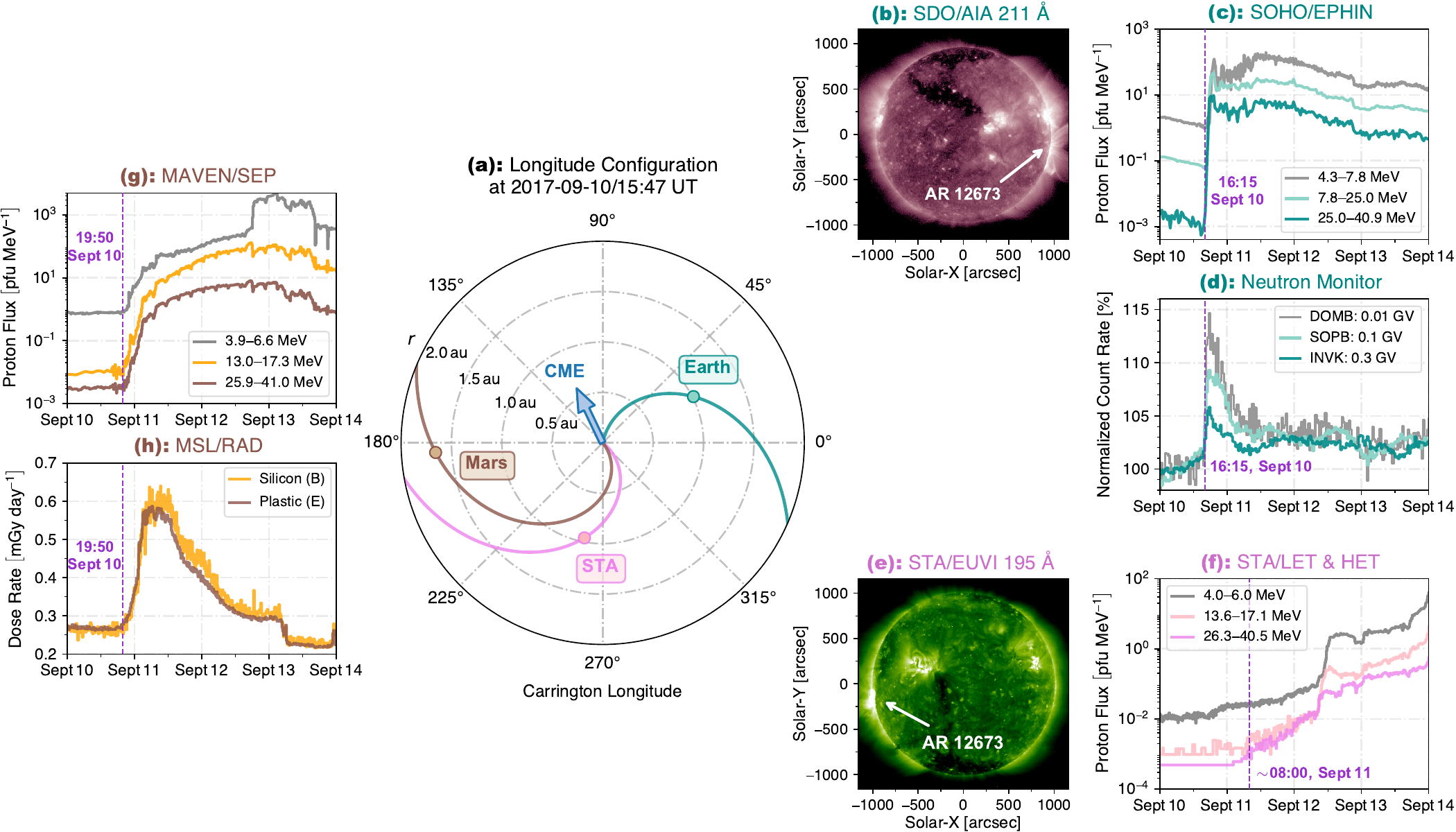}}
    \caption{Overview of the 2017 September 10 SEP event. 
    The observational panels are grouped clockwise around panel (a), following the observer locations from Earth (panels (b)--(d)), to STA (panels (e) and (f)), and then to Mars (panels (g) and (h)). 
    (a) Locations of Earth (green circle), STA (pink circle), and Mars (brown circle) at the CME eruption time (15:47 UT), shown in Carrington longitude and heliocentric distance ($r$). Curves in the corresponding colors show nominal magnetic field lines estimated from the local solar wind speeds (see Table~\ref{tab3:fp_comparison} below); the blue arrow indicates the CME propagation direction. 
    (b) SDO/AIA 211 \AA\ image at 15:54 UT, in which the CME eruption source region, AR 12673, is located at the west limb as viewed from Earth. 
    (c) Proton fluxes measured by SOHO/EPHIN in three energy channels, ranging from 4.3 to 40.9 MeV. Hereafter, ``pfu" denotes the particle flux unit, defined as $1\;\mathrm{pfu} = 1\;\mathrm{count \; cm^{-2}\; s^{-1}\; sr^{-1}}$. 
    (d) Count rates of three neutron monitors (DOMB, SOPB, and INVK), normalized to their 1-day pre-event averages, whose effective geomagnetic cutoff rigidities are 0.01, 0.1, and 0.3 GV, respectively. 
    (e) STA/EUVI 195 \AA\ image at 15:55 UT, in which AR 12673 is located at the east limb as viewed from STA. 
    (f) Proton fluxes measured by STA/LET and HET in energies ranging from 4.0 to 40.5 MeV. 
    (g) Proton fluxes measured by MAVEN/SEP in energies ranging from 3.9 to 41.0 MeV. 
    (h) Surface radiation dose rates recorded by the silicon detector (B) and the plastic scintillator (E) of MSL/RAD. 
    The vertical purple dashed lines mark the SEP onset times: 16:15 UT on September 10 at Earth (panels (c) and (d)), 19:50 UT on September 10 at Mars (panels (g) and (h)), and $\sim$08:00 UT on September 11 at STA (panel (f)).} \label{fig1:overview}
    \end{figure*}

    The SEP event on 2017 September 10 is one of the most intense eruptive events of solar cycle 24 that resulted in the GLE \#72 at Earth \citep[e.g.,][]{paassilta2018catalogue, waterfall2023high}\footnote{\url{https://cdaw.gsfc.nasa.gov/CME_list/sepe/} \label{url:cdawsep}}. It has been analyzed from various perspectives\footnote{See the special collection of papers at \url{https://agupubs.onlinelibrary.wiley.com/doi/toc/10.1002/(ISSN)1542-7390.SW-SEPT2017}. \label{url:201709paper}}, including the flare and its microwave, hard X-ray, and gamma-ray emission \citep{gary2018microwave, omodei2018fermi, chen2020measurement}; the CME-driven shock, its magnetic connectivity, and SEP modeling \citep{luhmann2018shock, ding2020modeling, kouloumvakos2020evidence, zhu2021shock}; the GLE recorded by ground-based neutron monitors, together with high-energy proton spectra and the radiation environment at Earth/L1 \citep{cohen2018ground, matthia2018solar, mishev2018first, obrien2018solar, bruno2019spectral, jiggens2019situ}; and the space weather effects at Mars \citep{guo2018modeling, hassler2018space, lee2018observations}. 
    It is associated with a fast halo CME launched by the eruption of a magnetic flux rope from NOAA active region (AR) 12673, located at Stonyhurst heliographic latitude $-8^{\circ}$ and longitude $+88^{\circ}$ (S08W88) as viewed from Earth, with the CME propagation direction indicated in Figure~\ref{fig1:overview}(a) \citep[see][for more details]{seaton2018observations, veronig2018genesis, warren2018spectroscopic, yan2018simultaneous}. 
    AR 12673 emerged with one of the fastest magnetic flux emergence rates on record \citep{sun2017super} and produced a sequence of major eruptions\footnote{See \url{https://svs.gsfc.nasa.gov/12706/} for a list of major flares and their visualizations. \label{url:svs}} during September 4--9 \citep[e.g.,][]{yang2017block, chamberlin2018solar, verma2018origin, yan2018successive, mitra2018successive, wu201904, scolini2020cme}. The September 10 eruption was associated with an X8.2-class X-ray flare\footnote{\url{https://ccmc.gsfc.nasa.gov/DONKI/events/13098/4/}}, which began at 15:35 UT, peaked at 16:06 UT, and ended at 16:31 UT in soft X-ray observations from the X-Ray Sensor \citep[XRS;][]{hanser1996design} on board the GOES-15 \citep{hillger2011goes}. 

    The CME was observed by the C2 and C3 coronagraphs of the Large Angle and Spectrometric Coronagraph \citep[LASCO;][]{brueckner1995lasco} on board the Solar and Heliospheric Observatory \citep[SOHO\footnote{The SOHO data used throughout this study are available at \url{https://soho.nascom.nasa.gov/data/catalogues.html} and/or \url{https://ssa.esac.esa.int/ssa/\#/pages/search}. \label{url:soho}};][]{domingo1995soho} from 15:59 UT onward, and by the COR1 and COR2 coronagraphs of the Sun Earth Connection Coronal and Heliospheric Investigation \citep[SECCHI;][]{howard2008sun} on board the Solar Terrestrial Relations Observatory-Ahead \citep[STA;][]{kaiser2008stereo} from 16:00 UT onward\footnote{See \url{https://cor1.gsfc.nasa.gov/catalog/cme/2017/COR1_preliminary_event_list_2017_09.html}. \label{ftn:coracme}}. 
    In the low corona, the eruption was observed in extreme ultraviolet (EUV) by the Atmospheric Imaging Assembly \citep[AIA\footnote{\url{https://aia.lmsal.com/} \label{url:aia}};][]{lemen2012atmospheric} on board the Solar Dynamics Observatory \citep[SDO;][]{pesnell2012solar} from Earth's view, revealing an extraordinary global EUV wave that propagated across the entire visible solar disk and around the solar limb \citep{liu2018truly}, and by the Extreme Ultraviolet Imager \citep[EUVI\footnote{The data are available at \url{https://stereo-ssc.nascom.nasa.gov/data/ins_data/secchi/L0/a/img/euvi/} and can also be accessed through the Virtual Solar Observatory (VSO) at \url{https://nso.virtualsolar.org/cgi/search}. \label{url:vso}};][]{wuelser2004euvi} of SECCHI from STA's view. The source AR location quoted above (S08W88 as viewed from Earth) corresponds to the flare position derived by the SolarSoft flare locator \citep{freeland1998data} using SDO/AIA 94~\AA\ running difference images\footnote{The record of this flare is available at \url{https://www.lmsal.com/solarsoft/latest_events_archive/events_summary/2017/09/10/gev_20170910_1535/index.html}. \label{url:solarsoft}}. This places the eruption site near the west limb as seen from Earth \citep[see Figure~\ref{fig1:overview}(b) and more details in][]{seaton2018observations, veronig2018genesis} and near the east limb as seen from STA (Figure~\ref{fig1:overview}(e)). 
    Plane-of-sky measurements indicated a CME speed of $\sim$3200 km$\;$s$^{-1}$ in Earth's view, with the eruption onset at 15:47 UT, as reported in the Coordinated Data Analysis Workshop (CDAW) SOHO/LASCO catalog\footnote{\url{https://cdaw.gsfc.nasa.gov/CME_list/} \label{url:cdawcme}} \citep{yashiro2004catalog, gopalswamy2009soho} and $\sim$1800 km$\;$s$^{-1}$ in STA's view as reported by the Computer Aided CME Tracking software (CACTus\footnote{\url{https://secchi.nrl.navy.mil/cactus/} \label{url:cactus}}) CME catalog \citep{robbrecht2004automated, robbrecht2009automated}. These plane-of-sky speeds are derived from the CME propagation within the respective coronagraph fields of view (FOVs), which cover 2--32 solar radii ($R_\mathrm{s}$) for SOHO/LASCO C2 and C3 coronagraphs combined and 2.5--15 $R_\mathrm{s}$ for STA/SECCHI COR2 coronagraph alone. 
    On the other hand, multi-spacecraft reconstructions yielded a three-dimensional (3D) CME speed of $\gtrsim 3700$ km$\;$s$^{-1}$ in the first 20 minutes of the CME eruption \citep[see Section 3.1 of][]{gopalswamy2018extreme}. The CME decelerated strongly about 30 minutes after the eruption, as shown in Figure 2(d) of \cite{gopalswamy2018extreme}. By using the NOAA Space Weather Prediction Center CME Analysis Tool \citep[SWPC-CAT\footnote{\url{https://ccmc.gsfc.nasa.gov/tools/SWPC-CAT/} \label{url:swpccat}};][]{millward2013operational}, the CME was estimated to have an average 3D speed of 2650 km$\;$s$^{-1}$ within 21.5 $R_\mathrm{s}$, as reported in the Space Weather Database Of Notifications, Knowledge, Information (DONKI) database\footnote{\url{https://ccmc.gsfc.nasa.gov/DONKI/events/} \label{url:donki}}. Further details on the eruption and its coronal signatures are given by \cite{veronig2018genesis} and \cite{hu2019effects}. 
    
    The eruption produced an intense SEP event that was observed at widely separated locations in the heliosphere, with strong particle flux enhancements observed at Earth and Mars and a weaker enhancement at STA \citep[e.g.,][]{guo2018modeling, lee2018observations}. Figure~\ref{fig1:overview}(a) shows the longitudinal configuration of the CME propagation direction and the observer locations. Below we summarize the \textit{in situ} particle and ground-based observations at each of these viewpoints, which sample very different magnetic connections to the eruption. 
    \begin{enumerate}
        \item With the source region located near the west limb as viewed from Earth, this event was favorably connected to Earth and produced a prompt enhancement of the particle flux shortly after the eruption, with their peak fluxes approximately 2--4 orders of magnitude above pre-event levels, as shown in Figure~\ref{fig1:overview}(c) for 4.3--40.9 MeV protons measured by the Electron Proton Helium INstrument \citep[EPHIN;][]{muller1995costep} on board SOHO. The event reached the S3 (Strong) level on the NOAA Solar Radiation Storm scale\footnote{See more details at \url{https://www.spaceweather.gov/noaa-scales-explanation}.}, corresponding to elevated radiation concerns for astronauts and high-altitude, high-latitude aviation \citep{redmon2018september}. 
        The expanding coronal shock also accelerated protons to relativistic energies \citep{kouloumvakos2020evidence, zhu2021shock}, and the highest-energy particles arrived at Earth within $\sim$30 minutes, as indicated by the purple dashed line in Figure~\ref{fig1:overview}(c), initiating secondary cascades in Earth's atmosphere that resulted in the GLE~\#72 \citep[see also][]{mavromichalaki2018real, mishev2018first}. Figure~\ref{fig1:overview}(d) shows the normalized count rates from three high-latitude ground-based neutron monitors obtained from the Neutron Monitor Database \citep[NMDB\footnote{\url{https://www.nmdb.eu/} \label{url:nmdb}};][]{mavromichalaki2011nmdb}, with effective geomagnetic cutoff rigidities of 0.01--0.3 GV. During the event, these neutron monitor count rates increased by $\sim$5\%--15\% above their pre-event levels. The nearly simultaneous sharp increases recorded at these stations mark the arrival of relativistic solar particles at ground level and provide the characteristic neutron monitor signature of GLE~\#72 \citep[e.g.,][]{poluianov2017gle, kurt2018some}. 
        Later, an interplanetary shock arrived at Earth around 20:00 UT on September 12 (see Figures~\ref{fig16:profile}(a)--(f) below), followed by a noticeable change in the decay rate of the energetic proton fluxes. 
        
        \item At 1 au, but separated from Earth by $\sim$128$^\circ$ eastward in longitude, STA observed a much weaker and delayed SEP enhancement. From the STA viewpoint, the active region was located near the east limb, resulting in a less favorable magnetic connection to the eruption than at Earth. Figure~\ref{fig1:overview}(f) shows proton fluxes from 4.0 to 40.5 MeV measured by the Low Energy Telescope \citep[LET\footnote{\url{https://izw1.caltech.edu/STEREO/Public/LET_public.html} \label{url:stalet}};][]{mewaldt2008low} and High Energy Telescope \citep[HET\footnote{\url{https://izw1.caltech.edu/STEREO/Public/HET_public.html} \label{url:stahet}};][]{von2008high} on board STA. The intensities rose much more gradually, with an onset near 08:00 UT on September 11, and peaked about two orders of magnitude below those observed at Earth. 
                
        \item Mars, located at $\sim$1.66 au and $\sim$156$^{\circ}$ west of Earth in longitude (Figure~\ref{fig1:overview}(a)), observed a strong and relatively prompt response to the event. Energetic particle measurements at Mars were obtained by the SEP instrument\footnote{\url{https://lasp.colorado.edu/maven/sdc/public/pages/datasets/sep.html} \label{url:mavensep}} \citep{larson2015maven} on board the Mars Atmosphere and Volatile EvolutioN spacecraft \citep[MAVEN\footnote{\url{https://pds-ppi.igpp.ucla.edu/mission/MAVEN} \label{url:maven}};][]{jakosky2015maven}. MAVEN/SEP directly measures the energy spectra of ions, predominantly protons, up to $\sim$6.6 MeV, while higher-energy proton arrival times and the peak of the event can be inferred from count rate measurements of penetrating particles through the detector \citep[e.g.,][]{lee2018observations, khoo2024multispacecraft}. The $\geqslant13$ MeV proton fluxes are a higher-level data product based on best-fit spectra derivations \citep[see][]{lee2023AGUobservations, lee2024AGUobservations}. The derivation of a single best-fit spectrum involves running an iterative routine that generates a set of random spectra and then multiplies these spectra with the SEP instrument response matrix to obtain the theoretical counts. These counts are then compared with the measured values and the reduced chi-squared analysis is used to determine the best fit. Details of the methodology and the data set will be provided in a forthcoming publication by C.~O.~Lee et al. (in preparation).
        
        Figure~\ref{fig1:overview}(g) shows the resulting proton fluxes in three representative energy channels from 3.9 to 41.0 MeV. The intensities began to rise at $\sim$19:50 UT on September 10, only $\sim$3.5 hr after the onset observed near Earth, and peaked about 2--3 orders of magnitude above the pre-eruption background. On the Martian surface, MSL/RAD \citep{hassler2012radiation} on board the Curiosity rover recorded a nearly simultaneous sharp increase in dose rate (Figure~\ref{fig1:overview}(h)), reaching 2--3 times its pre-eruption level, as noted in Section~\ref{sec1.1:sepsw}. These observations demonstrate the substantial radiation enhancement produced by the event at Mars \citep[see also][and references therein]{ehresmann2018energetic, hassler2018space, ramstad2018september, zeitlin2018analysis}. 
        The CME-driven shock subsequently arrived at Mars at about 02:30 UT on September 13 \citep[see Table 1 and Figure 3 of][]{lee2018observations}, accompanied by a moderate enhancement in the proton fluxes measured by MAVEN/SEP during the energetic storm particle (ESP) phase and a decrease of $\sim$20\% in the MSL/RAD dose rate associated with the Forbush decrease.
    \end{enumerate}
    
    In the following, we present numerical simulations of this event using the models described in Section~\ref{sec3:model} and compare them with various observations regarding the solar wind, CME, and SEPs.

\section{Numerical Models and Simulation Setup} \label{sec3:model}

    \subsection{The SOFIE Model Suite} \label{sec3.1:sofie}
    
    SOFIE\footnote{Available at \url{https://ccmc.gsfc.nasa.gov/ror/requests/SH/SWMF-AWSoM/swmfawsom_user_registration.php}. \label{url:sofieccmc}} is an integrated solar corona--interplanetary medium--CME--SEP model suite \citep{zhao2024solar}, built within the SWMF developed at the University of Michigan \citep{toth2005space}. It combines: (1) the stream-aligned \citep{sokolov2022stream} Alfv\'{e}n Wave Solar atmosphere Model-Realtime \citep[AWSoM-R;][]{sokolov2013magnetohydrodynamic, sokolov2021threaded, van2014alfven}, which provides the magnetogram-driven corona and solar wind; (2) the Eruptive Event Generator with the Gibson--Low \citep[GL;][]{gibson1998time} flux rope configuration \citep[EEGGL;][]{borovikov2017eruptive, jin2017data}, which parameterizes the analytical CME flux rope; and (3) the Multiple-Field-Line Advection Model for Particle Acceleration \citep[M-FLAMPA;][]{sokolov2004new, borovikov2018toward, liu2025physics}, which is currently able to follow particle acceleration and transport along advected field lines using a particle-number-conserved Poisson-bracket scheme \citep{sokolov2023high, liu2026implicit}. A complete description of the SOFIE model suite is given by \cite{zhao2024solar} and \cite{liu2025physics}. Below, we focus on the event-specific model setup, input parameters, and simulation procedures used in this study. 
    
    We use the GONG and ADAPT-GONG magnetograms, described and compared in Section~\ref{sec4:map}, as the photospheric magnetic field inputs to drive SOFIE. For each map, the Poynting flux parameter, $\left(S_\mathrm{A} / B\right)_\odot$, is adjusted based on the averaged unsigned radial magnetic field ($\langle \left|B_r\right| \rangle$), 
    % so that the same total Alfv\'{e}n wave energy is injected through the inner boundary in AWSoM-R. 
    so that the Alfv\'{e}n wave energy injection rate integrated over the solar surface is the same in the GONG- and ADAPT-GONG-driven runs. 
    We first adopt $\left(S_\mathrm{A} / B\right)_\odot = 0.550$ MW$\;$m$^{-2}\;$T$^{-1}$ for the GONG map, following the optimized value obtained by \cite{jivani2023global} and \cite{huang2024solar} for reproducing the observed global solar wind conditions, and then rescale it for the ADAPT-GONG map according to the ratio of the total unsigned magnetic fluxes of the two maps, yielding 0.564 MW m$^{-2}$ T$^{-1}$. This ensures that differences between the two simulations of the ambient solar wind arise primarily from the spatial distribution of the magnetic field rather than from a different energy budget. 
    
    EEGGL then uses the AR magnetic field from the corresponding magnetogram, together with the same input 3D CME speed of 2650 km$\;$s$^{-1}$, to determine the analytical GL flux rope parameters. Specifically, it first smooths the input magnetogram with a $5\times5$ pixel window to reduce the complexity of the photospheric field configuration \citep[e.g.,][]{jin2017data, liu2025physics}. Based on two user-selected points, one in each polarity, it then identifies  connected positive- and negative-polarity regions with $\left|B_r\right|\geqslant20$~G. The eruption center is located at the intersection of the PIL and the line joining the flux-weighted centers of these regions. 
    Since the AR magnetic fields differ between the two maps, the resulting inserted flux ropes also differ in their properties and energy content. Therefore, differences in the subsequent CME evolution reflect both the map-dependent background corona and solar wind and the corresponding map-dependent CME initialization. The resulting flux rope parameters for the two magnetograms are summarized in Table~\ref{tab1:param}. 
    These configurations are then used in the time-accurate magnetohydrodynamic (MHD) simulations, in which the evolving MHD parameters from AWSoM-R are coupled to M-FLAMPA every two minutes to model particle acceleration and transport processes, with the same parameters of M-FLAMPA adopted for both runs as detailed in Section \ref{sec3.2:setup}. 

    \begin{table*}[tp!]
    \begin{center}
    \caption{Input parameters of AWSoM-R, EEGGL, and M-FLAMPA for the GONG and ADAPT-GONG runs of the 2017 September 10 event.} \label{tab1:param}
    \setlength\tabcolsep{3pt}{
    \begin{tabular}{clcc}
        \toprule
        \multirow{2}{*}{Model} & \multirow{2}{*}{Parameter} & \multicolumn{2}{c}{Value} \\
        \cmidrule(lr){3-4}
        & & GONG run & ADAPT-GONG run \\
        \midrule % \cline{2-3} \\ \multirow{LINE}{*}{CONTENT} \multicolumn{NUMBER}{c}{xxx}
        AWSoM-R & Magnetogram time & 14:14 UT & 14:00 UT, realization 7 \\
         & Total unsigned magnetic flux & $3.84\times10^{23}\;\mathrm{Mx}$ & $3.74\times10^{23}\;\mathrm{Mx}$ \\
         & Averaged unsigned magnetic field ($\langle|B_r|\rangle$) & $6.31\;\mathrm{G}$ & $6.16\;\mathrm{G}$ \\
         & Poynting flux parameter ($\left(S_\mathrm{A}/B\right)_{\odot}$) & $0.550\; \mathrm{MW\; m^{-2}\; T^{-1}}$ & $0.564\; \mathrm{MW\; m^{-2}\; T^{-1}}$ \\
         & Total Alfv\'{e}n wave energy injection rate & $2.11\times10^{21}\;\mathrm{J\;s^{-1}}$ & $2.11\times10^{21}\;\mathrm{J\;s^{-1}}$ \\
        \midrule % \cline{2-3}
        EEGGL & CME speed & $2650\; \mathrm{km\; s^{-1}}$ & $2650\; \mathrm{km\; s^{-1}}$ \\
         & Inserted flux-rope type & Spheromak & Spheromak \\
         & Source AR center coordinates\tablenotemark{a} & ($115.50^{\circ}, -11.24^{\circ}$) & ($117.60^{\circ}, -10.50^{\circ}$) \\
         & Flux rope major axis orientation & $339.67^{\circ}$ & $90.20^{\circ}$ \\
         & Flux rope helicity & Positive & Positive \\
         & Flux rope radius ($r_0$) & 0.30 $R_\mathrm{s}$ & 0.80 $R_\mathrm{s}$ \\
         & Flux rope stretch & 0.60 $R_\mathrm{s}$ & 0.60 $R_\mathrm{s}$ \\
         & Flux rope apex height & 0.50 $R_\mathrm{s}$ & 1.00 $R_\mathrm{s}$ \\
         & Flux rope magnetic field strength ($B_0$) & 99.2 G & 35.3 G \\
        \midrule
        M-FLAMPA & Upstream parallel mean free path parameter\tablenotemark{b} & 0.1 au & 0.1 au \\
         & Injection scaling factor & 300 & 300 \\
        \bottomrule
    \end{tabular}}
    \end{center}
    \tablenotetext{a}{The coordinates are Carrington longitude and latitude.}
    \tablenotetext{b}{See Equation~(27) of \cite{li2003energetic} and Equation~(1) of \cite{liu2025physics}.}
    \end{table*}
    
    \subsection{Computational Grid and Simulation Setup} \label{sec3.2:setup}
    
    In AWSoM-R, the simulation domain consists of a threaded-field-line region from 1.0 to 1.1 $R_\mathrm{s}$, a block-adaptive 3D spherical grid in the solar corona (SC) from 1.1 to 24 $R_\mathrm{s}$, and a block-adaptive Cartesian grid in the inner heliosphere (IH) from 20 to 650 $R_\mathrm{s}$. In the innermost region between 1.0 and 1.1 $R_\mathrm{s}$, AWSoM-R employs the threaded-field-line model of \cite{sokolov2021threaded} and solves the field-aligned hydrodynamic equations, while in SC and IH, it solves the global extended MHD equations described in \cite{van2014alfven} and \cite{gombosi2018extended}. The SC and IH domains overlap between 20 and 24 $R_\mathrm{s}$, where a buffer region couples the solutions. 
    %, with the magnetic field extracted from the PFSS model expanded to spherical-harmonic degree 180  \citep{schatten1969model, altschuler1969magnetic, toth2011obtaining}}.
    
    The computational domain in SC is composed of grid blocks of $6 \times 8 \times 8$ cells (\textit{control volumes}), with a radially stretched grid constructed by adopting $\ln (r)$ as the coordinate rather than $r$. 
    The radial cell size ranges from 0.02 $R_\mathrm{s}$ near the Sun, which resolves the steep density and temperature gradients in the low corona, to $\sim$0.8 $R_\mathrm{s}$ at its outer boundary. The initial angular resolution is $1.4^\circ$ in both longitude and latitude, refined by one level to $0.7^\circ$ within $r = 1.7\; R_\mathrm{s}$ and coarsened by one level to $2.8^\circ$ beyond it. 
    The IH domain encloses the spherical SC domain and is composed of grid blocks of $8 \times 8 \times 8$ cells, with cell sizes from $\sim$0.3 $R_\mathrm{s}$ at its inner boundary to $\sim$20 $R_\mathrm{s}$ at its outer boundary. 
    Within both domains, we apply the adaptive mesh refinement \citep[AMR;][]{gombosi2003adaptive, toth2012adaptive} along the heliospheric current sheet (HCS), the CME propagation direction, and within a 30$^{\circ}$ cone toward each observer (Earth, STA, and Mars), so that the key coronal and heliospheric structures are resolved both globally and along the Sun--observer directions. 
    Specifically, the Mars-directed AMR cone extends to 1.7 au in the IH domain so as to better resolve the CME-driven shock and the magnetic field lines connected to Mars. Each run contains about 5 million cells in SC and 140 million in IH. 
    
    In M-FLAMPA, we initialize 648 magnetic field lines uniformly on a sphere at $r = 2.5\; R_\mathrm{s}$, covering 360$^{\circ}$ in longitude and $\pm85^{\circ}$ in latitude, and trace them inward to the SC inner boundary at 1.1 $R_\mathrm{s}$ and outward to the IH outer boundary at 650 $R_\mathrm{s}$. These field lines are represented by Lagrangian grid points, which are then advected through the coupled SC and IH domains, with a 2-minute coupling cadence between AWSoM-R and M-FLAMPA. 
    The cross-field diffusion and drift effects are not incorporated in the current M-FLAMPA simulations. Thus, the modeled differences in SEP intensities among different locations reflect differences in their magnetic connectivity to the evolving CME-driven shock. 
    The injected seed particle density at the shock front is prescribed from the suprathermal tail of the solar wind \citep[e.g.,][]{gloeckler2003ubiquitous, fisk2008acceleration} and is proportional to the local thermal pressure across the shock front \citep[e.g.,][]{zank2000particle, sokolov2004new}. The suprathermal tail extends from the plasma thermal energy to the particle injection energy of 10 keV, and we apply a multiplier of 300 to the seed particle population in order to bring the modeled SEP intensities to the observed order of magnitude. This factor compensates for uncertainties in the simplified seed-population prescription, and the same value is applied in both runs so that our comparison focuses on the relative differences in SEP intensities resulting from the magnetogram choice. 
    A detailed description of the latest default treatment of the upstream and downstream diffusion parameters governing particle acceleration and transport is given in Section~3.1 of \cite{liu2026counterintuitive}; see also Chapter~13 of \cite{gombosi1998physics} and references therein for the underlying transport theory. In this treatment, the only free parameter is the upstream parallel mean free path, parameterized following Equation~(27) of \cite{li2003energetic} and adopted in our model as Equation~(1) of \cite{liu2025physics}, whereas the downstream diffusion coefficient is determined directly from the MHD turbulence quantities provided by AWSoM-R. 
    
    In Table~\ref{tab1:param}, we summarize the input parameters of each model of SOFIE for the GONG- and ADAPT-GONG-driven runs of the 2017 September 10 SEP event.

\section{Magnetic Field Data} \label{sec4:map}
    
    In our simulations, the inner boundary magnetic field is prescribed by photospheric magnetic field maps. Two data products, GONG and ADAPT-GONG maps, are used in this study. 
    The hourly updated, zero-point corrected NSO/GONG synoptic magnetogram (\citealp{harvey1996global, hill2018global, jain2026three}; see footnote~\ref{url:gong} for its availability) is assembled from full-disk observations taken at different times, using weights proportional to $\cos^4(\Delta\phi)$, where $\Delta\phi$ is the longitude relative to the Earth-facing central meridian. This weighting favors measurements near the central meridian, where projection effects are reduced, allowing more reliable estimates of the radial magnetic field. Specifically, the magnetic field of AR 12673, located at W88 as viewed from Earth at the CME eruption time, represents a weighted average of observations during its disk passage, with the greatest weight assigned to measurements near its central meridian passage nearly one week earlier. 
    The ADAPT-GONG map (\citealp{arge2010air, arge2011improving, hickmann2015data}; see footnote \ref{url:adaptgong} for its availability) evolves the photospheric flux forward in time with a flux-transport model accounting for differential rotation, meridional circulation, and supergranular diffusion \citep{worden2000evolution}. New magnetic field observations, such as the zero-point corrected GONG data used here \citep[see also][]{li2021comparison, mays2025nasa}, are assimilated through an ensemble approach, providing an instantaneous estimate of the global photospheric field every 2 hours, including regions away from the central meridian and near the poles. 
    
\begin{figure*}[tp!]
    \centering {\includegraphics[width=0.9\hsize]{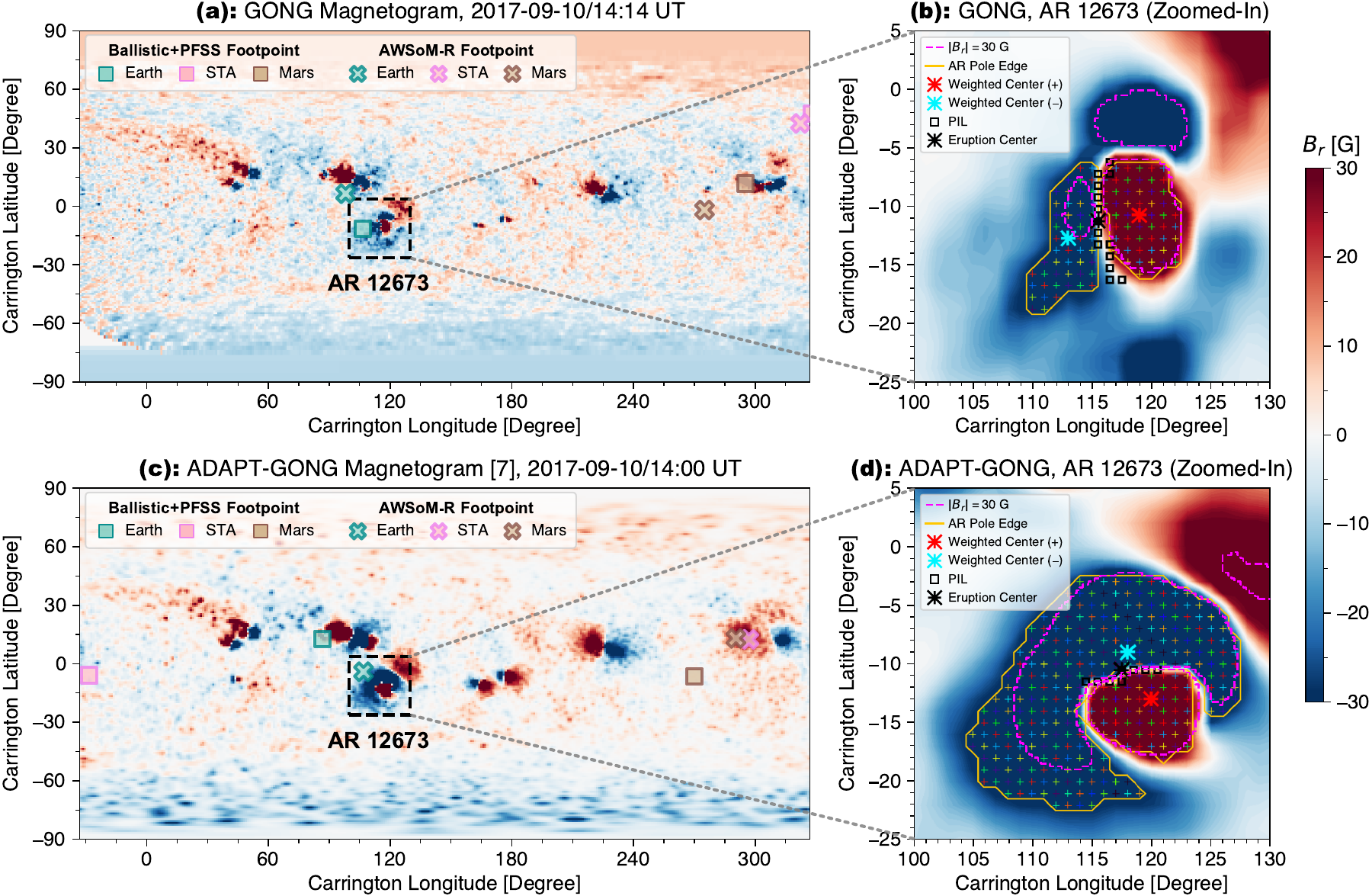}}
    \caption{Input photospheric magnetograms used to drive the two simulations: (a) the processed hourly updated, zero-point corrected GONG synoptic map at 14:14 UT on 2017 September 10; and (c) the \mbox{ADAPT-GONG} map (realization 7) at 14:00 UT on the same day. 
    In panels (a) and (c), the green, pink, and brown symbols mark the photospheric magnetic footpoints connected to Earth, STA, and Mars, respectively. Squares indicate footpoints obtained by ballistic back-mapping combined with the PFSS model, while cross marks indicate those obtained by tracing magnetic field lines through the steady-state AWSoM-R solutions. 
    The black dashed boxes indicate the CME source region, AR~12673, shown in enlarged views in panels (b) and (d), respectively. In these zoomed-in views, the magenta dashed contours indicate $\left|B_r\right|=30$~G for reference in the unsigned magnetic flux calculation, and the yellow outlines enclose the positive- and negative-polarity regions selected by EEGGL, with the selected grid points marked by small colored plus signs. The red and cyan asterisks indicate the weighted centers of the positive and negative polarities, respectively. The black squares mark the identified PIL, and the black asterisks mark the eruption centers determined by EEGGL.  
    All panels are shown in Carrington heliographic coordinates and share the same color scale for the radial magnetic field ($B_r$), saturated at $\pm30$~G.} \label{fig2:map}
    \end{figure*}

    For the run driven by the GONG map, we use the hourly updated synoptic map at 14:14 UT on 2017 September 10 (Figure~\ref{fig2:map}(a)), about 1.5 hours before the CME eruption onset at 15:47~UT. This interval reflects the timing of the available maps and the need to derive the steady-state solar wind background from pre-eruption conditions, as in our previous studies \citep{zhao2024solar, chen2025decent, liu2025physics, liu2026counterintuitive, manchester2025high, wraback20253d}. 
    For the ADAPT-GONG run, as the ADAPT-GONG maps are produced through a separate flux-transport and data-assimilation pipeline with its own output times at 2-hour intervals, we use realization 7 at 14:00 UT on the same day (Figure~\ref{fig2:map}(c)), the last ADAPT-GONG map available before the eruption onset. These two maps are close enough in time to sample essentially the same photospheric state. We select this realization of the ADAPT-GONG map (out of the 12 available realizations) following the approach adopted by \cite{huang2023modeling}, who simulated the solar wind during solar cycle 24. %This also limits our computational costs. 
    Using a single realization also keeps the computational cost manageable, as evaluating all 12 realizations would require repeating the full end-to-end solar wind, CME, and SEP simulations for each realization. The two products are provided on different latitudinal grids, uniform in the sine of latitude for GONG and in latitude for ADAPT-GONG, and we remap both onto a $1^{\circ}\times1^{\circ}$ grid in longitude and latitude before using them as the inner boundary condition. 
    The GONG map is further processed to account for the uncertainty of the synoptic magnetogram observations, especially the very weak fields at high latitudes. We set the radial magnetic field to $\mathrm{sgn}\left(B_r^\mathrm{GONG}\right) \times \min\left(3.75\,|B_r^\mathrm{GONG}|,\; |B_r^\mathrm{GONG}| + 5\; \mathrm{G}\right)$, where $B_r^\mathrm{GONG}$ is the observed radial field from the original GONG magnetogram. 
    This setting is consistent with previous studies by \cite{cohen2007semi}, \cite{jin2012global}, and \cite{huang2024solar}. Also, \cite{riley2007alternative} applies a similar factor to study the open flux problem. A comparison of CME evolution in simulations driven by original and processed GONG maps is presented by \cite{sachdeva2026evolution}, and the effects of this processing are not discussed further here. For the ADAPT-GONG map, however, we do not apply such a multiplicative factor, as a scaling has already been taken into account during its production \citep[see][]{hickmann2015data}. 
    
    All the remaining model input parameters are kept identical between the two runs, except for the Poynting flux parameter and the flux rope parameters derived from each map (Table~\ref{tab1:param} and Section~\ref{sec3.1:sofie}). 
    Specifically, we first compute the surface-averaged $\left|B_r\right|$ of each remapped input map, and then adjust the Poynting flux parameter, $\left(S_\mathrm{A}/B\right)_{\odot}$, so that the Poynting flux ($S_\mathrm{A}$), which is proportional to the total Alfv\'{e}n wave energy injection rate through the inner boundary, is identical ($2.11\times10^{21}$ J$\;$s$^{-1}$) in these two runs. 
    
    In Figure~\ref{fig2:map}, we show the two input photospheric magnetic field maps, both of which agree on the overall large-scale magnetic structure and contain strong magnetic fields in AR 12673, outlined by the dashed boxes around $(115^{\circ}, -10^{\circ})$ in Carrington coordinates. 
    In panels (a) and (c), we also mark the photospheric footpoints magnetically connected to Earth, STA, and Mars, derived using two approaches: ballistic back-mapping combined with the potential field source surface \citep[PFSS;][]{altschuler1969magnetic, schatten1969model} model and magnetic field line tracing through the steady-state AWSoM-R solutions. The Earth-connected footpoints lie close to AR 12673, whereas those connected to STA and Mars are located on the far side of the Sun, as discussed in detail in Section~\ref{sec5.1.4:fp}. 
    
    Nevertheless, these two magnetic field maps differ substantially in field strength and flux distribution, both within the AR and across the rest of the solar surface. In panels (b) and (d), we show zoomed-in views of the smoothed source region magnetic fields. 
    In the GONG synoptic map, the magnetic field at AR 12673 is weaker and more diffuse, and the two polarities are only marginally separated (panel (b)). The ADAPT-GONG map, which evolves the flux forward in time and assimilates the newest full-disk observations, yields a stronger and more compact AR, with the leading positive polarity wrapped by the trailing negative flux along a sharper polarity inversion line (PIL), as shown in panel (d). 
    Within the source region, we find another potential PIL in the GONG map near $(120^\circ,-5^\circ)$ (panel (b)). However, the adopted CME eruption center in GONG lies closer to the observed flare site (S08W88). We also test the alternative pairing involving the northern negative-polarity patch, but it yields an unrealistically large flux rope field strength ($\gtrsim170$~G) and is therefore not adopted. 
    Because EEGGL derives the flux rope orientation and field strength from the local AR field around the PIL, differences in the source region shape and polarity selection propagate directly into the inserted CME flux rope. The resulting parameters are listed in Table~\ref{tab1:param} and further discussed in Section~\ref{sec5.2:cme}. 
    The two maps also differ visibly in the weak-field regions, in the far-side ARs (e.g., around Carrington longitude, CL = 180$^{\circ}$, 240$^{\circ}$ and 300$^{\circ}$), and near the poles, where the ADAPT flux transport model supplies a physics-based estimate of the poorly observed polar field. 
    
    \begin{figure*}[tp!]
    \centering {\includegraphics[width=0.9\hsize]{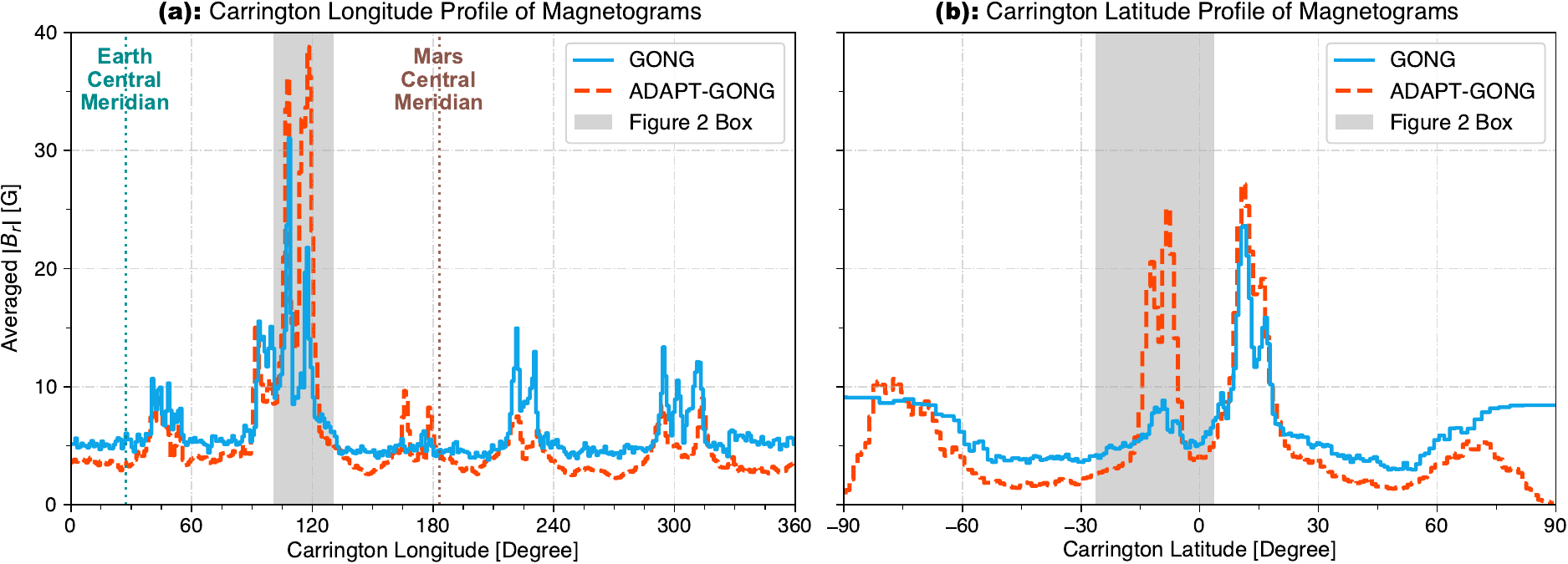}}
    \caption{Unsigned radial magnetic field, $\left|B_r\right|$, of the GONG (blue solid) and ADAPT-GONG (red dashed) maps shown in Figure~\ref{fig2:map}. (a) Latitude-averaged $\left|B_r\right|$ as a function of Carrington longitude, with the dotted vertical lines representing the central meridians as viewed from Earth (green) and Mars (brown) for reference. (b) Longitude-averaged $\left|B_r\right|$ as a function of Carrington latitude. The gray shaded bands mark the longitude and latitude ranges of the black dashed boxes in Figures~\ref{fig2:map}(a) and (c).} \label{fig3:mapavg}
    \end{figure*}
    
    These differences are quantified in Figure~\ref{fig3:mapavg} by the unsigned radial field $\left|B_r\right|$ averaged over latitude (panel (a)) and longitude (panel (b)). The largest difference in the longitude profile appears at $\sim$115$^{\circ}$, where the ADAPT-GONG peak ($\sim$39 G) exceeds the GONG peak ($\sim$31 G) by $\sim$26\%. The largest difference in the latitude profile appears near $-10^{\circ}$, where GONG shows only a weak shoulder of $\sim$9 G, while ADAPT-GONG peaks at $\sim$25 G, comparable to the northern peak near $+12^{\circ}$ that the two maps share. Considering these two profiles together, these locations correspond to the source region AR 12673. %, suggesting that the differences between the two maps arise primarily from how the source region field itself is represented. 
    We further calculate the unsigned magnetic flux within the source region for a more detailed quantitative comparison. Within a common area covering both black dashed boxes in Figures~\ref{fig2:map}(a) and (c), the unsigned flux is $2.4\times10^{22}$~Mx in the processed GONG map and $7.8\times10^{22}$~Mx in the ADAPT-GONG map, a factor of $\sim$3--4 higher in the latter. Restricting this comparison to grid points with strong magnetic field ($\left|B_r\right| > 30$~G) in this source region (as for the contours in Figures~\ref{fig2:map}(b) and (d)) increases this factor to $\sim$6. Within the AR edges identified by EEGGL (yellow outlines in the same panels), the corresponding unsigned fluxes are $7.6\times10^{21}$~Mx and $6.7\times10^{22}$~Mx, differing by almost an order of magnitude. Note that the AR boundary in the ADAPT-GONG map also encloses an area $\sim$4 times larger. Besides, across the 12 ADAPT-GONG realizations, the relative standard deviations of the unsigned flux are below 1\% in all three comparisons, indicating that the local flux differences mainly arise from the data products. 
    We note that the polar fields also differ: poleward of $\pm60^{\circ}$, the GONG map keeps a nearly uniform $\left|B_r\right|$ of 8--9 G, whereas the ADAPT-GONG map is asymmetric, with generally larger $\left|B_r\right|$ in the southern polar region than in the north and with $\left|B_r\right|$ decreasing toward both poles at different rates and to different values. 
    
    Despite these regional differences, the total unsigned magnetic flux is nearly identical between the two maps, $3.84\times10^{23}$ Mx for GONG and $3.74\times10^{23}$ Mx for ADAPT-GONG, a difference of only $\sim$2.6\%; the surface-averaged $\left|B_r\right|$ values are also close to each other, giving 6.31 G and 6.16 G, respectively. 
    Therefore, these two maps differ mainly in how the magnetic flux is distributed, especially in the source active region and the polar regions, rather than in its total amount. The consequences of this redistribution for the modeled corona, solar wind, CME, shock, and SEPs are comprehensively examined in the following section.

\section{Results} \label{sec5:result}

In the following, we present and compare the simulation results driven by the GONG and ADAPT-GONG maps. We describe the steady-state corona and solar wind (Section~\ref{sec5.1:bgsw}), the CME generation and propagation (Section~\ref{sec5.2:cme}), the shock front and magnetic connectivity (Section~\ref{sec5.3:shock}), and the SEP distribution (Section~\ref{sec5.4:sep}), illustrating how the effects of the magnetogram choice evolve through each stage and propagate through the whole modeling chain for the Sun--Earth--Mars system. 

\subsection{Steady-State Corona and Solar Wind} \label{sec5.1:bgsw}

    \subsubsection{Pre-eruption Coronal Structure} \label{sec5.1.1:bgcorona}

\begin{figure*}[tp!]
    \centering {\includegraphics[width=1.0\hsize]{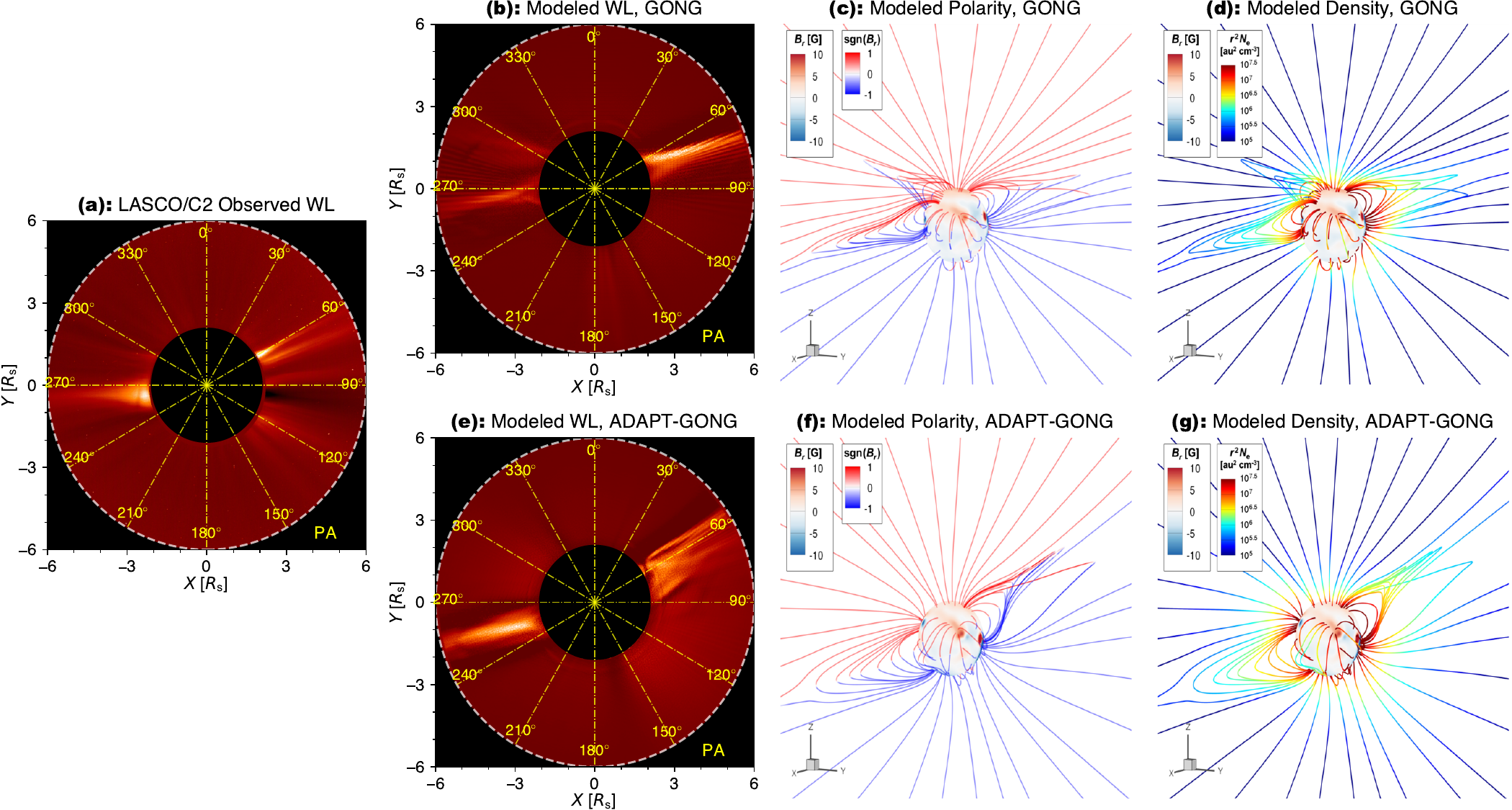}}
    \caption{Pre-eruption solar corona modeled using the two maps, as viewed from Earth. (a) WL image observed by LASCO/C2 shortly before the eruption (15:36 UT). (b), (e) Synthesized WL images from the GONG- and ADAPT-GONG-driven steady-state solutions of AWSoM-R simulations, respectively, with an FOV extending to $r = 6\; R_\mathrm{s}$. The position angle (PA), measured clockwise from the solar north in the plane-of-sky coronagraph images, is marked by the yellow dash-dotted lines, and the central black disks mask the C2 occulter region within 2 $R_\mathrm{s}$. (c), (f) Modeled magnetic field lines colored by their polarity, $\mathrm{sgn}(B_r)$, with the inner boundary surface of SC at $r = 1.1\; R_\mathrm{s}$ colored by $B_r$. (d), (g) Same as panels (c) and (f), except that the field lines are colored by the scaled electron density, $r^2N_\mathrm{e}$.} \label{fig4:bgwl}
    \end{figure*}

    We first examine the steady-state solar corona modeled using the processed GONG and ADAPT-GONG magnetograms. Figure~\ref{fig4:bgwl} shows the comparison of synthesized white-light (WL) images of the pre-eruption corona with the LASCO/C2 observations, together with the modeled 3D magnetic field and density structure. 
    The synthetic WL images are processed using an unsharp-mask filter and an exponential radial filter, following the procedure described in Section 3.1 of \cite{liu2026simulating}. We note that the filters applied to these synthetic WL images differ from those used for the LASCO/C2 observation, with the latter obtained directly from the Integrated Space Weather Analysis \citep[ISWA;][]{maddox2010utilizing} system\footnote{\url{https://iswa.ccmc.gsfc.nasa.gov/iswa_data_tree/observation/solar/soho/} \label{url:iswa}}. 
    
    As shown in Figure~\ref{fig4:bgwl}, both runs overall reproduce the observed east--west streamer structures within the FOV of LASCO/C2. We note that the streamer belt in the GONG-driven solution is flatter, tilted by $\sim$15$^{\circ}$--20$^{\circ}$ with respect to the solar equator (panel (b)), and exhibits larger latitudinal undulations (panels (c) and (d)), whereas the ADAPT-GONG-driven result shows a larger tilt ($\sim$30$^{\circ}$, panel (e)) and follows a smoother, more coherent path with smaller local variations in latitude (panels (f) and (g)). 
    The GONG-driven solution additionally produces a fainter secondary streamer in the northeast limb (PA $\sim$300$^{\circ}$ in panel (b)). This feature has no clear counterpart in the C2 image, or is quite weak there (panel (a)). 
    We find that this additional streamer in the GONG run arises from the far-side active regions near CL = 240$^{\circ}$ and 300$^{\circ}$, where the two maps differ substantially (Figures~\ref{fig2:map} and \ref{fig3:mapavg}). Such differences are expected to be more pronounced on the far side and near the east limb \citep[e.g.,][]{arge2013modeling, yeates2018global, perri2024impact, shi2025role, liu2026simulating}, especially for strong and rapidly evolving active regions, where the photospheric magnetic field is less directly constrained by the solar magnetic field observations. 
    Since the streamer belt and the associated HCS are closely related to the slow solar wind and define the large-scale magnetic configuration through which the CME propagates, the difference in their inclination and morphology further carries into the interplanetary structures discussed below.
    
    \subsubsection{High-Speed Streams} \label{sec5.1.2:bghss}
    
\begin{figure*}[tp!]
    \centering{
    \includegraphics[width=1.0\textwidth]{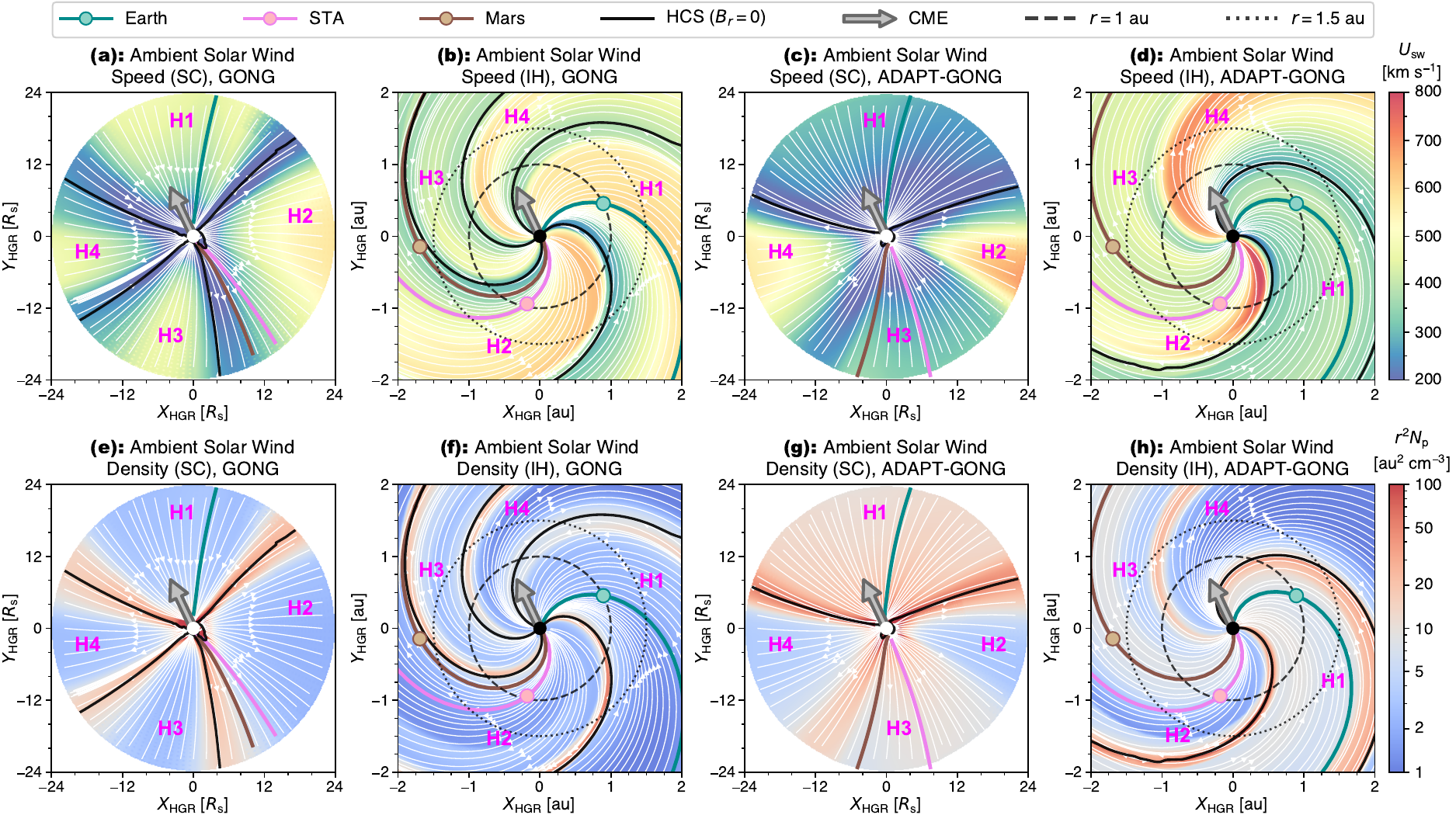}}
    \caption{Steady-state ambient solar wind driven by the two maps, shown in the solar equatorial plane. (a)--(d) Solar wind speed in the SC and IH domains in the heliographic rotating (HGR) coordinate system for the GONG- and ADAPT-GONG-driven runs, as labeled in the title of each panel. (e)--(h) The corresponding scaled proton number density, $r^2N_\mathrm{p}$. In each panel, white curves show the magnetic field lines, and the black curve marks the HCS ($B_r = 0$). Note that the field lines shown here and in Figures~\ref{fig11:cmesc}, \ref{fig12:cmeih}, and \ref{fig15:sepcont}(a)(b) below are traced using only the $x$ and $y$ components of the magnetic field in the equatorial plane and may therefore differ from the 3D field lines. The green, pink, and brown curves show the field lines connected to Earth, STA, and Mars, respectively, with the observer locations marked by the filled circles in the IH panels. The gray arrow indicates the CME propagation direction, and the black dashed and dotted circles mark $r = 1$ au and 1.5 au. The central white disk in the SC panels marks its inner boundary at $r = 1.1\; R_\mathrm{s}$, and the central black disk in the IH panels marks its inner boundary at $r = 20\; R_\mathrm{s}$. H1--H4 label the four major HSSs.} \label{fig5:swsUN}
    \end{figure*}

    Figure~\ref{fig5:swsUN} shows the ambient solar wind plasma speed and scaled density in the solar equatorial plane of the SC and IH domains. In both runs, we identify four major high-speed streams (HSSs) as regions where the solar wind speed at 1 au exceeds that of the adjacent slower wind by more than 100 km$\;$s$^{-1}$ over at least 30$^\circ$ in longitude \citep[adapted from][]{lindblad1981catalogue}. % In both runs, we identify four major high-speed streams (HSSs) as regions where the solar wind flows faster than in neighboring regions. 
    These streams, labeled H1--H4, originate around CL = 90$^{\circ}$, 0$^{\circ}$, 270$^{\circ}$, and 180$^{\circ}$, respectively. They are separated by slow and dense wind, which mostly lies around the HCS (indicated by black curves); the CME (gray arrow) propagates mainly through H1. However, the positions, angular widths, and strengths of these streams differ substantially between the two runs. 
    In the GONG-driven solution, H1 is faster, narrower, and associated with much lower plasma density ($U_\mathrm{sw} \approx 450$ km$\;$s$^{-1}$, $N_\mathrm{p} \approx 300$ cm$^{-3}$ at 0.1 au; $U_\mathrm{sw} \approx 600$ km$\;$s$^{-1}$, $N_\mathrm{p} \approx 2$ cm$^{-3}$ at 1 au). In the ADAPT-GONG-driven solution, H1 is slower and denser ($U_\mathrm{sw} \approx 300$ km$\;$s$^{-1}$, $N_\mathrm{p} \approx 1500$ cm$^{-3}$ at 0.1 au; $U_\mathrm{sw} \approx 400$ km$\;$s$^{-1}$, $N_\mathrm{p} \approx 10$ cm$^{-3}$ at 1 au), whereas the neighboring streams, H2 and H4, reach higher solar wind speeds compared to the GONG-driven solution. %, resulting in weaker density gradients between H1 and the neighboring slow-wind region around the HCS, which can facilitate the lateral expansion of the CME and its driven shock across the stream boundary. 
    On the opposite side of the CME source region, H3 also shows a slower and denser solar wind structure in the ADAPT-GONG solution ($U_\mathrm{sw} \approx 350$ km$\;$s$^{-1}$, $N_\mathrm{p} \approx 800$ cm$^{-3}$ at 0.1 au; $U_\mathrm{sw} \approx 500$ km$\;$s$^{-1}$, $N_\mathrm{p} \approx 8$ cm$^{-3}$ at 1 au) than in the GONG solution ($U_\mathrm{sw} \approx 450$ km$\;$s$^{-1}$, $N_\mathrm{p} \approx 400$ cm$^{-3}$ at 0.1 au; $U_\mathrm{sw} \approx 550$ km$\;$s$^{-1}$, $N_\mathrm{p} \approx 5$ cm$^{-3}$ at 1 au). 
    We next combine EUV observations with the modeled coronal magnetic structure to examine where these streams originate and why their properties differ between the two runs. 
    
\begin{figure*}[tp!]
    \centering{
    \includegraphics[width=1.0\textwidth]{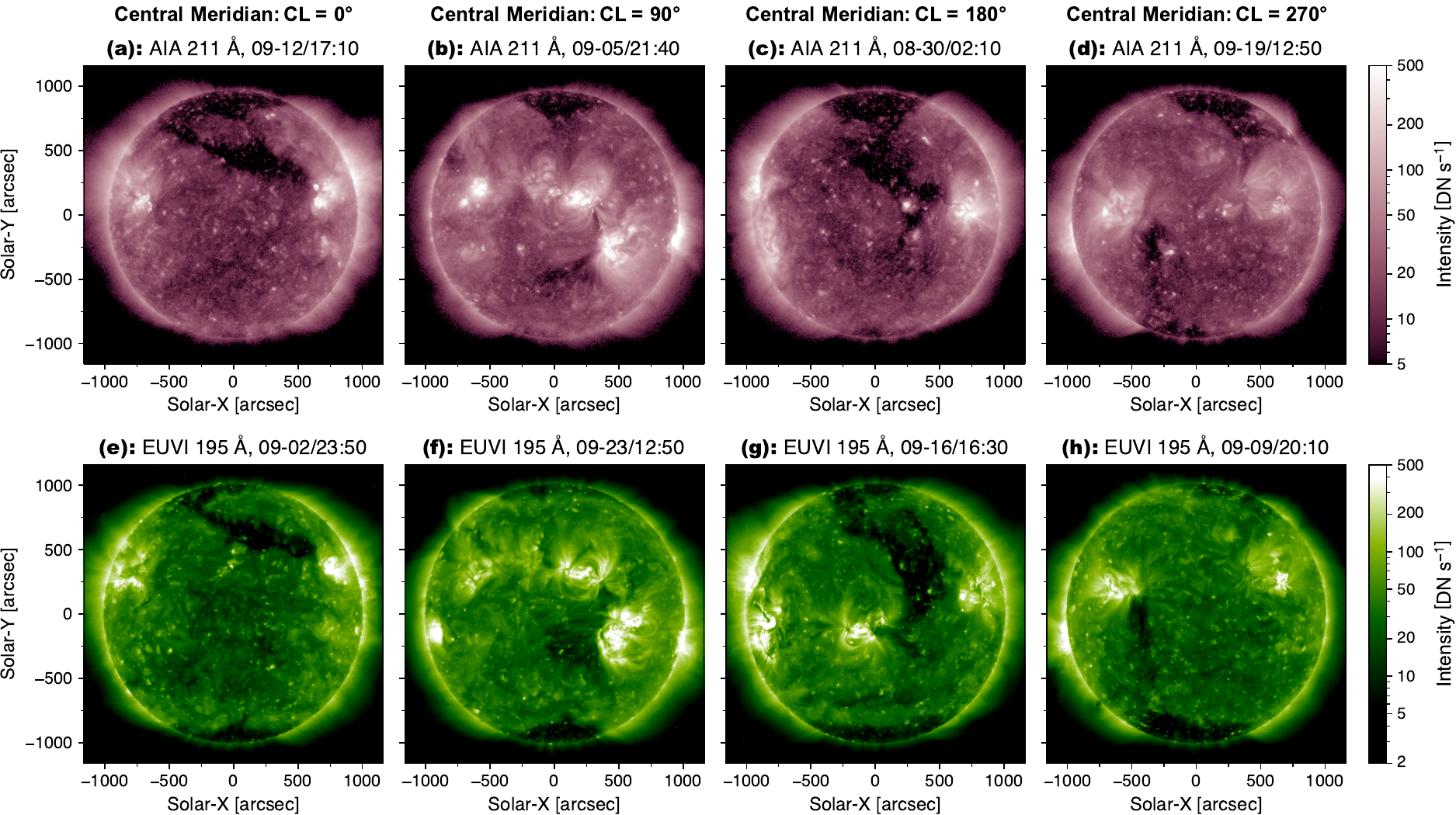}}
    \caption{EUV images of the solar corona obtained between 2017 August 30 and September 23 by SDO/AIA at 211 \AA\ (top row) and STA/EUVI at 195 \AA\ (bottom row). From left to right, the central meridians lie at CL = 0$^\circ$, 90$^\circ$, 180$^\circ$, and 270$^\circ$ and show the potential coronal source regions of the four HSSs, labeled H2, H1, H4, and H3 in Figure~\ref{fig5:swsUN}, respectively. Each column shows the same solar hemisphere as viewed by SDO and STA at different times, with the observation times indicated above each panel.} \label{fig6:fulleuv}
    \end{figure*}

    Figure~\ref{fig6:fulleuv} shows the coronal holes (CHs) that we regard as candidate sources of these four HSS structures, in SDO/AIA and STA/EUVI images centered around the corresponding source longitudes. H1 likely originates from small low-latitude CHs in the southern hemisphere next to AR~12673 near CL = 90$^{\circ}$ (panels (b) and (f)). H2 and H4 appear to correspond to two extensions of the northern polar CH, one reaching down to $\sim$20$^{\circ}$N near CL = 0$^{\circ}$ (panels (a) and (e)) and the other forming a broader extension reaching the equator near CL = 180$^{\circ}$ (panels (c) and (g)). H3 appears to originate from a narrow extension of the southern polar CH that approaches the equator near CL = 270$^{\circ}$ (panels (d) and (h)). Although the EUV images in Figure~\ref{fig6:fulleuv} sample different times and evolutionary stages within the 27-day interval, these CH structures can be identified in both the SDO and STA views. 
    
\begin{figure*}[tp!]
    \centering{
    \includegraphics[width=1.0\textwidth]{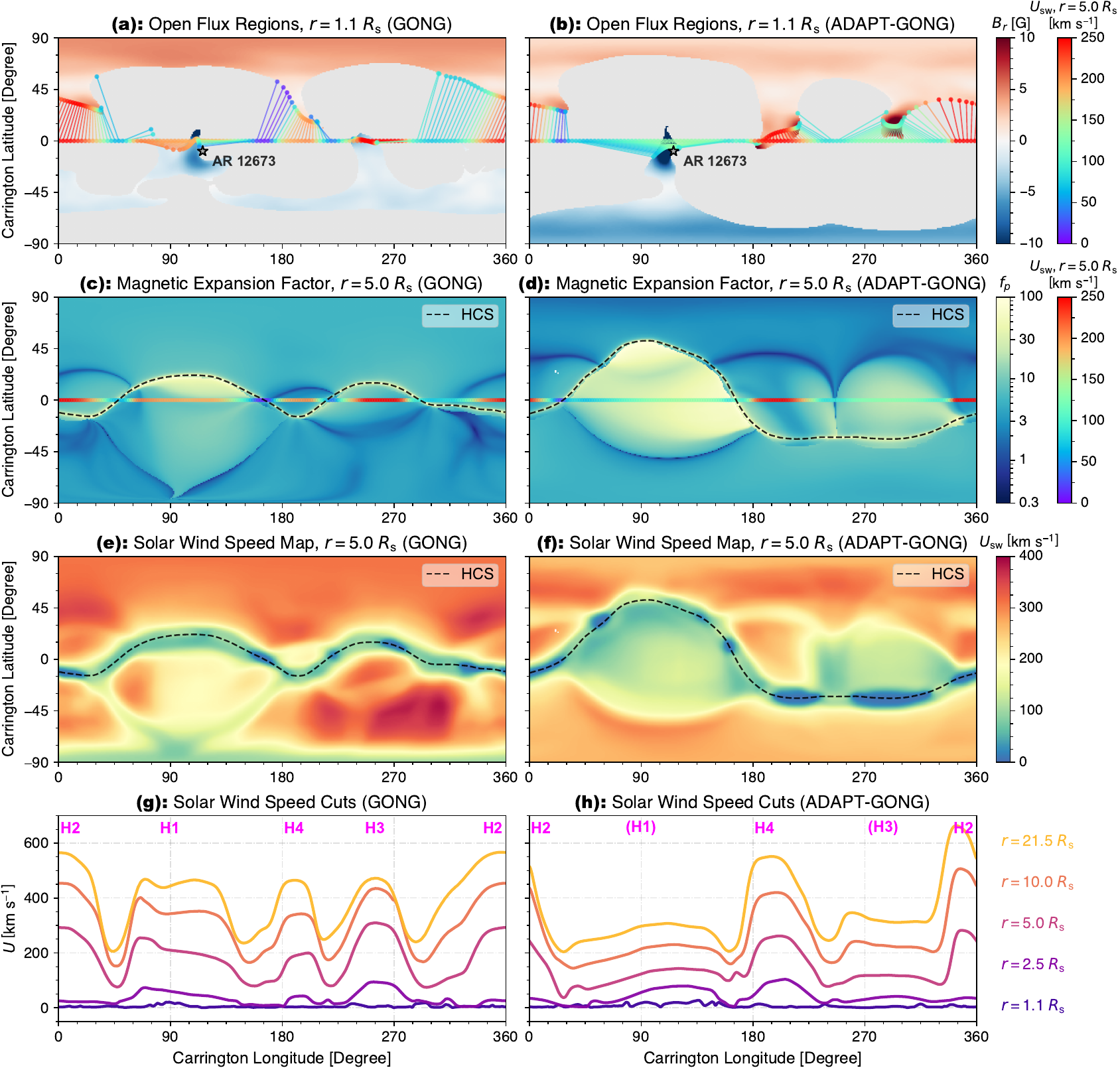}}
    \caption{Comparison of the coronal magnetic structure and solar wind in the steady-state GONG-driven (left column) and ADAPT-GONG-driven (right column) solutions.
    (a), (b) Colored points and connecting segments show the magnetic mapping from equatorial sampling locations at $r = 5\; R_\mathrm{s}$ to their footpoints at $r = 1.1\; R_\mathrm{s}$, with colors indicating the solar wind speed of the sampling locations at $r = 5\; R_\mathrm{s}$. Gray areas mark closed field regions at $r = 1.1\; R_\mathrm{s}$. The black star marks the flare site of AR~12673 (S08W88 as seen from Earth). 
    (c), (d) Magnetic flux tube expansion factor $f_p$ between $r = 1.1\; R_\mathrm{s}$ and 5 $R_\mathrm{s}$ (Equation~(\ref{eq:fp})), with the equatorial sampling locations colored by the solar wind speed as in panels (a) and (b). 
    (e), (f) Solar wind speed at $r = 5\; R_\mathrm{s}$. The black dashed curves in panels (c)--(f) mark the HCS ($B_r = 0$). 
    (g), (h) Equatorial solar wind speed profiles as functions of Carrington longitude at $r = 1.1$, 2.5, 5.0, 10.0, and 21.5 $R_\mathrm{s}$, with line colors noted on the right. Magenta labels indicate the longitudinal sectors of H1--H4 shown in Figure~\ref{fig5:swsUN}, with ``(H1)'' and ``(H3)'' denoting their weaker counterparts in the ADAPT-GONG-driven solution.} \label{fig7:openflux}
    \end{figure*}

    We then relate these observed CH structures to the modeled open flux regions and their corresponding HSSs in the equatorial plane. Figures~\ref{fig7:openflux}(a) and (b) show open flux regions at the SC inner boundary (1.1 $R_\mathrm{s}$). To identify which of these regions supply the equatorial solar wind, we overplot the footpoints of magnetic field lines traced from the equator at $r = 5\; R_\mathrm{s}$ to the inner boundary. The line segments connect each sampling location to its footpoint. The points and segments are colored by the wind speed at the equatorial sampling locations. 
    To characterize how the open flux tubes expand between these two radii, we plot their expansion factor in panels (c) and (d), defined by \cite{wang1990solar}: 
    \begin{equation}
        f_p = \frac{B(\bs{r}_\mathrm{f})}{B(\bs{r})} \left(\frac{r_\mathrm{f}}{r}\right)^2, \label{eq:fp}
    \end{equation}
    where $\bs{r}_\mathrm{f}$ is the position of a field line footpoint on the SC inner boundary at $r_\mathrm{f} = 1.1\; R_\mathrm{s}$, $\bs{r}$ is the sampling position where the same field line crosses the spherical surface at $r = 5\; R\mathrm{s}$, and $B(\bs{r}_\mathrm{f})$ and $B(\bs{r})$ denote the magnetic field strengths at these two positions, respectively. 
    A value of $f_p = 1$ corresponds to the cross-sectional area increase expected for radial expansion, whereas $f_p > 1$ indicates super-radial expansion between the two radii, and vice versa. We also overplot the solar wind speed at the equatorial sampling points to indicate the four HSS structures. 
    Panels (e) and (f) show the solar wind speed at $r = 5\; R_\mathrm{s}$ for direct comparison with the magnetic flux tube expansion in panels (c) and (d); the HCS is overlaid in panels (c)--(f) to indicate the magnetic polarity boundaries and their intersections with the equator. 
    Furthermore, we show in panels (g) and (h) the equatorial wind speed profile at five representative heights between $r = 1.1$ and 21.5 $R_\mathrm{s}$ to illustrate where the stream patterns form. 
    
    Comparing these open flux regions in Figures~\ref{fig7:openflux}(a) and (b) with the observed CHs in Figure~\ref{fig6:fulleuv}, we find that both runs contain open flux in the two northern CH extensions (near CL = 0$^\circ$ and 180$^\circ$ in Figures~\ref{fig6:fulleuv}(a), (c), (e), and (g)) and next to AR~12673 (near CL = 90$^\circ$ in Figures~\ref{fig6:fulleuv}(b) and (f)) in the southern hemisphere, but only the GONG-driven solution contains an open flux region matching the narrow southern CH extension near CL = 270$^\circ$ (Figures~\ref{fig6:fulleuv}(d) and (h)), and only the ADAPT-GONG-driven solution extends the northern open flux region near CL = 180$^\circ$ to the equator, consistent with the observed CH in Figures~\ref{fig6:fulleuv}(c) and (g). 
    These differences are consistent with previous studies showing that the open flux regions of the coronal models depend on both the input magnetic field maps and the model itself \citep[e.g.,][]{lee2011coronal, hayashi2016comparison, li2021comparison, huang2024adjusting}, and that models reproducing the observed CHs can still underestimate the open flux inferred from \textit{in situ} observations \citep{linker2017open}. 
    
    Moreover, as shown in Figures~\ref{fig7:openflux}(a) and (b), the equatorial sampling points at $r = 5\; R_\mathrm{s}$ connect to the open flux regions differently in the two runs. This difference arises because the HCS, which separates points linked to positive- and negative-polarity open flux, has a different shape in these two runs, as illustrated in Figures~\ref{fig7:openflux}(c)--(f) and also reflected in the corresponding streamer structures shown previously in Figure~\ref{fig4:bgwl}. 
    In the GONG-driven solution, the HCS stays within $\sim$20$^{\circ}$ of the equator and crosses it four times, so the equatorial plane contains four magnetic sectors, each filled by one HSS. In the ADAPT-GONG-driven solution, the HCS can rise to $\sim$45$^{\circ}$N near the longitudes of AR~12673, lies south of the equator at all other longitudes, and crosses the equator only twice. Its equatorial plane therefore contains only two sectors, a negative-polarity sector occupied by H1 and a positive-polarity sector shared by H4, H3, and H2. 
    
    In our simulations, the solar wind speed along an open flux tube is primarily determined by two quantities, its footpoint magnetic field strength and its expansion. The Alfv\'{e}n wave Poynting flux injected at the SC inner boundary scales with the local magnetic field strength through the ratio $\left(S_\mathrm{A}/B\right)_{\odot}$, which is constant over the inner boundary in each run (Section~\ref{sec4:map}), so a stronger footpoint field supplies more wave energy. This energy then spreads over the cross section of the flux tube, which grows as $r^2f_p$, so a larger expansion leaves less energy per unit area to heat and accelerate the solar wind \citep[e.g.,][]{wang1990solar, wang1991fast, cranmer2007self}. Consequently, strongly expanding flux tubes carry slower and denser solar wind. 
    The two input maps have different footpoint magnetic field strengths and flux tube expansions, which account for the differences in H1--H4 between the two runs. 
    \begin{enumerate}
        \item H1, located around CL = 90$^\circ$, originates from different open flux regions in these two simulations. 
        In the GONG-driven run, the equatorial field lines of H1 map back mostly to the weak-field, negative-polarity open flux east of AR~12673 (Figure~\ref{fig7:openflux}(a)), and these associated flux tubes expand moderately (Figure~\ref{fig7:openflux}(c)). 
        In the ADAPT-GONG run, regions around AR~12673 show stronger radial magnetic field strength (see more details in Section~\ref{sec4:map}), and field lines from nearly the whole negative sector (from CL = 45$^\circ$ to 180$^\circ$) converge onto the open flux rooted in the trailing negative polarity of this AR, where the magnetic field at the footpoint is strong (Figure~\ref{fig7:openflux}(b)). 
        The strong closed flux of AR 12673 inflates the overlying arcade and can force this open flux to expand into the large volume beneath the northward excursion of the HCS. Since this small footpoint area must fill a sector $\sim$135$^{\circ}$ wide, $f_p$ reaches $\gtrsim 100$, much larger than in the GONG-driven solution (Figure~\ref{fig7:openflux}). Although a stronger footpoint field indicates more wave energy injected, compared with the GONG run, the much larger expansion spreads it over a wider cross section and produces the slower and broader H1 as seen in Figures~\ref{fig5:swsUN}, \ref{fig7:openflux}(f), and \ref{fig7:openflux}(h). 
        
        \item H3, located around CL = 270$^\circ$, also originates from different open flux regions in the two runs. 
        In the GONG-driven solution, the HCS lies north of the equator over the longitude range spanned by H3, and the equatorial field lines of H3 map back to the narrow negative-polarity extension of the southern polar CH (Figure~\ref{fig7:openflux}(a)), with a moderate magnetic expansion factor (Figure~\ref{fig7:openflux}(c)). 
        In the ADAPT-GONG-driven solution, the southern polar open flux region stays at high latitudes around CL = 270$^\circ$, and the HCS lies well south of the equator. The equatorial plane instead samples positive-polarity open flux rooted in a compact, strong-field region next to an AR pair near 300$^{\circ}$ (see Figures~\ref{fig2:map}(c) and \ref{fig7:openflux}(b)). As with H1, the open flux from this compact source expands more strongly than that feeding H3 in the GONG-driven solution ($f_p$ of 10--30 versus 5--10; Figures~\ref{fig7:openflux}(c) and (d)) and produces the weaker H3. 
        We also note that this AR pair and the surrounding field lie on the far side as seen from Earth and are therefore less constrained by recent observations. 

        \item Unlike H1 and H3, H2 (near CL = 0$^{\circ}$) and H4 (near CL = 180$^{\circ}$) originate in both runs from the same open flux regions, the two equatorward extensions of the northern polar open flux region (Figures~\ref{fig7:openflux}(a) and (b)). 
        For H2, both maps have relatively weak fields around the corresponding footpoints near CL = 0$^{\circ}$ (Figures~\ref{fig2:map} and \ref{fig3:mapavg}(a)). Although these weak footpoint fields supply relatively little wave energy, the flux tubes expand only weakly in both runs ($f_p \lesssim 5$; Figures~\ref{fig7:openflux}(c) and (d)), leaving sufficient energy per unit area to accelerate a fast solar wind. 
        For H4, the expansion is similar in the two runs, but the ADAPT-GONG-driven footpoints lie closer to the equator and have about twice the field strength. 
        As a result, in Figures~\ref{fig7:openflux}(g) and (h), H2 forms a clear HSS of comparable speed in both runs at $r = 5\; R_\mathrm{s}$, whereas H4 is slightly faster in the ADAPT-GONG-driven solution.
    \end{enumerate}

    Figures~\ref{fig7:openflux}(g) and (h) further show the solar wind speed cuts at different heliocentric distances, which indicate the formation and outward evolution of the HSSs starting from $r = 1.1\; R_\mathrm{s}$ up to 21.5 $R_\mathrm{s}$ (almost the whole SC domain) with small longitudinal shifts of $\lesssim 10^{\circ}$, so the differences set up in the low corona carry into interplanetary space (Figure~\ref{fig5:swsUN}). 
    At $r = 21.5\; R_\mathrm{s}$, the GONG-driven profile has four distinct plateaus and/or peaks, separated by minima near the four HCS crossings. The ADAPT-GONG profile is dominated by the peaks of H2 and H4, whereas H1 and H3 form broad plateaus of slow wind. 
    We also find that in the ADAPT-GONG run, the local minimum of solar wind speed between H4 and H3 (around CL = 240$^{\circ}$) lies at the boundary between two open flux regions of the same positive polarity (Figures~\ref{fig7:openflux}(b) and (d)) rather than at the HCS, a configuration typical of pseudostreamers \citep[e.g.,][]{wang2007coronal, riley2012interplanetary}. Such boundaries form part of the separatrix web \citep{antiochos2011model, titov2011magnetic} and have been proposed as sources of the slow solar wind \citep[e.g.,][]{fisk2001behavior, antiochos2007structure, abbo2016slow}. 

    In summary, the magnetic field distributions in the two input maps show different open flux regions that govern the footpoint field strength, magnetic flux tube expansion, and variation in the HCS geometry, producing different HSSs and ambient solar wind structures that further affect the subsequent CME propagation, shock connectivity, and particle acceleration and transport, which are detailed in Sections~\ref{sec5.2:cme}--\ref{sec5.4:sep}.

    \subsubsection{Ambient Solar Wind at Observers} \label{sec5.1.3:bginstu}

    We compare both solutions with the \textit{in situ} measurements at Earth, STA, and Mars over the 27-day period centered on 14:00~UT on 2017 September 10. We use measurements from the Wind\footnote{\url{https://wind.nasa.gov/data.php} \label{url:wind}} spacecraft \citep{harten1995design, king2005solar} at Earth/L1 \citep[obtained from][]{cdaw_OMNIsw}, and the STA measurements are obtained from \cite{cdaw_STAsw}. At Mars, the upstream solar wind data are obtained from the MAVEN Solar Wind Ion Analyzer \citep[SWIA;][]{halekas2015solar} and Magnetometer \citep[MAG;][]{connerney2015maven}, with upstream intervals\footnote{Available at \url{https://homepage.physics.uiowa.edu/~jhalekas/drivers.html}. \label{url:mavensw}} identified following \cite{halekas2017seasonal} and \cite{halekas2017structure}, and also obtained from the Analyzer of Space Plasmas and Energetic Atoms \citep[ASPERA-3\footnote{\url{https://pds-geosciences.wustl.edu/missions/mars_express/aspera.htm} \label{url:aspera}};][]{barabash2006analyzer, ramstad2017mars} on board Mars Express \citep[MEX\footnote{\url{https://pds-ppi.igpp.ucla.edu/mission/Mars_Express} \label{url:mex}};][]{chicarro2004mars}. 
    
\begin{figure*}[tp!]
    \centering{
    \includegraphics[width=1.0\textwidth]{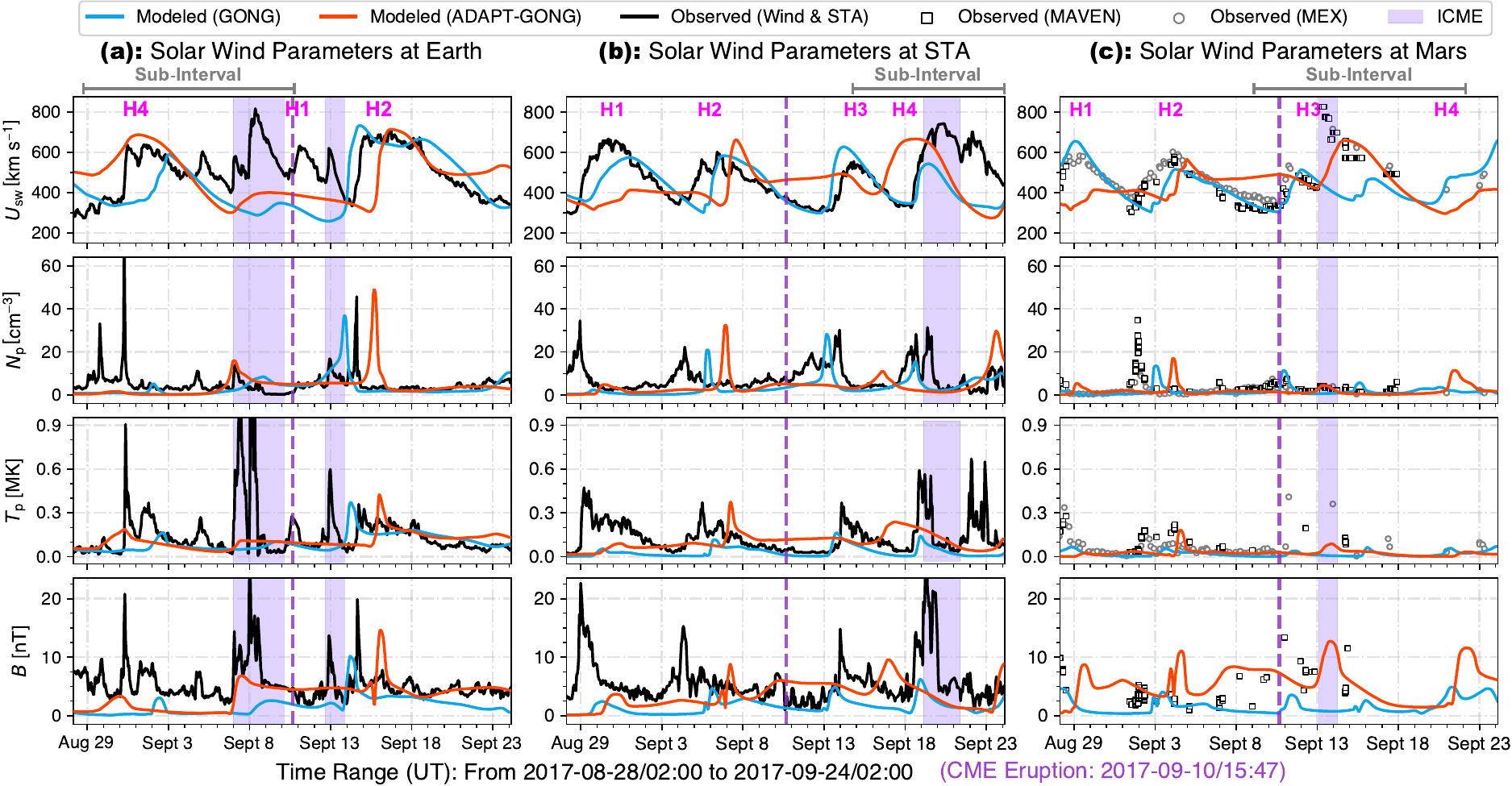}}
    \caption{Time-series comparison of the ambient solar wind parameters between the simulation results and \textit{in situ} observations over a 27-day period centered on the input magnetogram acquisition time at (a) Earth, (b) STA, and (c) Mars. 
    Rows from top to bottom show the solar wind speed ($U_\mathrm{sw}$), proton number density ($N_\mathrm{p}$), proton temperature ($T_\mathrm{p}$), and magnetic field strength ($B$). Black curves show the observations by Wind at L1 and by STA, and the black squares and gray circles show the MAVEN and MEX measurements upstream of the Martian bow shock, respectively. Blue and red curves represent the GONG- and ADAPT-GONG-driven simulation results, respectively. 
    A horizontal gray bracket above each top panel marks the sub-interval during which the longitude of each observer is within $\pm90^{\circ}$ of the CME source region, AR~12673, used for the subsequent quantitative analysis. The purple shaded periods mark intervals of interplanetary shock and/or ICME passages, which are excluded from the distance calculation. The vertical purple dashed line marks the CME eruption time, and labels H1--H4 at the top indicate the arrivals of the corotating HSSs labeled in Figure~\ref{fig5:swsUN}.} \label{fig8:swstime}
    \end{figure*}

    Figure~\ref{fig8:swstime} shows the time-series comparison of the simulated background solar wind with the \textit{in situ} observations at Earth/L1, STA, and Mars. Note that Figure~\ref{fig5:swsUN} shows the solar wind properties in the solar equatorial plane, whereas the actual observer trajectories lie a few degrees above or below this plane. The detailed HSS structure and associated solar wind conditions along these trajectories as shown in Figure~\ref{fig8:swstime} can therefore differ slightly from those shown in the equatorial slice. 
    Here, both runs capture the large-scale structure of the ambient solar wind, with the corotating HSSs H1--H4 arriving at each observer at approximately the observed times with reasonable and comparable magnitudes in both solutions. 
    However, the agreement with observations varies between the two runs, depending on the HSS and solar wind parameter considered. The main differences are as follows: %The differences lie in the features each run reproduces best: 
    \begin{enumerate}
        \item The GONG-driven run reproduces the timing and amplitude of the HSS fronts more closely at STA and Mars, especially the arrival of H1 and H2 relevant to the eruption source region. 
        
        \item The ADAPT-GONG-driven run tends to broaden the HSSs and shift their arrival times, as seen, for example, during the transition from H2 to H3 at STA and Mars. Nonetheless, it reproduces H4 better, particularly its amplitude at Earth and STA, and gives an overall stronger interplanetary magnetic field that is closer to the observed level at all three observers, while the GONG run systematically underestimates it. 
        % The stronger $B$ field in the ADAPT-GONG run may partly result from its stronger high-latitude magnetic fields, which can increase the open magnetic flux extending into the heliosphere \citep[e.g.,][]{linker2017open,riley2019can}.
        
        \item At Earth, both runs underestimate the peak speed of H1, with the ADAPT-GONG run closer to the observations. We also note that this stream arrives after a preceding ICME structure, which is not modeled in this study. 
    \end{enumerate}
    Among them, the first two differences can be traced back to the speed profiles within the SC domain (Figures~\ref{fig7:openflux}(g) and (h)), which propagate through interplanetary space to the observers. The four distinct plateaus and peaks of the GONG profile arrive as separate HSS fronts. The broad H1 and H3 plateaus of the ADAPT-GONG profile arrive as broader streams, and its strong H4 peak explains why this run could reproduce the H4 amplitude better at Earth and STA. 
    Hence, at the eruption time, the two runs yield different solar wind conditions at each observer %, especially within H1 at Earth and in the regions between H2 and H3 at STA and Mars 
    (see also Figure~\ref{fig5:swsUN}). We also note that the shaded intervals around September 13 at Earth and Mars, identified as described below, mark the passage of the interplanetary shock and/or ejecta associated with the CME eruption studied here, which are further discussed in Sections~\ref{sec5.2.2:cmepropagate} and \ref{sec5.4.2:profile}. 
    
    We then quantify the model--observation agreement using the data-distance metric introduced in Section~3.5 of \cite{sachdeva2019validation}, which measures the difference between the simulated and observed \textit{in situ} time profiles and is computed here for the four solar wind parameters. The intervals affected by interplanetary shock and/or interplanetary CME (ICME) passages, plotted as purple regions in Figure~\ref{fig8:swstime}, are excluded from the calculation, as such transient structures are not expected to be fully reproduced by a steady-state background solar wind simulation \citep[e.g.,][]{reiss2016verification, sachdeva2019validation, jivani2023global, koban2025dailyAGU}. 
    For Earth and STA, the ICME intervals are first obtained from publicly available catalogs for Wind\footnote{\url{https://wind.nasa.gov/ICME_catalog/ICME_catalog_viewer.php} \label{url:icmeearth}} at Earth/L1 \citep{nieves2018understanding} and STA\footnote{\url{https://stereo-ssc.nascom.nasa.gov/data/ins_data/impact/level3/} \label{url:icmesta}} \citep{jian2018stereo}, together with a comprehensive ICME catalog, ICMECAT\footnote{\url{https://helioforecast.space/icmecat} \label{url:helio4cast}} \citep{mostl2025icmecat, mostl2026magnetic}, and are then visually inspected and confirmed. At Mars, where no such ICME catalog is available and the \textit{in situ} measurements have relatively sparse temporal coverage, the ICME intervals are identified by visual inspection of the \textit{in situ} data, guided by \cite{lee2018observations} and \cite{ramstad2018september} for the 2017 September 10 event. 
    The resulting distance values for the two simulations are summarized in Table~\ref{tab2:distsw}. 
    
    \begin{table*}[tp!]
    \begin{center}
    \caption{Distance values between the simulated and observed solar wind parameters at Earth, STA, and Mars for the simulations driven by GONG and ADAPT-GONG maps.} \label{tab2:distsw}
    \setlength\tabcolsep{5.5pt}{
    \renewcommand{\arraystretch}{1.225}
    \begin{tabular}{cc ccccc ccccc}
        \toprule
        \multirow{2}{*}{Observer} & \multirow{2}{*}{Map} & \multicolumn{5}{c}{\shortstack{Over the full 27-day interval}} & \multicolumn{5}{c}{\shortstack{\footnotesize Over the source AR-centered sub-interval}} \\
        \cmidrule(lr){3-7} \cmidrule(lr){8-12}
        & & $U_\mathrm{sw}$ & $N_\mathrm{p}$ & $T_\mathrm{p}$ & $B$ & Mean & $U_\mathrm{sw}$ & $N_\mathrm{p}$ & $T_\mathrm{p}$ & $B$ & Mean \\
        \midrule \multirow{2}{*}{Earth} & GONG & 0.097 & 0.100 & 0.076 & 0.144 & \textbf{0.104} & 0.108 & 0.153 & 0.115 & 0.208 & 0.146 \\
        & ADAPT-GONG & 0.126 & 0.115 & 0.075 & 0.118 & 0.109 & 0.139 & 0.144 & 0.098 & 0.162 & \textbf{0.136} \\
        \midrule \multirow{2}{*}{STA} & GONG & 0.100 & 0.127 & 0.177 & 0.134 & \textbf{0.135} & 0.099 & 0.064 & 0.179 & 0.122 & \textbf{0.116} \\
        & ADAPT-GONG & 0.140 & 0.168 & 0.172 & 0.108 & 0.147 & 0.105 & 0.151 & 0.209 & 0.088 & 0.138 \\
        \midrule \multirow{2}{*}{Mars} & GONG & 0.134 & 0.131 & 0.125 & N.A. & \textbf{0.130} & 0.222 & 0.155 & 0.210 & N.A. & 0.196 \\
        & ADAPT-GONG & 0.136 & 0.147 & 0.127 & N.A. & 0.137 & 0.138 & 0.194 & 0.123 & N.A. & \textbf{0.152} \\
        \bottomrule
    \end{tabular}}
    \end{center}
    \tablecomments{The distance values are calculated following Section~3.5 of \cite{sachdeva2019validation}, excluding the ICME intervals shaded in Figure~\ref{fig8:swstime}. 
    The comparisons are performed over a 27-day interval (August 28--September 24) centered on the map time and over a sub-interval, during which each observer's longitude lies within $\pm90^{\circ}$ of the CME source region, AR 12673 (marked by the horizontal gray brackets in Figure~\ref{fig8:swstime}). 
    The distance value of the magnetic field strength at Mars is not calculated, because the MAVEN/MAG measurements upstream of the Martian bow shock provide too few points over this 27-day interval for a meaningful comparison, and MEX/ASPERA-3 carries no magnetometer. 
    For each observer and comparison interval, the smaller mean distance value, indicating better agreement with the observations, is shown in bold.}
    \end{table*}
    
    Previous validation studies have shown that well-reproduced ambient solar wind profiles typically yield distance values of $\lesssim 0.15$ over a 27-day window, while clear deviations correspond to $\gtrsim 0.2$ \citep[e.g.,][]{sachdeva2019validation, sachdeva2023solar, koban2025validation, liu2026counterintuitive}. In both runs, most of the resulting distance values fall below 0.15 over the full interval, comparable to those reported in the studies above, indicating that both solar wind backgrounds provide a reasonable basis for the CME and SEP modeling that follows. 
    Meanwhile, we note that the relative performance of the two runs varies among different solar wind parameters. The GONG-driven background reproduces the solar wind speed better at all three observers, with smaller distance values, which is consistent with its better agreement in the HSSs shown in Figure~\ref{fig8:swstime}. The ADAPT-GONG-driven run instead shows smaller distance values in the magnetic field strength at Earth and STA, consistent with its better reproduction of the observed $B$ magnitude in Figure~\ref{fig8:swstime}. 
    Averaged over all parameters during the 27 days, the GONG-driven run yields slightly smaller mean distances at all three observers. 
    
    More relevant to this eruptive event are the sub-intervals during which each observer's Carrington longitude lies within $\pm90^{\circ}$ of the CME source region, as indicated by the horizontal gray brackets in Figure~\ref{fig8:swstime}. The agreement degrades in this window for most cases, with some values even exceeding 0.2. The relative performance of the two runs also changes. The ADAPT-GONG run gives a slightly smaller mean distance at Earth (0.136 versus 0.146) and a more pronounced improvement at Mars (0.152 versus 0.196), the latter driven mainly by the solar wind speed (0.138 versus 0.222), whereas the GONG-driven run remains better at STA (0.116 versus 0.138). Thus, neither map is uniformly better across all different observers, parameters, or time intervals, while these two runs provide overall comparable ambient solar wind solutions.

    \subsubsection{Magnetic Footpoints of Observers} \label{sec5.1.4:fp}
    
    \begin{table*}[tp!]
    \begin{center}
    \caption{Magnetic connectivity of Earth, STA, and Mars at the CME eruption time.} \label{tab3:fp_comparison}
    \setlength\tabcolsep{3pt}{
    \footnotesize
    % \fontsize{10}{10}\selectfont
    \renewcommand{\arraystretch}{1.25}
    \begin{tabular}{cccc ccc ccc}
        \toprule
        \multirow{2}{*}{Observer} & \multirow{2}{*}{\shortstack{Location: $(r,\phi,\theta)$}} & \multirow{2}{*}{\shortstack{$U_\mathrm{sw}$}} & \multirow{2}{*}{Map} & \multicolumn{3}{c}{\shortstack{Ballistic and PFSS models}} & \multicolumn{3}{c}{\shortstack{AWSoM-R simulation}} \\
        \cmidrule(lr){5-7} \cmidrule(lr){8-10}
        & & & & $\phi_\mathrm{f}$ & $\theta_\mathrm{f}$ & $\Delta \phi_\mathrm{f,\,AR}$ & $\phi_\mathrm{f}$ & $\theta_\mathrm{f}$ & $\Delta \phi_\mathrm{f,\,AR}$ \\
        \midrule \multirow{2}{*}{Earth} & \multirow{2}{*}{$(1.01\;\mathrm{au}$, $27.2^\circ$, $7.2^\circ)$} & \multirow{2}{*}{$488\;\mathrm{km\;s^{-1}}$} & GONG & $107.2^\circ$ & $-11.4^\circ$ & $-7.8^\circ$ & $98.7^\circ$ & $6.5^\circ$ & $-16.3^\circ$ \\
        & & & ADAPT-GONG & $87.3^\circ$ & $12.8^\circ$ & $-27.7^\circ$ & $107.5^\circ$ & $-4.1^\circ$ & $-7.5^\circ$ \\
        \midrule \multirow{2}{*}{STA} & \multirow{2}{*}{$(0.96\;\mathrm{au}$, $259.3^\circ$, $-4.2^\circ)$} & \multirow{2}{*}{$378\;\mathrm{km\;s^{-1}}$} & GONG & $329.6^\circ$ & $47.4^\circ$ & $-145.4^\circ$ & $322.9^\circ$ & $42.5^\circ$ & $-152.1^\circ$ \\
        & & & ADAPT-GONG & $333.2^\circ$ & $-5.9^\circ$ & $-141.8^\circ$ & $297.5^\circ$ & $12.5^\circ$ & $-177.5^\circ$ \\
        \midrule \multirow{2}{*}{Mars} & \multirow{2}{*}{$(1.66\;\mathrm{au}$, $183.5^\circ$, $-4.9^\circ)$} & \multirow{2}{*}{$328\;\mathrm{km\;s^{-1}}$} & GONG & $295.8^\circ$ & $11.7^\circ$ & $-179.2^\circ$ & $275.6^\circ$ & $-1.9^\circ$ & $160.6^\circ$ \\
        & & & ADAPT-GONG & $270.7^\circ$ & $-6.5^\circ$ & $155.7^\circ$ & $291.1^\circ$ & $13.0^\circ$ & $176.1^\circ$ \\
        \bottomrule
    \end{tabular}}
    \end{center}
    \tablecomments{The observation locations and corresponding magnetic connectivity parameters are summarized here. The observer locations are given by heliocentric distance ($r$), Carrington longitude ($\phi$), and Carrington latitude ($\theta$) at the CME eruption time. The solar wind speed ($U_\mathrm{sw}$) is the one-day average of the \textit{in situ} measurements at each observer before the CME eruption and serves as the input for the ballistic back-mapping. The magnetic footpoint coordinates $(\phi_\mathrm{f}, \theta_\mathrm{f})$, also given in the Carrington heliographic coordinate system, are obtained either from ballistic back-tracing combined with the PFSS model or by tracing magnetic field lines through the steady-state AWSoM-R solution, each driven by the GONG or ADAPT-GONG synoptic map listed in the map column. The longitudinal separation from the source region, $\Delta\phi_\mathrm{f,\,AR}$, is the footpoint longitude measured relative to the Carrington longitude of AR~12673 ($\phi_\mathrm{AR}=115^\circ$) and wrapped to the range $[-180^\circ, 180^\circ)$. 
    Here, $\phi_\mathrm{AR}$ is obtained from the observed flare location (S08W88 as viewed from Earth), derived from SDO/AIA 94~\AA\ images (see footnote~\ref{url:solarsoft} in Section~\ref{sec2:overview}), and Earth's Carrington longitude ($\sim$27$^\circ$) at the eruption time, providing a consistent observational reference for both runs, independent of the AR boundaries or centroids in the input maps. 
    Negative and positive values of $\Delta\phi_\mathrm{f,\,AR}$ indicate that the footpoint lies eastward and westward of the source region, respectively.} 
    \end{table*}
    
    Another important aspect of the pre-eruption conditions is the magnetic connectivity of each observer to the source region. To characterize it, we derive the photospheric footpoints in association with Earth, STA, and Mars using two approaches \citep[see also][]{gieseler2023solar, kennis2024magnetic, biro2025developing}. 
    In the first approach, we take the \textit{in situ} solar wind speed averaged over the one-day window before the eruption as input, ballistically map each observer's location back to the prescribed source surface at $r = 2.5\; R_\mathrm{s}$, where the magnetic field is assumed to be radial and open, along the nominal Parker spiral, and then trace the field line to the photosphere through the PFSS model. In the second approach, the field lines are traced through the 3D, steady-state global MHD solution of AWSoM-R. Both approaches are applied to the two simulation results using the GONG and ADAPT-GONG maps. 
    Table~\ref{tab3:fp_comparison} summarizes the observer locations, magnetic footpoint coordinates, and their separations with respect to the source region at the CME eruption time. The corresponding footpoints are also marked on the input magnetograms in Figures~\ref{fig2:map}(a) and (c). 
    
    All four reconstructions place Earth's magnetic footpoint near the solar equator and close to the source region, only 7.5$^\circ$--27.7$^\circ$ eastward of AR~12673. Earth therefore has comparatively favorable initial magnetic connectivity to the source region. By contrast, the footpoints of STA and Mars lie far from the source longitude, with absolute longitudinal separations of 141.8$^\circ$--177.5$^\circ$ and 155.7$^\circ$--179.2$^\circ$, respectively, placing both on the far side of the Sun relative to the source region. The close proximity of Earth's footpoint to the source region is qualitatively consistent with its prompt SEP onset at 16:15 UT on September 10, whereas the onsets at Mars and STA are delayed to 19:50~UT on September 10 and $\sim$08:00~UT on September 11, respectively (see Figure~\ref{fig1:overview}). 
    These magnetic footpoints at the pre-eruption stage thus provide an initial reference for interpreting the observed SEP onset differences, although the detailed timing also depends on when each observer becomes connected to the evolving CME-driven shock and on the subsequent particle transport, as discussed in the shock-capturing and SEP analyses below (Sections~\ref{sec5.3:shock} and \ref{sec5.4:sep}). 
    Moreover, although all four reconstructions agree on this broad distinction, the detailed footpoint coordinates shown in Table~\ref{tab3:fp_comparison} vary with both the input magnetogram and the reconstruction method, with longitudinal spreads of $\sim$20$^{\circ}$--36$^{\circ}$ for Earth, STA, and Mars. These variations reflect the sensitivity of the inferred connectivity to the model assumptions and to the resulting coronal and heliospheric magnetic structure.

\subsection{CME Generation and Propagation} \label{sec5.2:cme}

    \subsubsection{Coronagraph Images and CME Kinematics} \label{sec5.2.1:wl}
    
    Based on the two background solar wind solutions, we then parameterize the CME flux rope using EEGGL and insert the resulting force-imbalanced flux rope into each steady-state corona. 
    Given the same observed CME speed of 2650 km$\;$s$^{-1}$, EEGGL derives different GL flux rope parameters from the AR magnetic fields of the two maps (see Table~\ref{tab1:param}): a compact flux rope with a lower apex height and a strong magnetic field (radius $r_0 = 0.30\; R_\mathrm{s}$, apex height 0.50 $R_\mathrm{s}$, and field strength $B_0 = 99.2$ G) for the GONG run, and a larger flux rope with a higher apex height and a weaker field (radius $r_0 = 0.80\; R_\mathrm{s}$, apex height 1.00 $R_\mathrm{s}$, and field strength $B_0 = 35.3$ G) for the ADAPT-GONG run, both anchored in AR~12673, with their footpoints differing slightly in longitude and latitude. %anchored at nearly the same source location. 
    The total poloidal magnetic fluxes are $1.45\times10^{22}$ Mx and $3.68\times10^{22}$ Mx for the GONG and ADAPT-GONG flux ropes, respectively, differing by a factor of $\sim$2.5 and further highlighting the distinct flux rope properties derived by EEGGL from the two input magnetograms. 
    Note that we do not model the CME initiation or energy buildup; instead, an analytical GL flux rope is inserted into the steady-state corona to launch the CME. The excess magnetic energy introduced by the flux rope is subsequently converted into kinetic, thermal, and gravitational energy, driving the expansion, heating, and acceleration of the CME \citep[e.g.,][]{roussev2003three, manchester2004three, manchester2008three, jin2013numerical, jin2017data, alvarado2022simulating, zhao2024solar, wraback20253d}. 
    
\begin{figure*}[tp!]
    \centering{
    \includegraphics[width=1.0\textwidth]{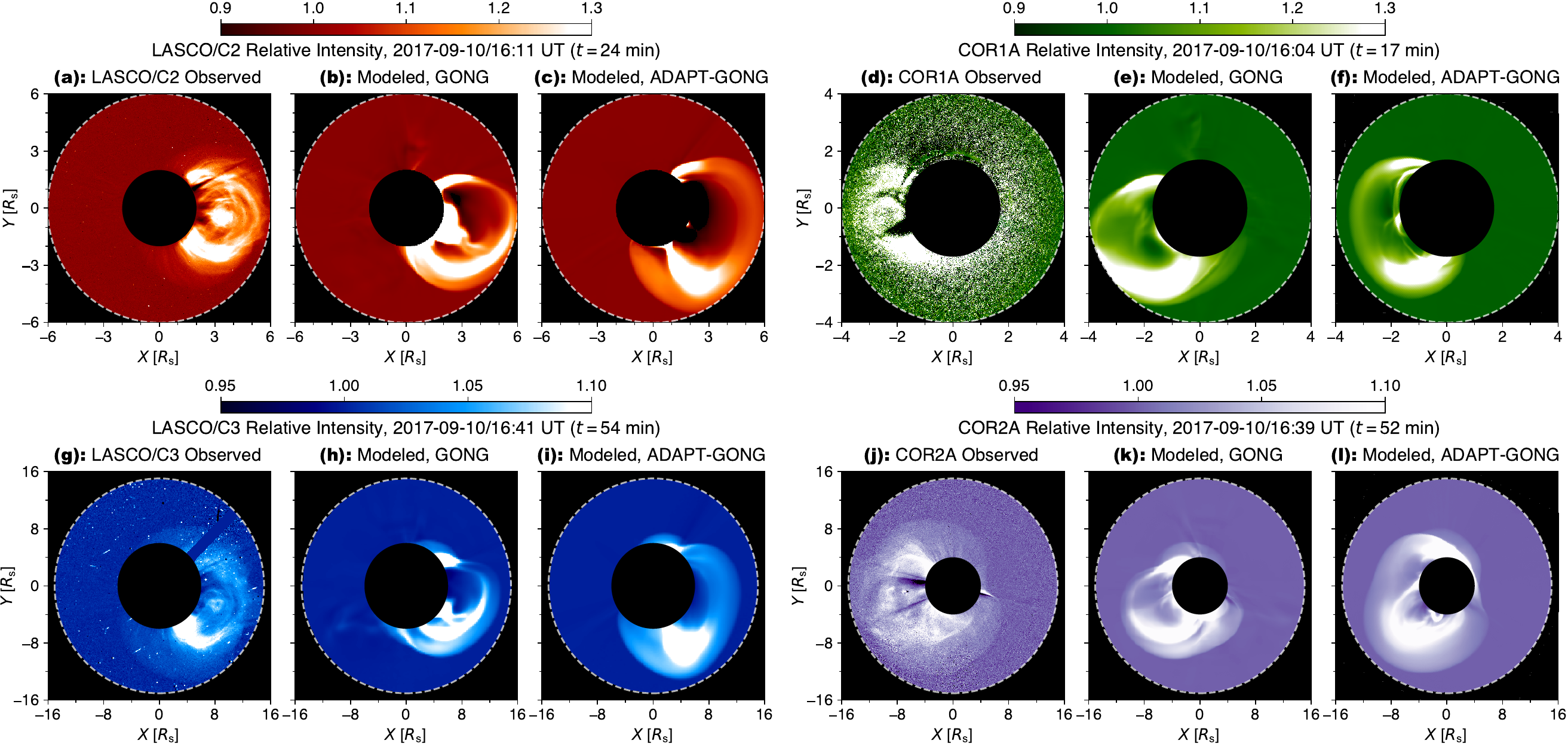}}
    \caption{Comparison of the observed and synthetic WL images of the CME, shown as the ratio of the CME image to the pre-event background. 
    (a)--(c) LASCO/C2 WL images at 16:11 UT ($t = 24$ min after eruption onset), showing the observation and the GONG- and ADAPT-GONG-driven simulations, respectively, with an FOV of 2--6 $R_\mathrm{s}$. (d)--(f) Corresponding STA/COR1 images at 16:04 UT ($t = 17$ min; FOV: 1.7--4 $R_\mathrm{s}$). (g)--(i) LASCO/C3 images at 16:41 UT ($t = 54$ min; FOV: 6--15 $R_\mathrm{s}$). (j)--(l) STA/COR2 images at 16:39 UT ($t = 52$ min; FOV: 4--15 $R_\mathrm{s}$). The central black disks mask the corresponding occulter regions.} \label{fig9:wl}
    \end{figure*}

    Figure~\ref{fig9:wl} shows the comparison of the observed and synthetic WL images of the CME during the early phase of its evolution, using the corresponding coronagraph FOVs from SOHO and STA. When calculating the relative intensities shown in the figure, we also account for the effect of the F corona far from the Sun \citep[set to be 2.3 $R_\mathrm{s}$; see][for details]{billings1966corona, manchester2008three, jin2017chromosphere}. 
    About 20 minutes after the eruption onset, both runs reproduce the position and morphology of the observed bright front toward the western limb in the LASCO/C2 view (panels (a)--(c)) and toward the eastern limb in the STA/COR1 view (panels (d)--(f)). 
    However, we can find clear differences between the two runs. 
    \begin{enumerate}
        \item At $t \approx 20$~min in the LASCO/C2 FOV, the GONG-driven CME more closely reproduces the observed angular width and shows a distinct bright core behind the leading front (panel (b)), whereas the ADAPT-GONG-driven CME expands over a broader latitudinal range (panel (c)). This broader expansion in the ADAPT-GONG run is consistent with the larger radius of its inserted flux rope, whose footpoints may span a wider region and thus allow interaction with more of the surrounding streamers. 
        
        \item At $t \approx 20$~min in the STA/COR1 view, the GONG-driven front (panel (e)) is displaced farther southward and reaches a slightly farther heliocentric distance than the ADAPT-GONG-driven front (panel (f)), indicating a faster early propagation in this view, which may partly reflect differences in the CME propagation direction between the two runs. We note that although the observed front is visible, the COR1 image here shows a high noise level owing to its exposed front lens, which introduces a high instrumental background and lowers the signal-to-noise ratio after background subtraction \citep{thompson2010background}. 
    
        \item Later, at around 50 minutes after the CME onset, both modeled fronts remain broadly consistent with the observed halo. 
        The GONG-driven CME, however, is narrower and more structured (panels (h) and (k)), whereas the ADAPT-GONG-driven CME undergoes greater lateral expansion and spans a wider range of PAs (panels (i) and (l)). In particular, the ADAPT-GONG run better reproduces the faint intensity enhancement along the southern flank seen in the observations, while this feature is much less evident in the GONG run. 
        This stronger lateral expansion is also consistent with the larger flux rope inserted in the ADAPT-GONG run. Since the magnetic energy of a flux rope scales approximately as $B_0^2 r_0^3$ \citep[e.g.,][]{borovikov2017eruptive}, the ADAPT-GONG flux rope carries about 2.4 times the energy of the more compact GONG flux rope%despite its weaker magnetic field
        , as also reflected in the poloidal magnetic flux difference noted above. 
    \end{enumerate}

    \begin{figure*}[tp!]
    \centering{
    \includegraphics[width=0.9\textwidth]{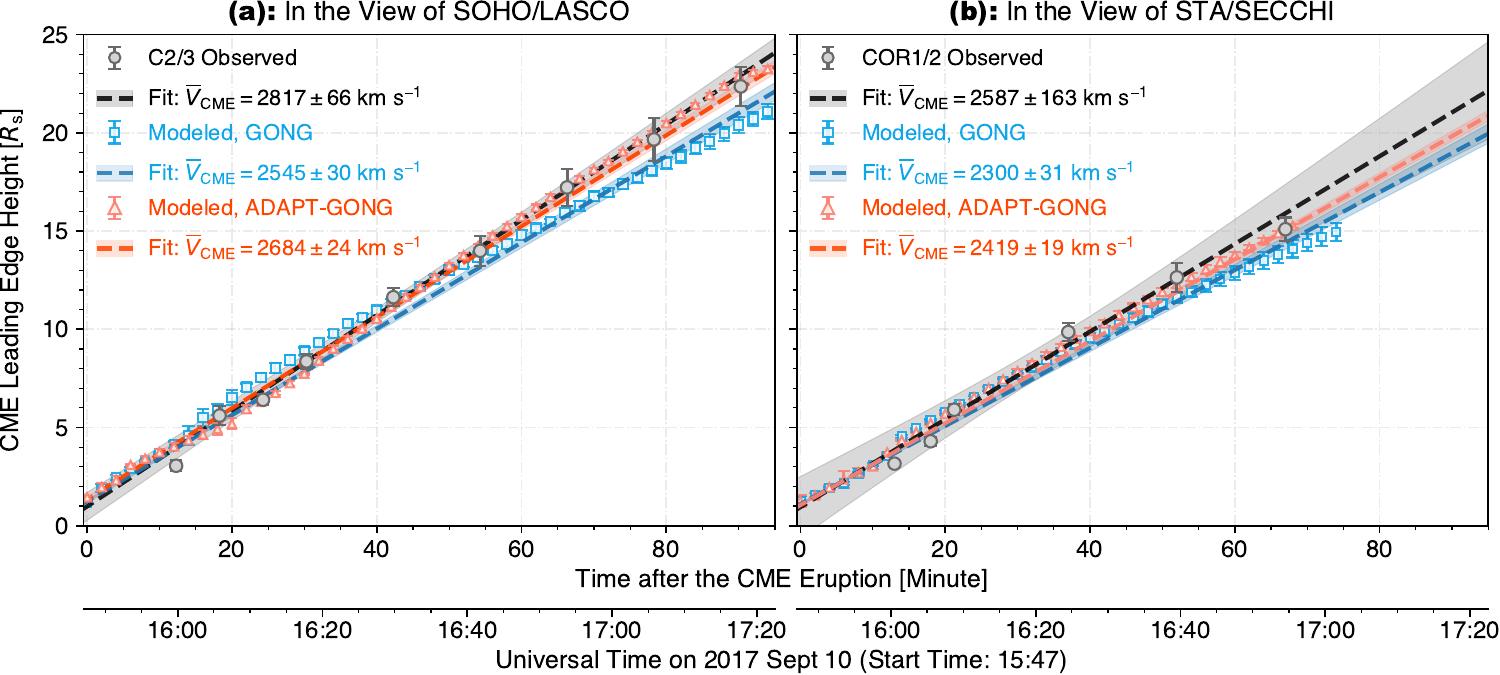}}
    \caption{CME leading-edge height as a function of time in the view of (a) SOHO/LASCO and (b) STA/SECCHI. 
    Gray circles show the leading-edge heights measured from the C2/C3 and COR1/COR2 images, respectively, while blue squares and red triangles show the heights extracted from the corresponding synthetic WL images of the GONG and ADAPT-GONG runs. The vertical bars indicate the uncertainties arising from the edge-detection algorithm of \cite{chen2025decent}. 
    The dashed lines in corresponding colors show the linear fits, with the fitted average speeds annotated in each panel and the shaded bands indicating the corresponding fit uncertainties. The bottom axis gives the universal time on 2017 September 10, with the CME eruption onset at 15:47 UT.} \label{fig10:cmeh}
    \end{figure*}
    
    Based on each frame of the observed and synthesized WL image series, at cadences of 12 minutes and 1 minute, respectively, we track the CME leading-edge height and derive the CME kinematics separately for each viewpoint. The CME heights are extracted with an edge-detection algorithm of \cite{chen2025decent}, which transforms each WL image into polar coordinates and applies K-means clustering to separate the CME enhancement from the background, allowing the leading edge to be identified as a function of PA. Hereafter, for each frame, the CME leading-edge height is taken as the maximum radial extent of the detected front over the selected PA range. The procedure includes several tunable parameters, including the PA range, the sizes of the median and Savitzky--Golay filters and the morphological structuring element, and the K-means clustering criterion. Rather than fixing these a priori, we run this algorithm over a set of combinations spanning a reasonable range for each parameter and take the median and standard deviation of the resulting leading-edge heights at each frame, shown in Figure~\ref{fig10:cmeh} as the symbols and vertical bars. 
    
    Figure~\ref{fig10:cmeh} shows the resulting CME leading-edge height--time profiles for the SOHO/LASCO and STA/SECCHI viewpoints, tracked out to 24 $R_\mathrm{s}$ in the LASCO view, the outer boundary of the SC domain, and to $\sim$15 $R_\mathrm{s}$ in the SECCHI view. 
    Linear fits over the first $\sim$1.5 hours yield average leading-edge speeds of $2817\pm66$, $2545\pm30$, and $2684\pm24$ km$\;$s$^{-1}$ in the SOHO/LASCO view for the observation, GONG, and ADAPT-GONG runs, respectively, and $2587\pm163$, $2300\pm31$, and $2419\pm19$ km$\;$s$^{-1}$ in the STA/SECCHI view. Both modeled CMEs propagate more slowly than observed, while the ADAPT-GONG-driven CME is consistently closer to the observed speed, differing by $\sim$5\% in the SOHO view and $\sim$7\% in the STA view, compared with $\sim$10\% for the GONG-driven CME in both views. 
    We also note that both runs track the observed heights better over the first $\sim$40 minutes, during which the two modeled fronts also remain comparable. Afterward, the ADAPT-GONG front continues to move ahead and follows the observed heights more closely, whereas the GONG-driven CME falls progressively farther behind (starting from $\sim$15 $R_\mathrm{s}$ and $\sim$10 $R_\mathrm{s}$ in the views of SOHO/LASCO and STA/SECCHI, respectively). These results indicate that the propagation speed difference between the two runs starts to become more apparent in the outer coronagraph FOVs, with a larger lag developing in the GONG run \citep[see also][]{vrvsnak2004kinematics, sachdeva2015cme}. 
    
    \subsubsection{CME Propagation} \label{sec5.2.2:cmepropagate}
    
\begin{figure*}[tp!]
    \centering{ 
    \includegraphics[width=1.0\textwidth]{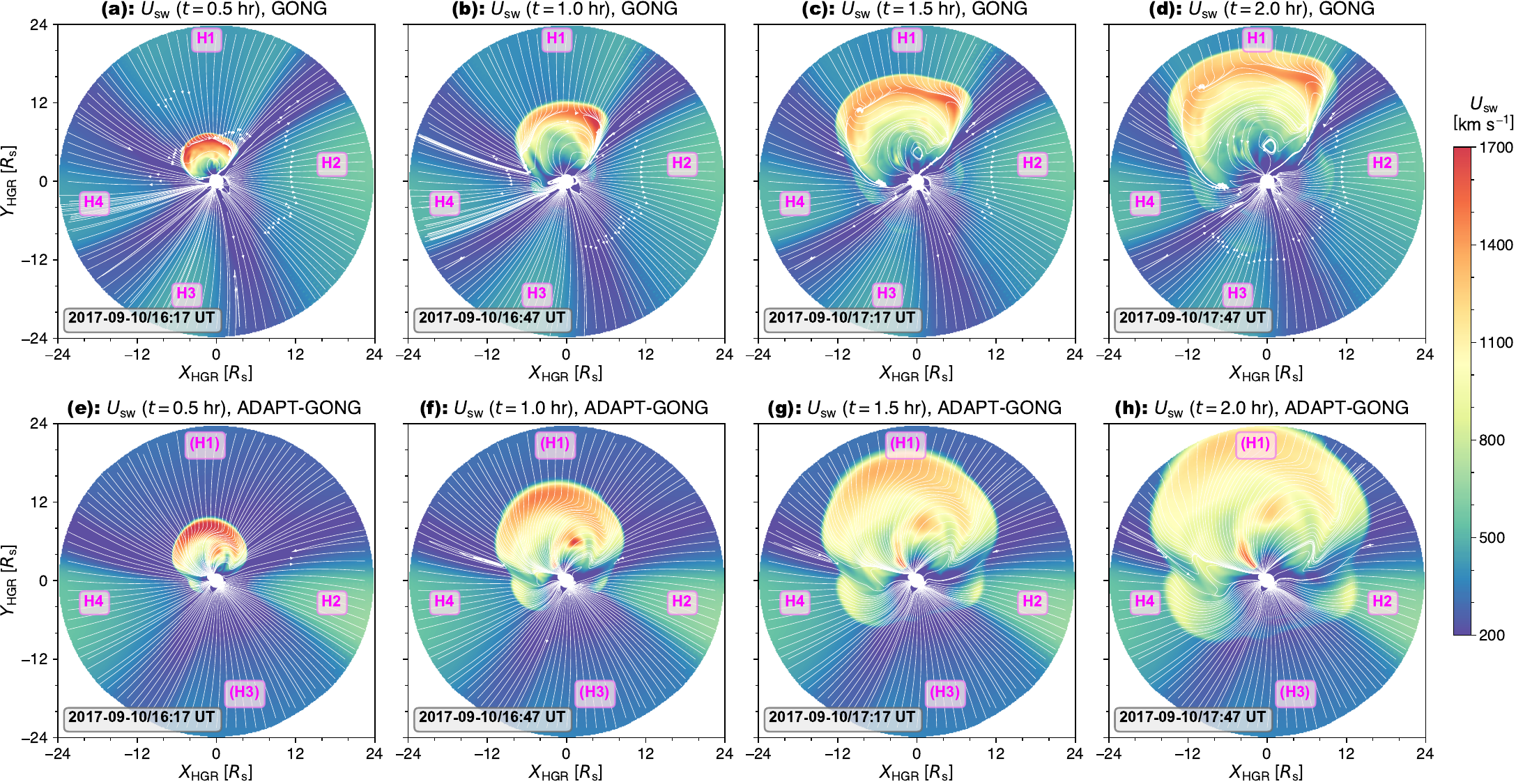}}
    \caption{CME propagation in the SC domain, shown by the solar wind speed in the solar equatorial plane at $t = 0.5$, 1.0, 1.5, and 2.0 hr after the CME eruption onset (columns from left to right, with the corresponding universal times annotated), for the GONG- and ADAPT-GONG-driven runs in the top and bottom rows, respectively. White curves show the magnetic field lines, and H1--H4 mark the HSSs labeled in Figure~\ref{fig5:swsUN}, with ``(H1)'' and ``(H3)'' in parentheses indicating the weaker counterparts in the upstream solar wind background of the ADAPT-GONG run.} \label{fig11:cmesc}
    \end{figure*}

\begin{figure*}[tp!]
    \centering{
    \includegraphics[width=1.0\textwidth]{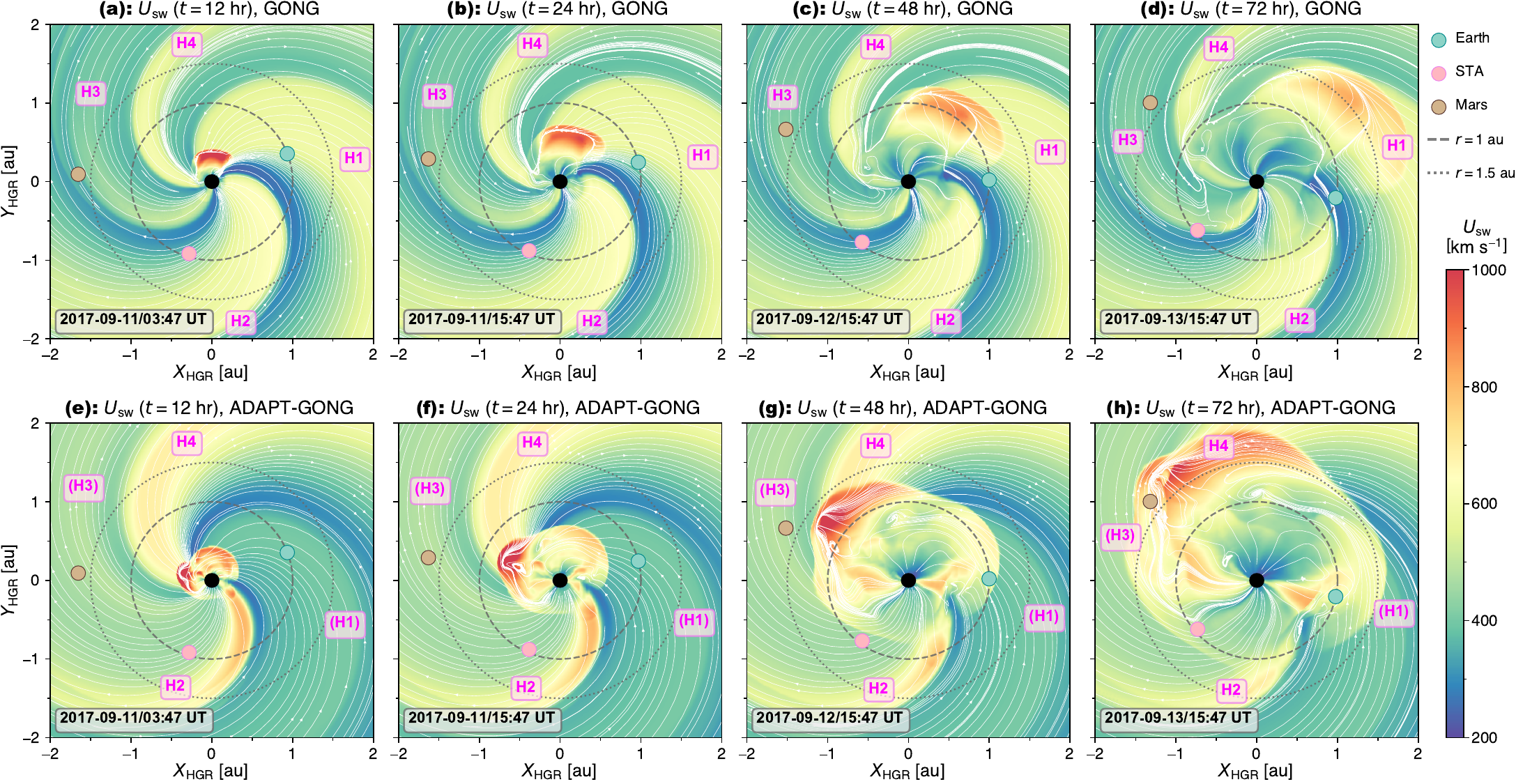}}
    \caption{Similar to Figure~\ref{fig11:cmesc}, but for the ICME propagation in the IH domain at $t=12$, 24, 48, and 72 hr after the CME eruption onset. The green, pink, and brown circles mark the locations of Earth, STA, and Mars, and the black dashed and dotted circles mark $r = 1$ and 1.5 au, respectively.} \label{fig12:cmeih}
    \end{figure*}

    Figures~\ref{fig11:cmesc} and \ref{fig12:cmeih} show the CME propagation in the solar equatorial plane through the SC and IH domains, where interactions with the structured ambient solar wind serve as a major source of the differences between the two solutions. 
    In the SC domain (Figure~\ref{fig11:cmesc}), the CME flux rope in each simulation is launched toward the same sector but propagates through different upstream backgrounds. 
    In the GONG run (Figures~\ref{fig11:cmesc}(a)--(d)), the CME encounters a strong but narrow HSS, H1. As the CME propagates, its flanks encounter slow, dense wind around the warped HCS on both sides of H1 while the relative speed between the front and the fast ambient wind is small (see also Figures~\ref{fig5:swsUN} and \ref{fig7:openflux}(e)), so that the front becomes flattened and its lateral expansion is confined. %The large speed difference between the CME flanks and the surrounding slow wind can also contribute to this confinement by enhancing the drag on the flanks \citep[e.g.,][]{zhou2017numerical, mayank2024swasti}. 
    By contrast, in the ADAPT-GONG run (Figures~\ref{fig11:cmesc}(e)--(h)), H1 is much weaker but wider, and the neighboring slow-wind belt is more inclined (see also the HCS in Figures~\ref{fig4:bgwl} and \ref{fig7:openflux}(f)). As H1 is wider, the slow, dense wind on both sides lies farther from the CME axis. The flanks of the wide ADAPT-GONG flux rope cross this slow wind by $t\approx1$ hr (Figure~\ref{fig11:cmesc}(f)) and then run into the faster H2 and H4 structures. Hence, the CME expands more freely and maintains a smoother and broader front. The front also propagates slightly faster than the GONG-driven result (see also Figure~\ref{fig10:cmeh}), in spite of a slower, denser upstream solar wind. 
    This faster propagation, as well as the wider lateral expansion, is mainly because the inserted flux rope in the ADAPT-GONG run carries about 2.4 times the magnetic energy of the GONG flux rope (Section~\ref{sec5.2.1:wl}), and this extra energy drives the CME mainly within the first few $R_\mathrm{s}$, where the Lorentz force dominates over the drag \citep[e.g.,][]{sachdeva2017cme, manchester2017physical, shen2022propagation}. %The drag becomes important only farther out, by which time the CME front in the ADAPT-GONG run is already $\sim$3~$R_\mathrm{s}$ ahead at $t = 1$ hr (Figures~\ref{fig11:cmesc}(b) and (f)). 
    These early interactions also shape the shock geometry discussed below in Section~\ref{sec5.3:shock}, where the broader flux rope expanding into the more tilted streamer belt produces a shock that extends to higher latitudes in the ADAPT-GONG-driven run. 
    These differences reflect the combined effects of the ambient coronal and solar wind structures and the properties of the inserted flux ropes. 

    As the CME propagates into the IH domain as an ICME (Figure~\ref{fig12:cmeih}), its interaction with the large-scale stream structure determines whether and when the disturbance reaches each observer. The two runs diverge sharply here. 
    In the GONG run, the ICME propagates into the stream interaction region (SIR) formed ahead of H1. The slow, dense wind on both flanks suppresses the lateral expansion and confines the disturbance to a narrow longitudinal sector, continuing the confinement already seen in the SC domain (Figures~\ref{fig11:cmesc}(a)--(d)). The sustained momentum exchange with this dense ambient plasma also exerts strong drag on the ICME, contributing to the weakening of the driven shock \citep[see also][]{vrvsnak2013propagation, dumbovic2018drag, temmer2023cme}. %The sustained momentum exchange with the denser ambient plasma then erodes the driven shock, as expected from the drag exerted by the ambient wind. 
    In the ADAPT-GONG run, the wider and faster ICME instead overtakes the upstream slower stream (H1) over a broad range of longitudes, %so it keeps expanding laterally rather than being confined. Its eastern flank expands into H4, which is faster in this run, and the boundary region between H1 and H4 presents a weaker density gradient, allowing the flank to expand through the CME-solar wind interaction. 
    because its flux rope is more energetic (Section~\ref{sec5.2.1:wl}) and H1 is slower and wider (Section~\ref{sec5.1:bgsw}). As the CME/ICME flanks have already crossed the slow, dense wind in the SC domain (Figures~\ref{fig11:cmesc}(f)--(h)), the solar wind drag decreases \citep[e.g.,][and references therein]{zhou2017numerical, kay2022beyond, mayank2024swasti}. The ICME therefore keeps expanding laterally rather than being confined. Its eastern flank expands into H4, which is faster in this run, and this flank carries the fastest sector of the ICME at 48 and 72 hr (Figures~\ref{fig12:cmeih}(g)--(h)). 
    Later, the ICME-driven shock front in the ADAPT-GONG run crosses $r = 1$ au near Earth's longitude at $t\approx2$ days and the ejecta sweep the sector toward Mars at $r\approx1.5$--1.7 au around 2.5 days, consistent with the \textit{in situ} observations of shock arrivals discussed below in Figure~\ref{fig16:profile}. In both runs, the ICME does not reach STA, also consistent with the observations and other model predictions for this event \citep[e.g.,][]{guo2018modeling, luhmann2018shock}. 
    
    In addition to the difference in lateral expansion described above, the ambient HSSs further modulate the local shock strength in both runs: the compression ratio is larger where the ICME propagates into the slow, dense wind whereas it is smaller within the rarefied fast streams. This variation further contributes to the asymmetries of the shock surfaces shown in Figure~\ref{fig13:shk3d}, thereby affecting the longitudinal extent of the SEP source region discussed later in Section~\ref{sec5.4:sep}. We further note that such interactions between CMEs/ICMEs and the structured ambient solar wind, including the HSSs, SIRs, and the HCS, and their effects on CME deformation, propagation, and shock evolution have been studied extensively in previous numerical and observational works \citep[e.g.,][]{odstrvcil1999three, winslow2016longitudinal, heinemann2019cme, palmerio2022cme, geyer2023interaction, wijsen2023effect, manchester2025high, liu2026counterintuitive}. 

    % Finally, we note that two smaller eruptions ($\lesssim 1000$ km$\;$s$^{-1}$) occurred from the same active region about 16--24 hr before the 15:47 UT eruption and are not included in our simulations. These preceding eruptions were substantially weaker than the main event and were therefore less likely to drive strong shocks or produce a large suprathermal seed population. 
    % Being weaker and well separated in time, they likely left relatively few seed particles near the region where the main shock forms, and the separation is longer than that typically considered for the twin-CME scenario \citep[e.g.,][]{li2012twin, ding2020modeling}. We therefore do not expect a strong direct CME--CME interaction in the low corona, although some preconditioning of the ambient heliosphere by the earlier eruptions may still be possible and is not examined in this work.     
    We note that two smaller eruptions ($\lesssim 1000$ km$\;$s$^{-1}$) occurred from the same active region about 16--24 hr before the 15:47 UT eruption, first observed by SOHO/LASCO C2 at 16:24 UT and 23:12 UT on September 9 with linear speeds of $\sim$470 and $\sim$1020 km$\;$s$^{-1}$, respectively (CDAW catalog; see footnote~\ref{url:cdawcme}). 
    In DONKI, the corresponding CMEs\footnote{The DONKI entries for the September 9 CMEs with start times of 16:48~UT and 23:12~UT are available at \url{https://ccmc.gsfc.nasa.gov/DONKI/events/13089/1/} and \url{https://ccmc.gsfc.nasa.gov/DONKI/events/13095/2/}, respectively.} are listed with start times of 16:48 UT and 23:12 UT on September 9, average speeds of 465 and 333 km$\;$s$^{-1}$ within $r = 21.5\; R_\mathrm{s}$, initial propagation directions of S09W82 and N02W92, and angular half-widths of $35^\circ$ and $33^\circ$, respectively. 
    They were substantially weaker than the main event, less likely to drive strong shocks, and separated by a longer interval than typically considered for the twin-CME scenario \citep[e.g.,][]{li2012twin, ding2020modeling}. %We therefore focus on the September 10 15:47 UT eruption and do not model the preceding eruptions or direct CME--CME interactions. 
    Nevertheless, these two preceding CMEs may have disturbed the ambient solar wind through which the main CME propagated, potentially affecting its expansion, shock evolution, and magnetic connectivity. Our present simulations focus on the September 10 eruption at 15:47 UT and do not include these preceding CMEs, their effects on the ambient solar wind, or direct CME--CME interactions. 
    Another consideration is the residual energetic particle population from even earlier SEP events on September 4 and 6. The associated CMEs were first observed at 20:36~UT on September 4 and 12:24~UT on September 6, with average speeds of 1114 and 1238~km s$^{-1}$ within $r = 21.5\; R_\mathrm{s}$, initial propagation directions of S10W10 and S15W23, and angular half-widths of $46^\circ$ and $45^\circ$, respectively, as reported in DONKI\footnote{Details about the September 4 (20:36~UT) and September 6 (12:24~UT) CMEs are available at \url{https://ccmc.gsfc.nasa.gov/DONKI/events/12989/3/} and \url{https://ccmc.gsfc.nasa.gov/DONKI/events/13013/2/}, respectively.}. 
    Although their corresponding interplanetary shocks passed Earth on September 6 and 8, respectively, elevated low-energy proton intensities associated with these earlier events were still observed at Earth and STA before the September 10 eruption \citep{bruno2019spectral}. These residual particles may have provided additional seed particles for reacceleration by the September 10 CME-driven shock \citep[see also other case studies by][]{gomez2017sunward, palmerio2021cme, niemela2023advancing}. Our simulations do not explicitly account for this particle population as a reservoir. Together with the omission of the preceding CMEs/ICMEs and their effects on the ambient solar wind discussed above, this represents a limitation that may affect the modeled shock evolution, magnetic connectivity, and SEP intensities. 
    % The ambient heliosphere was nevertheless disturbed beforehand. SEP events associated the September 4 and 6 eruptions, left elevated low-energy proton intensities observed at Earth and STA \citep{bruno2019spectral}, while the corresponding interplanetary shocks passed Earth on September 6 and 8, respectively, with disturbed solar wind conditions persisting afterward. These earlier events may therefore have preconditioned the solar wind and contributed to the suprathermal seed population \citep[see also other case studies by][]{gomez2017sunward, palmerio2021cme, niemela2023advancing}. Our simulations do not include these preceding CMEs/ICMEs or the resulting seed particles, which represents a major limitation that may affect the modeled shock evolution, magnetic connectivity, and SEP intensities. 

\subsection{Shock Front and Magnetic Connectivity} \label{sec5.3:shock}
    
    To characterize the CME-driven shock, we apply the shock-capturing tool to both simulations, which is described in Section~4.3 of \cite{liu2025physics} and has been utilized in studies by \cite{chen2025evidence}, \cite{liu2026counterintuitive} and \cite{zhang2026radio}. Along each radial direction of a longitude--latitude grid, the shock front is at the largest speed jump, the local shock normal is then determined from the captured shock surface geometry, and the upstream and downstream states are sampled to derive the shock parameters. 
    In this work, we adopt a speed-jump threshold of $-50$~km$\;$s$^{-1}$ across the shock front, which is the default value of the shock-capturing tool and allows the shock front to be clearly identified in the present analysis, but can be adjusted for better visualization if needed. We then track the shock continuously from the low corona into interplanetary space until this criterion is no longer satisfied. 
    
\begin{figure*}[tp!]
    \centering{
    \includegraphics[width=1.0\textwidth]{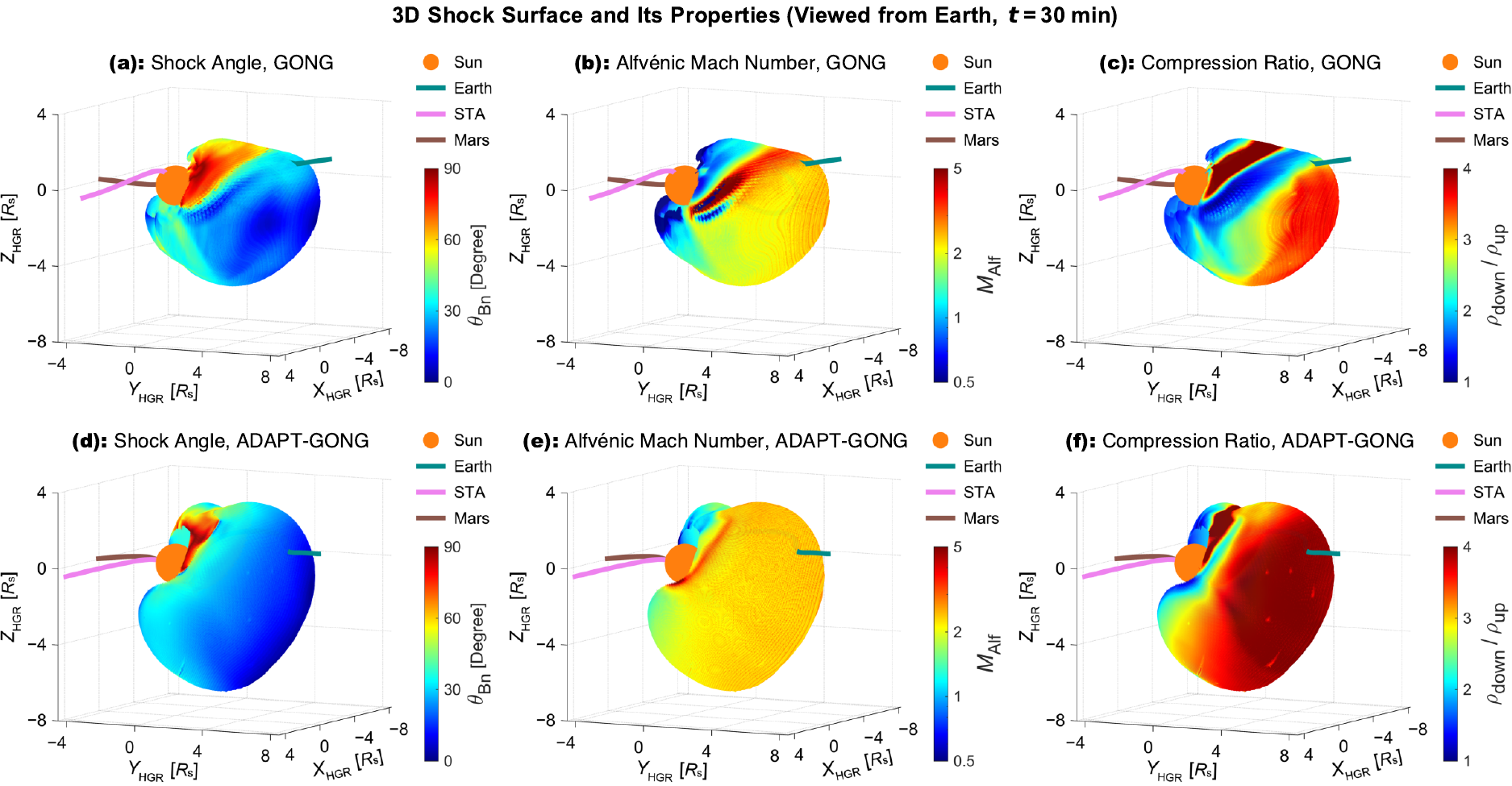}}
    \caption{Captured 3D shock surfaces at $t = 30$ minutes after the CME eruption onset, as viewed from the Earth direction, for the GONG- and ADAPT-GONG-driven runs in the top and bottom rows, respectively. The surfaces are colored by the shock angle ($\theta_{Bn}$; left column), the Alfv\'{e}nic Mach number ($M_\mathrm{Alf}$; middle column), and the density compression ratio ($\rho_\mathrm{down}/\rho_\mathrm{up}$; right column). In each panel, the orange sphere marks the Sun, and the green, pink, and brown curves show the magnetic field lines connected to Earth, STA, and Mars, respectively.} \label{fig13:shk3d}
    \end{figure*}

    At $t = 30$ minutes, Figure~\ref{fig13:shk3d} shows the captured 3D shock fronts and their magnetic connections to Earth, STA, and Mars. At this time, the shock fronts in both runs are quasi-parallel ($\theta_{Bn}\lesssim 30^{\circ}$) over most of their surfaces, with quasi-perpendicular bands near the flanks, and $M_\mathrm{Alf}$ typically ranges from 2 to 3. 
    However, the two runs differ in both geometry and strength. The ADAPT-GONG-driven shock extends to higher latitudes, consistent with its wider propagated flux rope and the more tilted HCS and streamer belt in its solar wind background. Its compression ratio approaches 4 over a large fraction of the surface, including the nose. In the GONG run, the shock is generally weaker, with the compression ratio peaking at $\sim$3--3.5 near the nose and falling below 2.5 toward the flanks.
    In both runs, we also find a narrow band in the northern hemisphere where the compression ratio temporarily exceeds 4 and $M_\mathrm{Alf}$ is locally enhanced. This region corresponds to a strong interaction of the shock with the slow, dense plasma surrounding the HCS. The relatively weak magnetic field and high plasma density in this region reduce the local Alfv\'{e}n speed, resulting in a larger $M_\mathrm{Alf}$ and an enhanced compression ratio. 
    We note that the occurrence of compression ratios exceeding the adiabatic strong-shock limit of 4 is also attributed to nonideal processes, such as heat conduction, included in the current MHD simulation, as discussed by \cite{jin2022assessing}. %In addition, the flux rope inserted for the CME eruption is initially out of force balance, so the shock parameters during the early stage of the simulation can be extremely large. 
    
    \begin{figure*}[tp!]
    \centering{
    \includegraphics[width=1.0\textwidth]{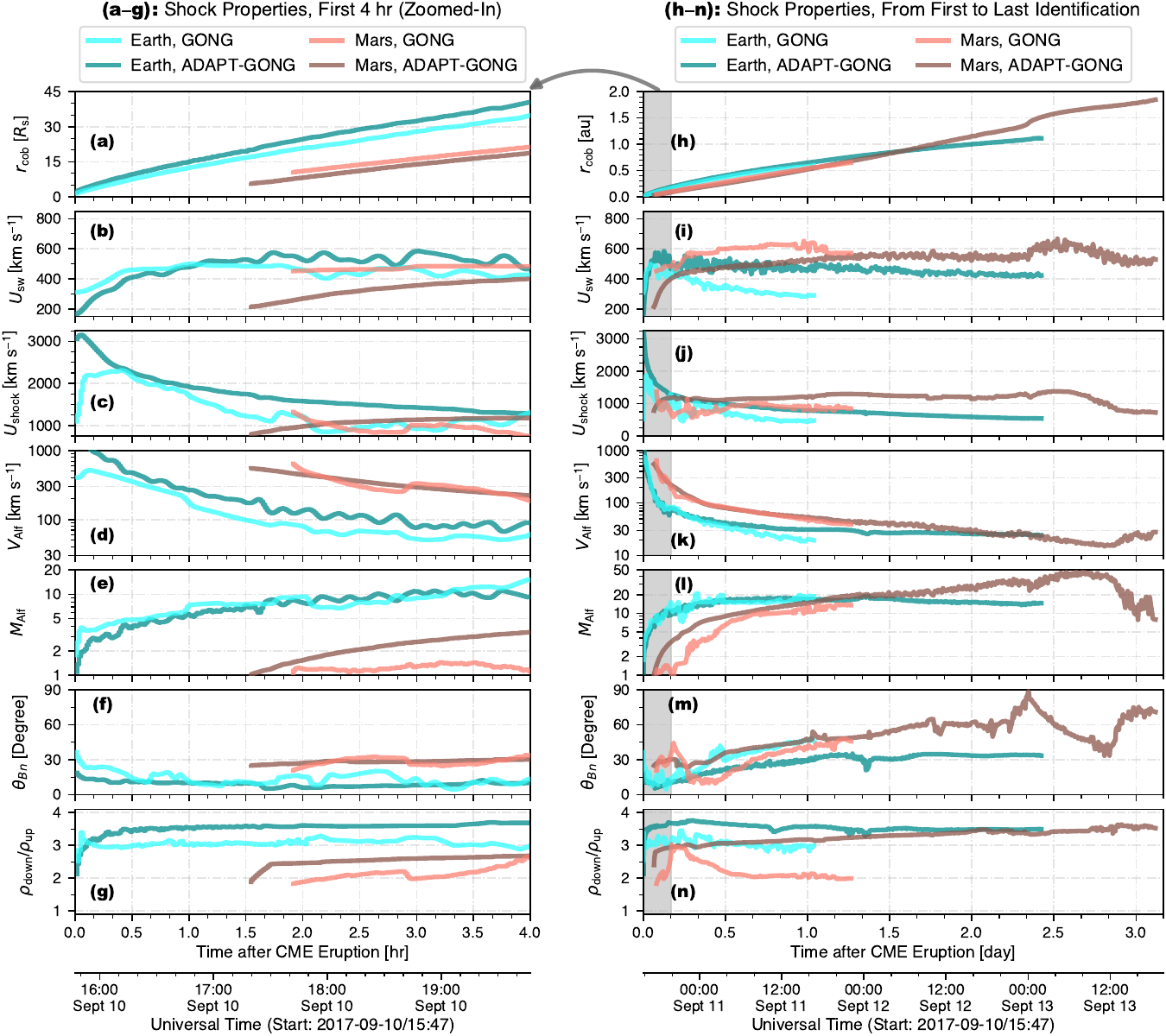}}
    \caption{Time evolution of the shock properties at the cobpoints of Earth (cyan and green) and Mars (red and brown) in the GONG (lighter colors) and ADAPT-GONG (darker colors) runs. Panels (a)--(g) show the properties during the first 4 hours after the CME eruption onset, and panels (h)--(n) show the entire evolution from the time when the shock is first captured until the identification threshold is no longer satisfied, with the gray shading marking the first 4-hour window shown in panels (a)--(g). 
    Rows from top to bottom show the cobpoint heliocentric distance ($r_\mathrm{cob}$), the upstream solar wind speed ($U_\mathrm{sw}$), the shock speed ($U_\mathrm{shock}$), the local Alfv\'{e}n wave speed ($V_\mathrm{Alf}$), the Alfv\'{e}nic Mach number ($M_\mathrm{Alf}$), the shock angle ($\theta_{Bn}$), and the density compression ratio ($\rho_\mathrm{down}/\rho_\mathrm{up}$). Note that the $y$-axis ranges may differ between the left and right panels within the same row.} \label{fig14:shkcob}
    \end{figure*}
    
    We also trace the magnetic field lines in the AWSoM-R solution from Earth, STA, and Mars back to the shock surface to determine their cobpoints, i.e., the points where observer-connected field lines and the shock front intersect, following the concept of ``Connecting with the OBserver Point'' introduced by \cite{heras1995three}. 
    In both runs, we find no direct magnetic connection between STA and the shock, so we focus on the cobpoints of Earth and Mars below. Figure~\ref{fig14:shkcob} shows the time evolution of the shock properties at these cobpoints, from the time the shock is first captured until it weakens below the adopted shock identification threshold. Magnetic connection between Earth and the shock is established within minutes of the eruption in both runs, while the Mars-connected cobpoint appears at $t\approx1.5$ hr in the ADAPT-GONG run and $t \approx 1.9$ hr in the GONG run, when the expanding shock flank reaches the field lines connected to Mars, located $\sim$156$^{\circ}$ west of Earth. We find the following differences in the shock properties between the two runs: 
    \begin{enumerate}
        \item During the first 4 hours (Figures~\ref{fig14:shkcob}(a)--(g)), Earth's cobpoints show overall similar evolution in the two runs, but systematic differences emerge with time. The cobpoint in the ADAPT-GONG-driven run gradually propagates to a slightly larger heliocentric distance and is associated with a higher shock speed, especially after the first $\sim$1--2 hr. 
        The compression ratio also remains systematically higher, at $\gtrsim 3.5$ compared with $\sim$3.0--3.4 in the GONG run. In contrast, the two runs show broadly comparable Alfv\'{e}nic Mach numbers, increasing from 1 to $\sim$10, and both shocks remain quasi-parallel, consistent with the shock properties studied by \cite{liu2019geometry}, \cite{kouloumvakos2020evidence} and \cite{zhu2021shock}. %The upstream proton thermal pressures are also similar between the two runs over the first 4 hours, suggesting comparable seed populations available for particle acceleration based on the formulation in \cite{zank2000particle} and \cite{sokolov2004new}. 
        According to standard diffusive shock acceleration theory, the higher compression ratio in the ADAPT-GONG run leads to a harder particle spectrum \citep[e.g.,][]{drury1983introduction, jones1991plasma}. 
        
        \item At Mars, whose initial photospheric footpoint lies on the far side of the source region (Table~\ref{tab3:fp_comparison}), the difference between the two runs is more pronounced because of differences in the expansion of the CME and its driven shock shown in Figure~\ref{fig11:cmesc}. 
        The Mars-connected cobpoints sample different upstream solar wind conditions, with the cobpoint in the GONG run generally encountering a faster ambient flow (panels (b) and (i)), whereas the corresponding shock fronts initially propagate at similar speeds but later become slower in the ADAPT-GONG run (panels (c) and (j)). %with the GONG run generally encountering a slower ambient flow and the corresponding shock front maintaining a lower speed than in the ADAPT-GONG run (see Figure~\ref{fig5:swsUN}). 
        In the GONG run, $M_\mathrm{Alf}$ remains at only $\lesssim 1.5$ during the first 4 hours, indicating a very marginally super-Alfv\'{e}nic shock, while the compression ratio initially rises from $\lesssim 2$ to $\lesssim 3$ (panel (g)) and later decreases to $\sim$2 (panel (n)). 
        In the ADAPT-GONG run, by contrast, the Alfv\'{e}nic Mach number increases from 1 to $\sim$4 in the first 4 hours, while the compression ratio approaches $\sim$3. 
        The Mars-connected shock is also first identified $\sim$25 min earlier and closer to the Sun, at $r_\mathrm{cob}\sim 5\; R_\mathrm{s}$ compared with $\sim10\; R_\mathrm{s}$ in the GONG run (panel (a)). This earlier connection at a smaller heliocentric distance in the initial stage can provide access to a larger seed population, which, together with the higher Alfv\'{e}nic Mach number, provides considerably more favorable conditions for efficient particle acceleration along Mars-connected field lines during the first few hours in the ADAPT-GONG run. 
        %The upstream proton thermal pressure is also substantially higher when the Mars-connected shock is first captured here and then rapidly decreases as the cobpoint moves outward. Because the seed population is proportional to $P_\mathrm{th}$ in our SEP simulations \citep[e.g.,][]{sokolov2004new, liu2025physics}, this much higher early thermal pressure corresponds to a larger population of seed particles available for acceleration. 
        % Together, the higher Alfv\'{e}nic Mach number and larger early seed population in the ADAPT-GONG run provide considerably more favorable conditions for efficient particle acceleration along the Mars-connected field lines during the first few hours. 
        
        \item During the later evolution (Figures~\ref{fig14:shkcob}(h)--(n)), the two runs diverge further in their shock properties. 
        In the GONG run, the shocks at both the Earth- and Mars-connected cobpoints weaken below the identification threshold within $\sim$1.0--1.3 days, at heliocentric distances of only $\sim$0.7--0.8 au, before reaching either observer (panel (h)). Although interplanetary shocks generally weaken with heliocentric distance \citep[e.g.,][]{salman2020radial, kay2022beyond}, the shock in this run dissipates well inside 1 au, while observations show clear shock arrivals at both planets \citep[see][and Figure~\ref{fig16:profile} below]{guo2018modeling, lee2018observations, schwadron2018update}. 
        In the ADAPT-GONG run, the shock at Earth's cobpoint persists to $\sim$1.1 au at $t\approx2.4$ days and that at Mars' cobpoint continues beyond $\sim$1.9 au at $t\approx3.1$ days (panel (h)). Along both connections, the density compression ratio remains at $\sim$3--3.5 (panel (n)) and the Alfv\'{e}nic Mach number ranges from 10 to 50 (panel (l)). This large $M_\mathrm{Alf}$ value results from the combination of the persistently large shock speed (panel (j)) and the decreasing local Alfv\'{e}n speed (panel (k)), which drops from $\sim150$ km$\;$s$^{-1}$ near the inner boundary of the IH domain to $\lesssim30$ km$\;$s$^{-1}$ farther into the heliosphere. Meanwhile, the shock angle evolves considerably (panel (m)): Earth's cobpoint stays quasi-parallel at $\theta_{Bn}\approx30^{\circ}$, whereas Mars' cobpoint changes from quasi-parallel ($\sim$30$^{\circ}$) to quasi-perpendicular ($\gtrsim 60^{\circ}$) within the first $\sim$2.4 days. 
        At $t\approx2.3$ days, Mars' cobpoint in the ADAPT-GONG run shows a distinct excursion in all parameters, with $\theta_{Bn}$ peaking near 90$^{\circ}$ before returning to $\sim$30$^{\circ}$ and fluctuating, accompanied by transient enhancements in the upstream solar wind speed, shock speed, and Alfv\'{e}nic Mach number. This feature arises as the shock flank propagates into H4 (see Figure~\ref{fig12:cmeih}), where the interaction with the HSS reorients the local shock normal relative to the ambient magnetic field and alters the shock properties sampled at the cobpoint. 
        The ADAPT-GONG run therefore maintains shock conditions favorable for continued particle acceleration along the field lines connected to both observers, with important consequences for the modeled SEP fluxes discussed below. 
    \end{enumerate}
    
\subsection{Particle Acceleration and Propagation} \label{sec5.4:sep}

We next present the SEP results from the two simulations in SOFIE/M-FLAMPA, including the global radial--longitudinal distribution of proton fluxes, the time--intensity profiles, and the energy spectra. 

\subsubsection{Spatial Distribution of SEPs} \label{sec5.4.1:spacesep}

    \begin{figure*}[tp!]
    \centering{
    \includegraphics[width=1.0\textwidth]{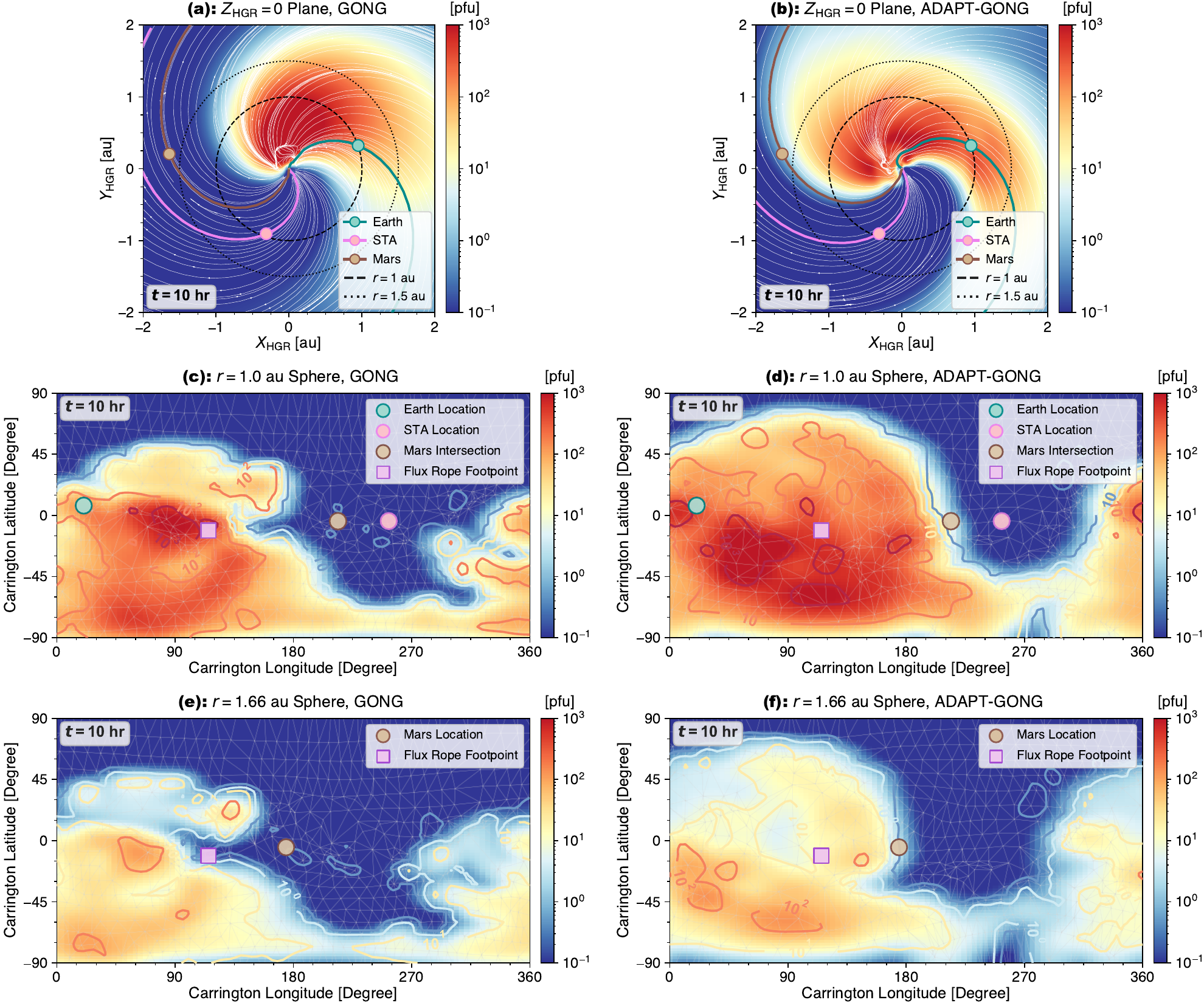}}
    \caption{Modeled $> 10$ MeV proton fluxes at $t = 10$ hr after the CME flux rope is inserted, in the GONG-driven run (left column) and the ADAPT-GONG-driven run (right column). 
    Panels (a) and (b) show the proton fluxes in the solar equatorial plane ($Z_\mathrm{HGR} = 0$). Multiple white curves indicate the magnetic field lines, and the green, pink, and brown curves are those connected to Earth, STA, and Mars, respectively. The black dashed and dotted circles mark $r = 1$ and 1.5 au. 
    Panels (c) and (d) show longitude--latitude maps of the fluxes on the $r = 1.0$ au sphere, with the locations of Earth and STA, the intersection of the Mars-connected field line with the sphere, and the flux rope footpoint in Table~\ref{tab1:param} marked as labeled. Panels (e) and (f) show the same quantities as panels (c) and (d), but on the $r = 1.66$ au sphere with the Mars location and the flux rope footpoint marked. 
    The thin gray vertices in panels (c)--(f) correspond to the intersections of the M-FLAMPA field lines with the spherical shell, and the edges illustrate the triangulation skeleton constructed with the Delaunay approach \cite[see][]{delaunay1934sphere} and used for the interpolation.} \label{fig15:sepcont}
    \end{figure*}
    
    Panels (a) and (b) of Figure~\ref{fig15:sepcont} show the modeled $>10$ MeV proton flux in the solar equatorial plane at $t = 10$ hr after the CME eruption onset for the GONG- and ADAPT-GONG-driven runs, respectively. The white curves indicate multiple magnetic field lines, with some distorted patterns near the Sun associated with the CME flux rope. 
    In both runs, the enhanced fluxes fill the sector of interplanetary magnetic field lines rooted around the source region, and Earth lies within this enhanced sector, magnetically connected to the strong western portion of the shock (see also Figures~\ref{fig12:cmeih}, \ref{fig13:shk3d}, and \ref{fig14:shkcob}(g)). 
    STA, whose footpoint lies far to the east of the source region (see also Table~\ref{tab3:fp_comparison}), remains outside the enhancement in both runs and shows no flux increase in our model. In contrast, STA observed a significant, delayed, and long-lasting SEP enhancement \citep[see Figure~\ref{fig1:overview}(f) and][]{bruno2019spectral}. This discrepancy may partly result from the omission of cross-field diffusion in the present calculations \citep[see Section~\ref{sec3.2:setup} and][]{sokolov2004new}, as this process can transport particles to longitudes poorly connected to the shock \citep[e.g.,][]{giacalone2012longitudinal, gomez2015circumsolar, laitinen2016solar, ding2020modeling, husidic2024cross, lario2026major, young2026quantifying}. 
    Besides, as discussed in Section~\ref{sec5.2.2:cmepropagate}, the preceding eruptions are not included in the present simulations, and their preconditioning of the ambient heliosphere may also affect the magnetic connectivity and SEP distribution 
    % We also note that the two smaller eruptions from the same active region about one day earlier are not included in the simulations. As discussed in Section~\ref{sec5.2.2:cmepropagate}, although we do not expect them to contribute substantially to the seed particle population, they may still precondition the ambient plasma and magnetic field and thereby affect the magnetic connectivity of this CME and its associated shock, as well as the resulting SEP flux distribution 
    % and these preceding events may modify the ambient plasma and magnetic field through which this CME eruption propagates and enhance the seed particle population available for shock acceleration 
    \citep[e.g.,][]{gopalswamy2004intensity, li2012twin, ding2015seed, zhuang2020role, palmerio2021cme}. 
    The two runs differ mainly in the longitudinal extent of the enhanced sector: it is comparatively narrow in the GONG run, leaving the Mars-connected field lines along its weak eastern edge, whereas in the ADAPT-GONG run, even though Mars' initial photospheric footpoint lies at 176.1$^\circ$ west of the source region (Table~\ref{tab3:fp_comparison}), the sector spans a wider range of longitudes due to the expansion of the CME/ICME and ultimately encompasses the Mars-connected field lines. 
    
    Panels (c)--(f) of Figure~\ref{fig15:sepcont} display the corresponding longitude--latitude distributions of $>10$ MeV protons on the $r = 1.0$ and $1.66$ au spheres at the same time. 
    At 1 au, both runs place Earth within the enhanced-flux region eastward of the eruption site, with proton fluxes of the same order of magnitude, while STA remains in the low-flux background. 
    At 1.66 au, however, the two runs diverge substantially. In the GONG run, Mars lies within a low-flux zone separating the enhanced regions, whereas in the ADAPT-GONG run the enhancement extends across the Mars longitude, reaching $\sim$10--$10^2$ pfu compared with only $\lesssim 1$ pfu in the GONG run. The ADAPT-GONG enhancement also extends to higher northern latitudes on both spheres, consistent with the greater high-latitude extent of its shock surface (Figure~\ref{fig13:shk3d}). 
    We also note that near the boundaries of these enhanced-flux regions, the cross-field diffusion and particle drifts, which are not included in the current simulations, could alter particle access to the observer. These effects may therefore modify the global flux distribution, the magnitude of the flux contrast between the two runs, and the detailed time profiles and spectra presented below \citep[see also][]{marsh2013drift, laitinen2013energetic, laitinen2016solar, battarbee2017solar, waterfall2022modeling, moradi2025impact, young2026quantifying}, although these effects are not quantified here and will be examined in future work. 

\subsubsection{Time--Intensity Profiles} \label{sec5.4.2:profile}
    
\begin{figure*}[tp!]
    \centering{
    \includegraphics[width=1.0\textwidth]{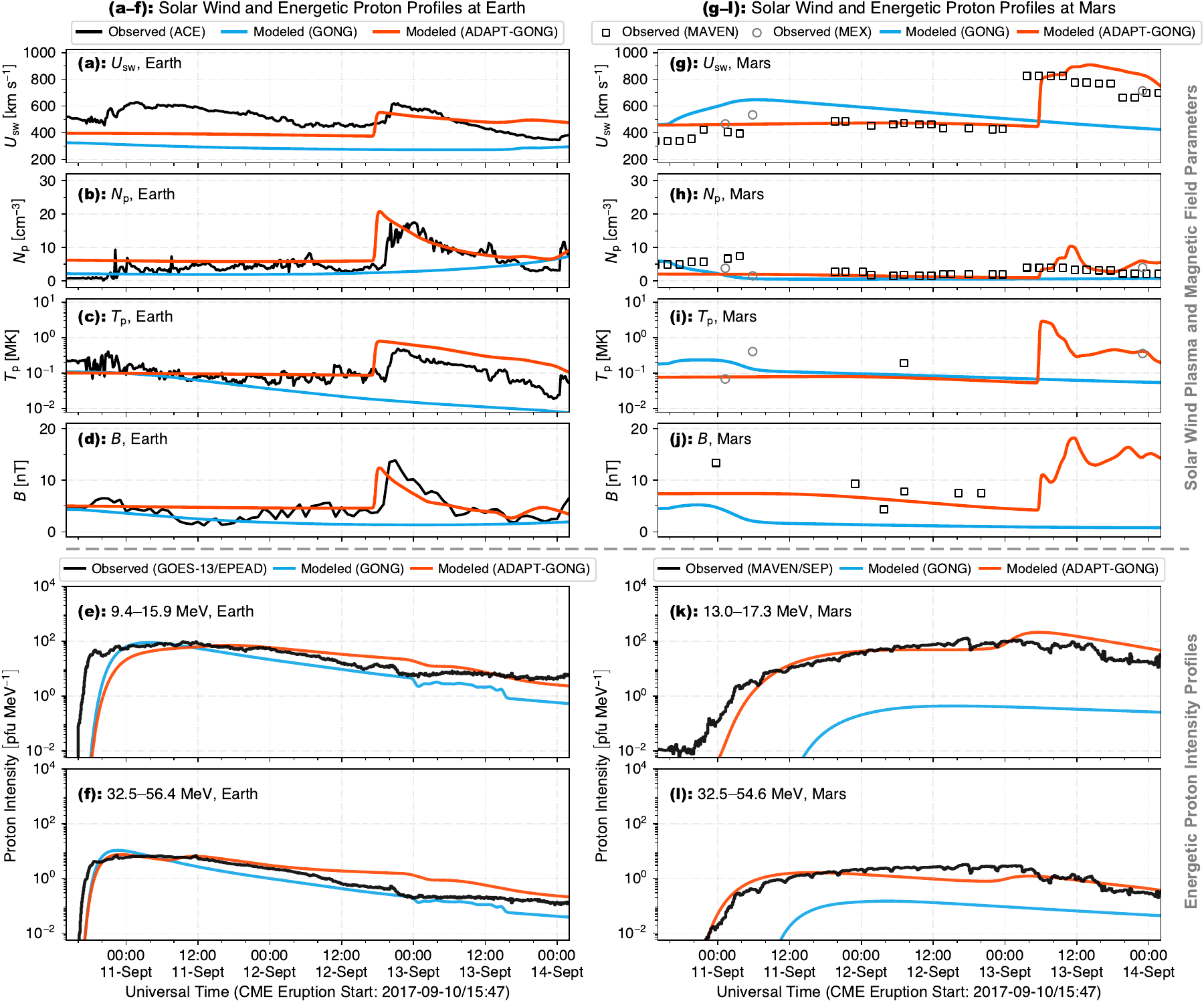}}
    \caption{Modeled and observed time profiles of the solar wind plasma parameters and energetic protons at Earth (left column) and Mars (right column). The top four rows show the solar wind speed, proton number density, proton temperature, and magnetic field strength: black curves show the ACE observations at Earth, black squares and gray circles show the MAVEN and MEX measurements at Mars, and blue and red curves show the GONG- and ADAPT-GONG-driven simulation results, respectively. The bottom two rows show the energetic proton intensities in two representative energy channels, following the same style for the observations (black curve) and the two runs (blue and red curves): 9.4--15.9 MeV and 32.5--56.4 MeV protons observed by GOES-13/EPEAD at Earth, and 13.0--17.3 MeV and 32.5--54.6 MeV protons observed by MAVEN/SEP at Mars.} \label{fig16:profile}
    \end{figure*}

    Figure~\ref{fig16:profile} shows the comparison of the modeled and observed solar wind plasma, magnetic field strength, and energetic proton profiles at Earth and Mars. In the solar wind profiles observed by the Solar Wind Electron, Proton and Alpha Monitor \citep[SWEPAM\footnote{\url{https://izw1.caltech.edu/ACE/ASC/level2/lvl2DATA_SWEPAM.html} \label{url:acesw}};][]{mccomas1998solar} and the Magnetic Field Experiment \citep[MAG\footnote{\url{https://izw1.caltech.edu/ACE/ASC/level2/lvl2DATA_MAG.html} \label{url:acemag}};][]{smith1998ace} on board the Advanced Composition Explorer \citep[ACE;][]{stone1998advanced} at L1, an interplanetary shock arrived at Earth late on September 12 ($\sim$20:00 UT), indicated by the simultaneous jumps in speed, density, temperature, and magnetic field strength, with the ADAPT-GONG run reproducing these signatures $\sim$3 hours earlier than observed. 
    At Mars, the same run also reproduces the observed shock arrival at $\sim$02:30 UT on September 13, with a delay of $\sim$4 hours, when MAVEN recorded a sudden increase of the upstream solar wind speed to $\sim$800 km$\;$s$^{-1}$ \citep{lee2018observations, ramstad2018september}. 
    These opposite timing offsets of the shock arrival at Earth and Mars arise because they encounter different sectors of the shock front (see Figure~\ref{fig12:cmeih}), which undergo different levels of deceleration and deformation as they interact with the ambient solar wind. Discrepancies in the modeled solar wind speed and density along their respective propagation paths can therefore contribute to the slightly earlier shock arrival at Earth and the later arrival at Mars. 
    However, the GONG run shows no distinct signature of shock arrival at either Earth or Mars, consistent with the shock weakening below the adopted identification threshold (Figure~\ref{fig14:shkcob}) and the confined lateral expansion of the ICME during its propagation (Figure~\ref{fig12:cmeih}). 
    
    In Figures~\ref{fig16:profile}(e) and (f), we show two representative proton energy channels at Earth, corresponding to lower and higher energies, respectively. The observations are from the westward-viewing telescope of the Energetic Proton, Electron, and Alpha Detector (EPEAD) on board GOES-13 \citep{onsager1996operational}, with the effective energy values based on the calibration of \cite{sandberg2014cross}. 
    We notice that data from the Energetic and Relativistic Nuclei and Electron \citep[ERNE;][]{torsti1995energetic} experiment on board SOHO are unavailable during this event, while the relatively broad energy bins of SOHO/EPHIN may limit our ability to resolve the energy dependence of the onset, compared with GOES-13/EPEAD. We therefore adopt the GOES-13 SEP measurements with calibrated effective channel energies to analyze the SEP profiles here and the spectra below in Section~\ref{sec5.4.3:spectra}. 
    
    At Earth, as illustrated in Figures~\ref{fig16:profile}(e) and (f), both runs reproduce the observed profiles within the same order of magnitude using the same prescribed injection scaling factor of 300, which is first set based on the observations at Earth and then applied unchanged at Mars. The modeled Mars SEP intensities therefore provide an independent prediction rather than a separately tuned result. 
    % The GONG run rises more promptly and closely matches the observed rapid onset (see also Figure~\ref{fig1:overview}(c)) while overestimating the onset peak by a factor of $\sim$2--3. The ADAPT-GONG run rises more gradually over a few hours and then tracks the observed intensities and decay. 
    The GONG run rises more steeply than the ADAPT-GONG run and comes closer to the observed rapid onset (see also Figure~\ref{fig1:overview}(c)), although both runs begin rising $\sim$1--2 hours later than observed. The GONG run matches the observed peak at 9.4--15.9~MeV and exceeds it by a factor of $\sim$2 at 32.5--56.4~MeV, but then decays faster, falling below the observations by a factor of $\gtrsim$2 after $t \approx 24$~hr. The ADAPT-GONG run stays below the observations for the first $\sim$24~hr at 9.4--15.9~MeV and tracks them within a factor of $\sim$1.5 until $t \approx 36$~hr, but then decays more slowly and exceeds the observations by a factor of $\sim$2--3 around the shock arrival. 
    We also note that, during the rapidly evolving onset phase, the lower-energy GOES channels (e.g., panel (e)) may contain non-negligible contributions from higher-energy protons owing to the energy response of the detector, which can lead to an apparently earlier onset that does not necessarily reflect the actual onset of protons \citep[e.g.,][]{rodriguez2014intercalibration, rodriguez2017validation, jiggens2019situ, kress2021observations, liu2024AGU}. 
    
    At Mars, Figures~\ref{fig16:profile}(k) and (l) show two proton energy channels of MAVEN/SEP with energies similar to those shown at Earth; however, the two runs differ drastically. 
    The ADAPT-GONG run reproduces the observed onset about 4--8 hours after the CME eruption onset, followed by the subsequent gradual rise and plateau within a factor of $\sim$2--3 during the ESP phase, even though the pre-eruption photospheric footpoint of Mars lies on the far side of the source region (see also Figure~\ref{fig2:map} and Table~\ref{tab3:fp_comparison}). This run also produces a modest SEP enhancement around the ICME shock arrival at $\sim$04:00 UT on September 13, consistent with additional local acceleration near the shock. A smaller enhancement appears shortly before the shock arrival, around 00:00 UT on September 13, due to a combination of local acceleration near the approaching ICME-driven shock and modification of the shock properties during its interaction with the preceding SIR (see Figure~\ref{fig14:shkcob}). 
    In contrast, in the GONG run, SEP flux onset at Mars is delayed by more than 12 hours and the peak intensities are underestimated by $\sim$1--2 orders of magnitude, as a result of its later magnetic connection, much smaller early injected seed population, and much weaker flank shock along the Mars-connected field lines, followed by further weakening during propagation (Sections~\ref{sec5.2.2:cmepropagate} and \ref{sec5.3:shock}). 
    We also note that the MAVEN/SEP proton fluxes above 13 MeV shown here and later in Figure~\ref{fig17:specmat} are derived from the penetrating proton counts measured by MAVEN/SEP using reduced chi-squared analysis to determine the best-fit spectra \citep[see the description in Section~\ref{sec2:overview} and][]{lee2023AGUobservations, lee2024AGUobservations}, and therefore carry uncertainties. 
    Taken together, these results show that the choice of the input magnetogram for SOFIE, for the proton energy channels investigated here (Figures~\ref{fig15:sepcont} and \ref{fig16:profile}(e)(f)(k)(l)), has a relatively modest effect where magnetic connectivity is favorable (e.g., Earth in this event), but a much stronger influence where the connectivity is poor or limited to the shock flank (e.g., Mars in this event). 

\subsubsection{Energy Spectra} \label{sec5.4.3:spectra}
    
\begin{figure*}[tp!]
    \centering{
    \includegraphics[width=0.5\textwidth]{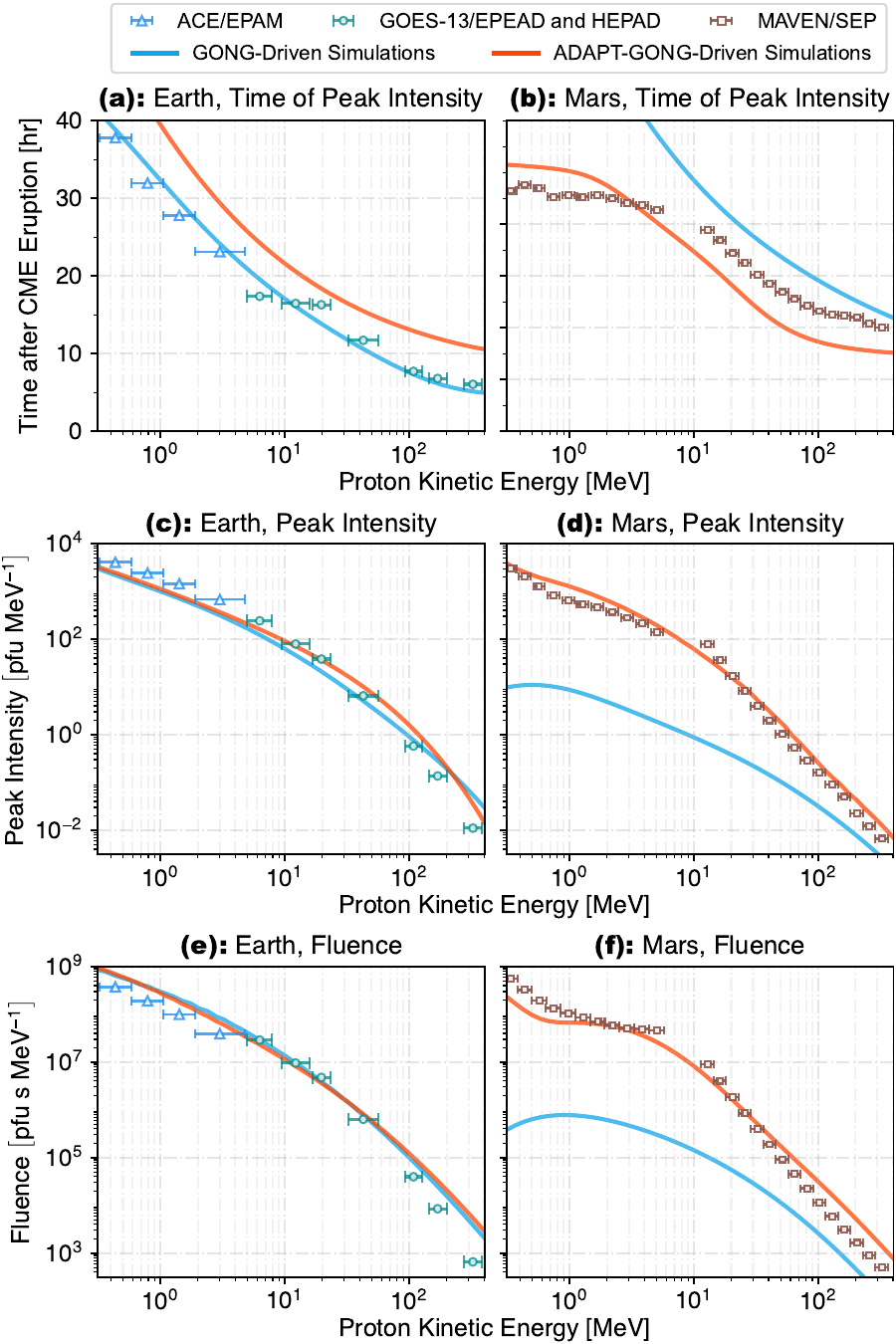}}
    \caption{Energy-dependent comparison of the M-FLAMPA simulations with energetic proton observations at Earth (left column) and Mars (right column). Panels (a)--(b), (c)--(d), and (e)--(f) show the time of peak intensity after the CME eruption, peak intensity, and time-integrated fluence, respectively, as functions of the proton kinetic energy ranging from $\sim$0.4 MeV to $\sim$400 MeV. Blue and red curves represent the GONG- and ADAPT-GONG-driven simulations, respectively. Observational data are shown by symbols: ACE/EPAM/LEMS120 (blue triangles) and GOES-13/EPEAD and HEPAD (green circles) at Earth, and MAVEN/SEP (brown squares) at Mars, with horizontal bars indicating the corresponding energy-bin widths.} \label{fig17:specmat}
    \end{figure*}

    We further compare the modeled and observed energy dependence of the peak time, peak intensity, and time-integrated fluence at both Earth and Mars, as shown in Figure~\ref{fig17:specmat}. The energetic proton observations at Earth combine the lower-energy measurements from the Low-Energy Magnetic Spectrometer 120 (LEMS120) of the Electron, Proton, and Alpha Monitor \citep[EPAM\footnote{\url{https://izw1.caltech.edu/ACE/ASC/level2/lvl2DATA_EPAM.html} \label{url:aceepam}};][]{gold1998electron} on board ACE with the higher-energy measurements from the westward-viewing telescope of EPEAD and the High Energy Proton and Alpha Detector (HEPAD) on board GOES-13 \citep{onsager1996operational}. For the GOES-13 data\footnote{The original data are accessible at \url{https://www.ngdc.noaa.gov/stp/satellite/goes/dataaccess.html}. \label{url:goessep}}, we adopt the effective energy ranges calibrated by \cite{sandberg2014cross} below 80 MeV and by \cite{bruno2017calibration} above 80 MeV, with the background quiet-time pre-event contribution subtracted. The observed peak time decreases with increasing energy at both locations, as higher-energy particles arrive and maximize earlier. Consequently, the peak-intensity spectrum is constructed from the maximum intensity attained in each energy channel at different times. For the fluence data, we integrate the particle intensity over a 4-day window following the CME eruption. 
    
    At Earth, where the field lines are well connected to the shock nose (Figures~\ref{fig13:shk3d} and \ref{fig14:shkcob}), the two runs differ only modestly and the GONG-driven simulation is slightly better. As shown in panel (a), the GONG run reproduces the observed energy dependence of the peak time over the full observed range, whereas the ADAPT-GONG run peaks 4--8 hours later. %, as also seen in the plateaus in Figures~\ref{fig16:profile}(e)--(f) reflecting the evolving shock connectivity and particle acceleration efficiency. 
    This delay is not due to a later onset, since the SEP flux starts to rise at nearly the same time in both runs (Figures~\ref{fig16:profile}(e)--(f)). Instead, the GONG run reaches a relatively sharp peak and then decays, while the ADAPT-GONG run rises into a broad plateau lasting $\sim$12 hours. This is because the ADAPT-GONG shock at Earth's cobpoint continuously maintains a higher compression ratio and shock speed and remains identified for $\sim$2.4 days, compared with $\sim$1 day in the GONG run (Figure~\ref{fig14:shkcob}). The continuous acceleration at the shock keeps injecting particles onto Earth's field line, so the particle intensity stays close to its maximum for longer and peaks later. 
    In panel (c), both runs reproduce the observed spectral shape and stay within a factor of a few of the ACE/EPAM and GOES-13 measurements from $\sim$0.4 MeV to $\sim$400 MeV, with the GONG run lying slightly higher and closer to the observations below $\sim$30 MeV, above which the ADAPT-GONG spectrum steepens more rapidly and aligns better. 
    The two fluence spectra in panel (e) are nearly indistinguishable and agree with observations to within a factor of $\lesssim 3$ in the range of $\sim$0.4--400 MeV. 
    
    At Mars, where the particles arrive through the evolving shock flank (Figure~\ref{fig12:cmeih}), the two runs instead diverge sharply. 
    In panel (b), the ADAPT-GONG run matches the observed peak times to within a few hours, running slightly late below $\sim$5 MeV and slightly early above $\sim$20 MeV, whereas the GONG run peaks more than 10 hours late at low and intermediate energies and approaches the observations only above $\sim$20 MeV. 
    In panel (d), the ADAPT-GONG run likewise follows both the magnitude and slope of the MAVEN/SEP peak-intensity spectrum, whereas the weak and briefly connected shock flank in the GONG run results in intensities 1--2 orders of magnitude lower over most of the energy range. 
    A similar contrast appears in the fluence spectrum shown in panel (f), where the GONG run falls short by up to $\sim$2--3 orders of magnitude at the lowest energies. The ADAPT-GONG run therefore reproduces not only the peak intensities at Mars but also the 4-day fluence, including the plateau at $\sim$1--4 MeV, which we find is primarily associated with the enhanced proton intensities during the ESP phase.

\section{Summary and Conclusions} \label{sec6:summary}

\begin{table*}[tp!]
\begin{center}
\caption{Comparison of the GONG and ADAPT-GONG simulations for the 2017 September 10 event.} \label{tab4:run_comparison}
\setlength\tabcolsep{3.5pt}{
\fontsize{9}{9}\selectfont
\renewcommand{\arraystretch}{1.025}
\begin{tabular}{llll}
\toprule

\colhead{Property} & \colhead{GONG-Driven Run} & \colhead{ADAPT-GONG-Driven Run} & \colhead{Compared with Observations} \\
\midrule

\multicolumn{4}{c}{\textbf{Ambient corona and solar wind}} \\ 
\multirow{2}{*}{\shortstack[l]{\textbf{Coronal streamer}\\ (Figure~\ref{fig4:bgwl})}} &
\multirow{2}{*}{\shortstack[l]{Flatter, with an extra faint\\ northeastern streamer}} &
\multirow{2}{*}{\shortstack[l]{More inclined and smoother}} &
\multirow{2}{*}{\shortstack[l]{Both comparable; extra streamer\\ in GONG run not clearly observed}} \\ \\
\cdashline{1-4}

\multirow{3}{*}{\shortstack[l]{\textbf{Solar wind: 27-day}\\ \textbf{interval}
                (Figure~\ref{fig8:swstime};\\ Table~\ref{tab2:distsw})}} &
\multirow{3}{*}{\shortstack[l]{Smaller mean data distance\\ 
                values at all three observers;\\ closer solar wind speeds}} &
\multirow{3}{*}{\shortstack[l]{Closer magnetic field\\strengths at Earth and STA}} &
\multirow{3}{*}{\shortstack[l]{The GONG run better overall}} \\ \\ \\
\cdashline{1-4}

\multirow{3}{*}{\shortstack[l]{\textbf{Solar wind: source AR-}\\\textbf{centered sub-interval} \\ (Figure~\ref{fig8:swstime}; Table~\ref{tab2:distsw})}} &
\multirow{3}{*}{\shortstack[l]{Smaller mean data distance\\ value at STA}} &
\multirow{3}{*}{\shortstack[l]{Smaller mean data distance\\ values at Earth and Mars}} &
\multirow{3}{*}{\shortstack[l]{The ADAPT-GONG run\\ slightly better overall}} \\ \\ \\
\cdashline{1-4}

\multirow{3}{*}{\shortstack[l]{\textbf{Photospheric}\\ \textbf{footpoints}\\
                (Table~\ref{tab3:fp_comparison})}} &
\multirow{3}{*}{\shortstack[l]{Earth: Close to the source\\ region ($16.3^\circ$ eastward);\\ STA and Mars: far side}} &
\multirow{3}{*}{\shortstack[l]{Earth: Close to the source\\ region ($7.5^\circ$ eastward);\\ STA and Mars: far side}} &
\multirow{3}{*}{\shortstack[l]{Both qualitatively consistent with\\
                prompt SEP onset at Earth and \\
                delayed onsets at STA and Mars}} \\ \\ \\
\midrule

\multicolumn{4}{c}{\textbf{CME and shock evolution}} \\ 
\multirow{3}{*}{\shortstack[l]{\textbf{Coronagraph}\\ \textbf{morphology}\\ (Figure~\ref{fig9:wl})}} &
\multirow{3}{*}{\shortstack[l]{Smaller initial angular width;\\ 
                less evident southern-flank\\enhancement later}} &
\multirow{3}{*}{\shortstack[l]{Broader expansion; more\\ 
                evident southern-flank\\enhancement later}} &
\multirow{3}{*}{\shortstack[l]{GONG run closer in initial width;\\ 
                ADAPT-GONG run closer for the\\later southern flank and expansion}} \\ \\ \\
\cdashline{1-4}

\multirow{3}{*}{\shortstack[l]{\textbf{Leading-edge}\\ \textbf{kinematics}\\(Figure~\ref{fig10:cmeh})}} &
\multirow{3}{*}{\shortstack[l]{Average speed $\sim$10\%\\below observations}} &
\multirow{3}{*}{\shortstack[l]{Average speed $\sim$5--7\%\\below observations}} &
\multirow{3}{*}{\shortstack[l]{ADAPT-GONG run closer to\\ observed CME heights and speeds\\ in views from Earth and STA}} \\ \\ \\
\cdashline{1-4}

\multirow{4}{*}{\shortstack[l]{\textbf{CME propagation}\\ \textbf{and shock arrivals}\\ 
                (Figures~\ref{fig11:cmesc}, \ref{fig12:cmeih}, and \\ \ref{fig16:profile}(a)--(d), (g)--(j))}} &
\multirow{4}{*}{\shortstack[l]{Confined expansion shaped\\ by nearby streams; no clear\\ 
                signature of shock arrival\\ at either observer}} &
\multirow{4}{*}{\shortstack[l]{Broader expansion with\\ CME--HSS interactions;\\ 
                stronger shock persists\\ beyond Earth and Mars}} &
\multirow{4}{*}{\shortstack[l]{ADAPT-GONG run reproduces\\ shock arrivals at Earth and\\ 
                Mars, with a timing error of\\ $\sim$3--4 hr; GONG run does not}} \\ \\ \\ \\
\cdashline{1-4}

\multirow{7}{*}{\shortstack[l]{\textbf{Magnetic}\\ \textbf{connectivity}\\
                (Figures~\ref{fig13:shk3d} and \ref{fig14:shkcob})}} &
\multirow{7}{*}{\shortstack[l]{Earth: Prompt, compression\\ 
                ratio $\sim$3.0--3.4; connection\\ persists to $\sim$0.7 au;\\ 
                Mars: Later, weak flank\\ 
                ($M_{\rm Alf}\lesssim1.5$); connection\\ persists to $\sim$0.7--0.8 au;\\ 
                STA: No connection}} &
\multirow{7}{*}{\shortstack[l]{Earth: Prompt, compression\\ 
                ratio $\sim$3.5; connection\\ persists to $\sim$1.1 au;\\ 
                Mars: $\sim$1.5~hr, stronger flank\\ 
                ($M_{\rm Alf}\sim4$); connection\\ persists beyond $\sim$1.9 au;\\ 
                STA: No connection}} &
\multirow{7}{*}{\shortstack[l]{No direct shock property\\ observables;\\ 
                ADAPT-GONG run more\\ 
                consistent with shock conditions\\ 
                favorable for particle acceleration\\ 
                over a broader range of longitudes\\ 
                and latitudes}} \\ \\ \\ \\ \\ \\ \\
\midrule

\multicolumn{4}{c}{\textbf{SEP distributions, profiles, and spectra}} \\[2pt]
\multirow{4}{*}{\shortstack[l]{\textbf{Global distribution}\\ \textbf{of SEP fluxes}\\ (Figure~\ref{fig15:sepcont})}} &
\multirow{4}{*}{\shortstack[l]{More confined enhanced-\\flux region; much weaker\\
                SEP fluxes around Mars}} &
\multirow{4}{*}{\shortstack[l]{Broader enhanced-flux region\\
                around both Earth and Mars}} &
\multirow{4}{*}{\shortstack[l]{No observed global SEP flux map;\\
                ADAPT-GONG run more\\
                consistent with SEP observations\\
                at both Earth and Mars}} \\ \\ \\ \\
\cdashline{1-4}

\multirow{6}{*}{\shortstack[l]{\textbf{Time--intensity}\\ \textbf{profiles}\\ 
                (Figures~\ref{fig16:profile}(e), (f),\\ (k), and (l))}} &
\multirow{6}{*}{\shortstack[l]{Earth: Steeper rise;\\ 
                peak $\sim$2$\times$ high,\\ then faster decay;\\ 
                Mars: Onset $>12$~hr late,\\ with fluxes much lower}} &
\multirow{6}{*}{\shortstack[l]{Earth: Prompt rise; flux\\ 
                within $\sim$1.5$\times$ until $\sim$36~hr,\\ then $\sim$2--3$\times$ high later;\\ 
                Mars: Reproduces the onset,\\ peak, decay, and ESP phases}} &
\multirow{6}{*}{\shortstack[l]{Earth: The GONG run closer\\  at onset and late decay, the\\ ADAPT-GONG run in the\\ intermediate phase;\\ Mars: The ADAPT-GONG run\\ closer; STA: Neither reproduces}} \\ \\ \\ \\ \\ \\
\cdashline{1-4}

\multirow{5}{*}{\shortstack[l]{\textbf{Peak flux and}\\ 
                \textbf{fluence spectra}\\ (Figure~\ref{fig17:specmat})}} &
\multirow{5}{*}{\shortstack[l]{Earth: Within a factor of\\ $\sim$3 of observations;\\ 
                Mars: Substantially\\ underestimated,\\ especially at low energies}} &
\multirow{5}{*}{\shortstack[l]{Earth: Within a factor of\\ $\sim$3 of observations;\\ 
                Mars: Peak and fluence\\ spectra overall reproduced}} &
\multirow{5}{*}{\shortstack[l]{Earth: Comparable peak flux\\ and fluence spectra in both runs;\\ 
                Mars: The ADAPT-GONG run\\ closer in both amplitude and\\ spectral shape}} \\ \\ \\ \\ \\

\bottomrule
\end{tabular}}
\end{center}
\tablecomments{The comparisons are specific to this event and the selected GONG and ADAPT-GONG input magnetic field maps. Relevant figures and/or tables are included for each item; see the corresponding discussion in the main text for details.}
\end{table*}

We investigate how the choice of photospheric magnetic field map affects physics-based simulations of the solar wind, CME, and SEPs from the Sun to Earth and Mars, using the 2017 September 10 event as a case study. This event was associated with an X8.2-class flare and a fast halo CME from AR 12673 and produced GLE \#72, the first GLE recorded on the surfaces of both Earth and Mars. 
We perform two simulations within the SOFIE model suite, one driven by a GONG synoptic magnetogram and the other by an ADAPT-GONG map taken at nearly the same time. We keep the total Alfv\'{e}n wave energy injection identical in these two runs by adjusting the input Poynting flux parameter for AWSoM-R, and use EEGGL to derive map-dependent flux rope parameters from the corresponding source region magnetic fields, while keeping all other model settings the same. M-FLAMPA is then used to model particle acceleration and transport along the evolving magnetic field lines. The two simulations follow the event from the low corona to beyond the orbit of Mars, enabling us to trace how differences in the photospheric magnetic field propagate through the ambient solar wind, CME and shock evolution, and ultimately the SEP distributions. Table~\ref{tab4:run_comparison} summarizes the main differences between the two simulations and their comparisons with observations, suggesting that neither magnetogram performs uniformly better across all quantities and locations, with their relative performance depending on the aspect considered. 

Although the two maps have similar total unsigned magnetic flux, differing by only $\sim$2.6\%, they distribute the magnetic flux differently, especially in the rapidly evolving source region AR 12673, the far-side ARs, and the high-latitude regions. These differences produce distinct streamer-belt and HCS geometries, as well as different HSS structures in the SC and IH domains. 
The HSS structures are shaped by the open flux regions, magnetic connectivity, footpoint field strengths, and flux tube expansions in each SC solution. In the ADAPT-GONG run, the HCS rises to $\sim$45$^{\circ}$N near the longitudes of AR~12673. The open flux tubes rooted in regions of strong magnetic field next to this AR have large expansion factors ($\gtrsim$100) and fill a $\sim$135$^{\circ}$ sector in longitude. This produces a slower and wider stream (H1) ahead of the CME source region. In the GONG run, the HCS stays within $\sim$20$^{\circ}$ of the equator, and H1 is faster and narrower. These streams form below 21.5~$R_\mathrm{s}$, extend into interplanetary space, and provide different background conditions for CME propagation and SEP acceleration and transport. 

We also derive the photospheric footpoints of the field lines passing through Earth, STA, and Mars. Despite the differences in the background structures, both runs show a consistent magnetic connectivity pattern, with Earth connected close to the source region and STA and Mars connected to its far side. 
At these three observer locations, we further compare the background solar wind solutions with the \textit{in situ} observations. Both solutions are also broadly comparable to the observed solar wind at Earth, STA, and Mars. Over the full 27-day interval, the GONG-driven solution shows overall better agreement at all three observers, whereas during the source AR-centered sub-interval, the relative performance depends on the specific observer, with the ADAPT-GONG run performing slightly better at Earth and Mars and the GONG run at STA. 

The differences become more noticeable after the CME eruption. The map-dependent source region fields produce different analytical flux rope parameters, which then evolve within their respective structured solar wind backgrounds. For the same observed CME speed, EEGGL provides a compact, strong-field flux rope for the GONG map and a larger, weaker flux rope for the ADAPT-GONG map. In the latter case, however, the initial magnetic energy is $\sim$2.4 times that in the GONG run because of the difference in the magnetic field of the active region and the flux rope radius. 
In the GONG-driven simulation, the CME undergoes more confined lateral expansion and its subsequent interaction with the ambient solar wind channels the disturbance along an SIR ahead of H1, where the drag is strong. The associated shock substantially weakens and falls below the identification threshold at $\sim$0.7--0.8 au, before reaching either Earth or Mars. 
In contrast, the ADAPT-GONG-driven CME expands over a broader longitudinal range and drives a stronger shock that survives to larger heliocentric distances, remaining connected to Earth out to $\sim$1.1 au and to Mars beyond $\sim$1.9 au. This is because its flux rope is more energetic and H1 is slower and wider, with the surrounding slow, dense wind lying farther from the CME axis. The flanks cross this wind within the first hour and run into the faster H2 and H4, so the drag is small and the ICME overtakes H1 over a broad range of longitudes. This CME/ICME evolution is also more consistent with the observed interplanetary shock arrivals at both Earth and Mars, with absolute errors of $\sim$3--4 hours in the modeled shock arrival times. 
These differences in the evolving shock also propagate directly into the SEP distribution function simulations. At Earth, with favorable magnetic connectivity to the shock in both runs, SEP profiles and energy spectra are reproduced reasonably well, with the GONG run capturing the prompt onset more closely. 
At Mars, where the particles are accelerated primarily at the evolving shock flank and subsequently transported to the observer, the simulation results are much more sensitive to the choice of input magnetic field map. The ADAPT-GONG run reproduces the observed SEP timing, intensities, and the 4-day fluence reasonably well, including the enhancement during the ESP phase, whereas the GONG run yields a substantially delayed onset and much lower intensities, and entirely misses the ESP enhancement. 
Together, these differences show that the choice of input map changes both the ambient solar wind and the CME flux rope, since EEGGL parameterizes the flux rope based on each map. Both then shape the shock evolution and magnetic connectivity, so the impact of the input magnetogram varies substantially among different aspects of the CME--SEP simulations and across heliospheric locations. 

It should be noted that these results do not establish a general ranking between the GONG and ADAPT-GONG magnetic field maps in solar wind--CME--SEP simulations, as this work is based on one single event and one realization of each map product. Rather, they demonstrate that differences in the input magnetogram can propagate through the coupled modeling chain of the ambient solar wind, CME, shock, and SEP evolution, and that SOFIE can capture this evolution from the low corona to Mars and beyond. 
For space weather prediction, these results suggest that the treatment of the photospheric magnetic field can affect multiple stages of this modeling chain and, consequently, CME and SEP predictions at different heliospheric locations. In particular, good agreement with the observed ambient solar wind does not necessarily translate into better performance in predicting the CME and SEP intensity evolutions, since this involves complex interactions across multiple stages of the modeling chain, as demonstrated in this study. Future work will extend this analysis to a broader set of events and heliospheric configurations to assess whether the differences found here are systematic and whether particular map products lead to more reliable CME and SEP predictions.

\begin{acknowledgments}
The authors express their sincere gratitude to the anonymous reviewer for the time and effort devoted to improving this paper. 
This work is supported by the NASA Space Weather Center of Excellence program under award 80NSSC23M0191 (CLEAR), the NASA Living With a Star (LWS) Strategic Capability project under NASA grant 80NSSC22K0892 (SCEPTER), and the NSF ANSWERS grant GEO-2149771. 
In addition, N. Sachdeva acknowledges the NASA LWS grant 80NSSC24K1104; C. Lee and P. Dunn acknowledge support from the NASA LWS grant 80NSSC21K1325 and the NASA Mars Data Analysis Program grant 80NSSC19K1224; and A. Bruno acknowledges support from NASA under award 80GSFC24M0006. 
The authors express their sincere gratitude to the SOHO (a project of international cooperation between ESA and NASA; footnote~\ref{url:soho}), SDO/AIA (footnote~\ref{url:aia}), Wind (footnote~\ref{url:wind}), and GOES (footnote~\ref{url:goessep}) teams; the STA/EUVI (footnote~\ref{url:vso}), COR1, COR2, LET (footnote~\ref{url:stalet}), and HET (footnote~\ref{url:stahet}) instrument teams; the ACE/SWEPAM (footnote~\ref{url:acesw}), MAG (footnote~\ref{url:acemag}), and EPAM (footnote~\ref{url:aceepam}) instrument teams; the MAVEN/SWIA, MAG, and SEP instrument teams (footnotes~\ref{url:mavensep}, \ref{url:maven}, and \ref{url:mavensw}); the MEX/ASPERA-3 team (footnotes~\ref{url:aspera} and \ref{url:mex}); and the MSL/RAD team for providing valuable observational data. 
The authors acknowledge the use of the GONG synoptic maps (footnote~\ref{url:gong}), the ADAPT-GONG maps (footnote~\ref{url:adaptgong}), the SWPC-CAT (footnote~\ref{url:swpccat}), the ISWA system (footnote~\ref{url:iswa}), the GLE list (footnote~\ref{url:gle}), the CDAW SEP event list (footnote~\ref{url:cdawsep}) and CME catalog (footnote~\ref{url:cdawcme}), the CACTus CME list (footnote~\ref{url:cactus}), the SolarSoft flare locator (footnote~\ref{url:solarsoft}), the DONKI database (footnote~\ref{url:donki}), the neutron monitor database (footnote~\ref{url:nmdb}), and the ICME catalogs of Wind (footnote~\ref{url:icmeearth}), STA (footnote~\ref{url:icmesta}), and ICMECAT (footnote~\ref{url:helio4cast}). 
The authors also thank the VSO project (footnote~\ref{url:vso}) at NSO for facilitating access to the solar and heliospheric data. 
Computational resources supporting this work are provided by the NASA High-End Computing (HEC) Program through the NASA Advanced Supercomputing (NAS) Division at the Ames Research Center\footnote{\url{https://www.nas.nasa.gov/hecc/}}, and by the high-performance computing support from the Texas Advanced Computing Center (TACC) Frontera\footnote{\url{https://tacc.utexas.edu/systems/frontera/}} at the University of Texas at Austin \citep{stanzione2020frontera}. 
\end{acknowledgments}

\bibliography{ref_paper}
\bibliographystyle{aasjournalv7.1}

\end{document}